\documentclass[a4paper,reqno]{amsart}  

\usepackage[T1]{fontenc}         
\usepackage[applemac]{inputenc} 
\usepackage{amsthm}
\usepackage{amsmath}
\usepackage{amsfonts}
\usepackage{amssymb}
\usepackage[arrow, matrix, curve]{xy}
\usepackage{picinpar, graphicx}
\usepackage{geometry}
\usepackage{epic}
\usepackage{color}
\usepackage{multicol}
\usepackage{multirow}
\usepackage{wasysym}
\usepackage{url}
\usepackage{float}

\newtheorem{Def}{Definition}

\newtheorem{Prop}{Proposition}

\theoremstyle{remark}
\newtheorem{Rem}{Remark}
\newtheorem{Ex}{Example}

\newcommand{\mean}{\mathrm{mean}}
\newcommand{\med}{\mathrm{median}}

\begin{document}

\renewcommand{\thefootnote}{}

\title[Variations on 2-parameter families of  forecasting functions and applications]{Variations on two-parameter families of  forecasting functions: seasonal/nonseasonal Models, comparison to the exponential smoothing and ARIMA models, and applications to stock market data}

\author{Nabil Kahouadji}
\date{}

\begin{abstract}  We introduce twenty four two-parameter families of advanced time series forecasting functions using a new and nonparametric approach. We also introduce the concept of powering and derive nonseasonal and seasonal models with examples in education, sales, finance and economy. We compare the performance of our twenty four models to both Holt--Winters and ARIMA models for both nonseasonal and seasonal times series. We show in particular that our models not only do not require a decomposition of a seasonal time series into trend, seasonal and random components, but leads also to substantially lower sum of absolute error and a higher number of closer forecasts than both Holt--Winters and ARIMA models. Finally, we apply and compare the performance of our twenty four models using five-year stock market data of 467 companies of the S\&P500. \\

\end{abstract}

\maketitle

\section{Introduction} Time series forecasting is an important, required and a common task in business to help inform future decision making and planning. There is a wide range of time series forecasting methods, often developed within specific disciplines for specific purposes, each of which has its own level of performance and cost. Exponential smoothing \cite{B, H, W} and ARIMA (Auto Regressive Integrated Moving Average) \cite{AP} methods are the two most widely used approaches to time series forecasting, and provide complementary approaches to the problem. While exponential smoothing models are based on a description of the trend and seasonality in the data, ARIMA models aim to describe the autocorrelations in the data \cite{HA}. For seasonal time series forecasting, both the exponential smoothing and ARIMA models require a decomposition of the seasonal time series into a trend, seasonal and random components. We introduce in this paper a new and different approach/method to advanced time series forecasting: given a time series $X$ and a time series $Y$ of the same size $n$, we define twenty four two-parameter functions that output the forecasted/predicted $(n+1)$ data entry. These functions are then used as building blocs for various models for both nonseasonal and seasonal time series. Our approach is nonparametric and rely on the computations of estimates via three methods,  and then choosing one of the estimate via a series of optimization processes using eight different optimization criteria. The functions, models, examples and applications are introduced and covered in the next five sections.   In section 2, we define both the rate of interest and the rate of discount times series of a given time series $Y$, both of which are used to define $\sharp-$ and $\flat$- mean and median forecast estimates of a given order for a given time series $X$. We then introduce eight optimization criteria $\kappa = 1, \dots, 8$, which allow us to choose a preferred maximum order and a preferred optimization length, both of which  are used to forecast the next (unknown) observation for the time series $X$. Combining both the mean and median forecasts enables us to define the $\sharp$- and $\flat$-forecasts, and then combining the $\sharp$ and $\flat$  estimates enables us to define the $\natural$-forecast, leading to twenty four new two-parameter families of forecasting functions.  We illustrate in Example 1 all the introduced functions and objects,  and we compute all the different types of estimates and forecasts.  In section 3, we introduce non-seasonal balanced forecasting models by suggesting a maximum order and a length for the twenty four advanced forecasting functions. Moreover, we introduce a concept of powering by considering a power of the time series $Y$ used to derive the rate of interest and the rate of discount times series. The latter is used to define two optimized models: a non-seasonal mean-optimized sum of absolute error balanced $\alpha$-power, and a non-seasonal latest-optimized sum of absolute error balanced $\alpha$-power models.  In Example 2, we consider nine time series of total fall enrollment in postsecondary institutions, and we compute for each time series the sequence of forecasts using all twenty four balanced forecasting functions, and we compare the results and functions using the sum of absolute errors. In Example 3, we illustrate the powering by computing the $\alpha$-power forecasts for the same nine time series used in Example 2, and we compare the sum of absolute errors. In Example 4, we illustrate the non-seasonal mean- and latest-optimized balanced $\alpha$-power models. We find, among other things, that the sum of absolute errors are improved.  In section 4, we introduce a non-seasonal balanced $\delta$-forecasting models for long time series by using only the last $\delta$ observations of a time series $X$. In Example 5, we compute the sequence of the latest 100 balanced $\delta$-forecasts for a monthly time series of new one family homes for sales in the United States from August 2001 to  July 2021, i.e., 240 months. Moreover, we compare our twenty four models to both Holt--Winters and ARIMA models. In particular, we compare the sum of absolute errors, the sum of square errors, and we count the number of times each of the models lead to the best forecast. In section 5, we define seasonal functionally balanced forecasting models, and we introduce a seasonal power time series mapping. We then introduce a seasonal per-period sum of absolute error stochastic latest-optimized $\alpha$-power model. In Example 7, we illustrate the seasonal  per-period sum of absolute error stochastic latest-optimized $\alpha$-power model for the classic Box and Jenkins airline data by computing the sequence of twelve-step-ahead forecasts for the last ten years (120 months). We also compare our forecasts to both Holt--Winters and ARIMA Models. In particular, we compute the sum of absolute errors, and we count the number of times (out of 120 monthly forecasts) that each of the models lead to the best monthly and yearly forecasts. In Example 8, we also illustrate the seasonal  per-period sum of absolute error stochastic latest-optimized $\alpha$-power model for quarterly total retails sales in the United States. In both Examples 7 and 8, we find that our twenty four seasonal  per-period sum of absolute error stochastic latest-optimized $\alpha$-power models not only lead to a significantly lower sum of absolute errors, but lead also to a high number of best forecasts, thus outperforming both Holt--Winters and ARIMA models. Finally, in section 6, we present an application of the non-seasonal balanced $\delta$-forecasting functions to stock market data using daily stock prices of 467 companies of the S\&P 500. For each of these 467 companies, we forecast the latest 100 daily stock prices using 72 different models, i.e., a total of 3,362,400 daily forecasted stock prices, and we count the number of times a forecast lie between the lowest and highest price of the next day. Moreover, we perform nine principal component analyses of the nine multivariate score data (467 observations and 8 variables), and we compare the performance of all the 72 models, and provide the bi-plots of the nine principal component analyses. 

\section{Two-Parameter Families of Forecasting Functions}
Let $X = (x_1, x_2, \dots, x_n)$ and $Y = (y_1, y_2, \dots, y_n)$  be two  time series with $n\geq 2$ observations each. We define the rate of interest time series $R_Y$ associated to $Y$ as the time series $R_Y = (r_1, r_2, \dots, r_{n-1})$ of size $n-1$ defined by
\begin{equation}
 r_i := \dfrac{y_{i+1} - y_i}{y_i} \quad \text{ for } i=1, \dots, n-1
\end{equation}
We define also the rate of discount time series $D_Y$ associated to $Y$ as the time series $D_Y = (d_1, d_2, \dots, d_{n-1})$ of size $n-1$ defined by
\begin{equation}
 d_i := \dfrac{y_{i+1} - y_i}{y_{i+1}} \quad \text{ for } i=1, \dots, n-1
\end{equation}
Given the rate of interest time series $R_Y$, we define the mean (resp., median) $i$th rate of interest $\overline{r}_{i, \lambda}$ (resp., $\widetilde{r}_{i, \lambda}$) of order $\lambda$ as follows:
\begin{eqnarray}
\overline{r}_{i, \lambda} =& \mean(r_{i-\lambda+1}, \dots, r_{i}) \quad &\text{ where } i=1, \dots, n-1 \text{ and } \lambda = 1,\dots, i - 1\\
\widetilde{r}_{i, \lambda} =& \med (r_{i-\lambda+1}, \dots, r_{i}) \quad &\text{ where } i=1, \dots, n-1 \text{ and } \lambda = 1,\dots, i - 1
\end{eqnarray}
Similarly, given the rate of discount time series $D_Y$, we define the mean (resp., median) $i$th rate of discount $\overline{d}_{i, \lambda}$ (resp., $\widetilde{d}_{i, \lambda}$) of order $\lambda$ as follows:
\begin{eqnarray}
\overline{d}_{i, \lambda} =& \mean(d_{i-\lambda+1}, \dots, d_{i}) \quad &\text{ where } i=1, \dots, n-1 \text{ and } \lambda = 1,\dots, i - 1\\
\widetilde{d}_{i, \lambda} =& \med (d_{i-\lambda+1}, \dots, d_{i}) \quad &\text{ where } i=1, \dots, n-1 \text{ and } \lambda = 1,\dots, i - 1
\end{eqnarray}

Given a time series $X$ and a rate of interest times series $R_Y$, we define two types of forecast estimates of order $\lambda$ for $x_{i+1}$. The $\sharp$-mean  (resp., $\sharp$-median) forecast estimate of $x_{i+1}$ of order $\lambda$, denoted $^\sharp\widehat{x}_{\overline{i+1}, \lambda}$ (resp., $^\sharp\widehat{x}_{\widetilde{i+1}, \lambda}$), is defined as
\begin{eqnarray} \label{meanforest}
^\sharp\widehat{x}_{\overline{i+1}, \lambda}  = x_{i} \cdot (1+ \overline{r}_{i-1, \lambda} )\quad \text{ where } i=2, \dots, n \text{ and } \lambda = 1,\dots, i \\\label{medianforest}
^\sharp\widehat{x}_{\widetilde{i+1}, \lambda}  = x_{i} \cdot (1+ \widetilde{r}_{i-1, \lambda} )\quad \text{ where } i=2, \dots, n \text{ and } \lambda = 1,\dots, i 
\end{eqnarray}

Similarly, given a time series $X$ and a rate of discount times series $D_Y$, we define another two types of forecast estimates of order $\lambda$ for $x_{i+1}$. The $\flat$-mean  (resp., $\flat$-median) forecast estimate of $x_{i+1}$ of order $\lambda$, denoted $^\flat\widehat{x}_{\overline{i+1}, \lambda}$ (resp., $^\flat\widehat{x}_{\widetilde{i+1}, \lambda}$), is defined as
\begin{eqnarray} \label{meanDforest}
^\flat\widehat{x}_{\overline{i+1}, \lambda}  = x_{i} \div (1- \overline{d}_{i-1, \lambda} )\quad \text{ where } i=2, \dots, n \text{ and } \lambda = 1,\dots, i \\\label{medianDforest}
^\flat\widehat{x}_{\widetilde{i+1}, \lambda}  = x_{i}\div(1- \widetilde{d}_{i-1, \lambda} )\quad \text{ where } i=2, \dots, n \text{ and } \lambda = 1,\dots, i 
\end{eqnarray}

In what follows, the symbol $\star$ stands for either $\sharp$ or $\flat$. Given the $\star$-mean (resp.,  $\star$-median) estimate forecasts $^\star\widehat{x}_{\overline{i}, \lambda}$ (resp.,  $^\star\widehat{x}_{\widetilde{i}, \lambda}$) of order $\lambda$ for $i=3, \dots, n$ and $\lambda = 1, \dots, i-1$, we define eight optimization criteria of sum of $\star$-mean (resp., $\star$-median) estimate forecast errors to forecast $x_i$ of length $\nu$, denoted $^\star\Sigma^\kappa_{\overline{i}, \lambda, \nu}$  (resp., $^\star\Sigma^\kappa_{\widetilde{i}, \lambda, \nu}$), where $\nu = 1, \dots, n-1$, $\lambda + \nu < n$ and $\kappa=1, \dots, 8$, by :

\begin{equation}\label{Sigma1}
^\star\Sigma^1_{\overline{i}, \lambda, \nu}:= \displaystyle{\sum_{j=i-\nu+1}^i }\bigg(\dfrac{^\star\widehat{x}_{\overline{j}, \lambda}  - x_j}{^\star\widehat{x}_{\overline{j}, \lambda} }\bigg)^2, \qquad  ^\star\Sigma^1_{\widetilde{i}, \lambda, \nu}:= \displaystyle{\sum_{j=i-\nu+1}^i }\bigg(\dfrac{^\star\widehat{x}_{\widetilde{j}, \lambda}  - x_j}{\widehat{x}_{^\star\widetilde{j}, \lambda} }\bigg)^2 ,
\end{equation}
\begin{equation}\label{Sigma2}
^\star\Sigma^2_{\overline{i}, \lambda, \nu}:=\displaystyle{\sum_{j=i-\nu+1}^i } \bigg|\dfrac{^\star\widehat{x}_{\overline{j}, \lambda}  - x_j}{^\star\widehat{x}_{\overline{j}, \lambda} }\bigg|, \qquad ^\star\Sigma^2_{\widetilde{i}, \lambda, \nu}:= \displaystyle{\sum_{j=i-\nu+1}^i } \bigg|\dfrac{^\star\widehat{x}_{\widetilde{j}, \lambda}  - x_j}{^\star\widehat{x}_{\widetilde{j}, \lambda} }\bigg|,
\end{equation}
\begin{equation}\label{Sigma3}
^\star\Sigma^3_{\overline{i}, \lambda, \nu}:= \displaystyle{\sum_{j=i-\nu+1}^i }\dfrac{(^\star\widehat{x}_{\overline{j}, \lambda}  - x_j)^2}{|^\star\widehat{x}_{\overline{j}, \lambda} |},   \qquad ^\star\Sigma^3_{\widetilde{i}, \lambda, \nu}:= \displaystyle{\sum_{j=i-\nu+1}^i }\dfrac{(^\star\widehat{x}_{\widetilde{j}, \lambda}  - x_j)^2}{|^\star\widehat{x}_{\widetilde{j}, \lambda} |},
\end{equation}
\begin{equation}\label{Sigma4}
^\star\Sigma^4_{\overline{i}, \lambda, \nu}:= \displaystyle{\sum_{j=i-\nu+1}^i }\bigg(\dfrac{^\star\widehat{x}_{\overline{j}, \lambda}  - x_j}{x_j }\bigg)^2,  
 \qquad  ^\star\Sigma^4_{\widetilde{i}, \lambda, \nu}:=\displaystyle{\sum_{j=i-\nu+1}^i }\bigg(\dfrac{^\star\widehat{x}_{\widetilde{j}, \lambda}  - x_j}{x_j }\bigg)^2, 
\end{equation}
\begin{equation}\label{Sigma5}
^\star\Sigma^5_{\overline{i}, \lambda, \nu}:=\displaystyle{\sum_{j=i-\nu+1}^i } \bigg|\dfrac{^\star\widehat{x}_{\overline{j}, \lambda}  - x_j}{x_j }\bigg|,  \qquad ^\star\Sigma^5_{\widetilde{i}, \lambda, \nu}:=\displaystyle{\sum_{j=i-\nu+1}^i } \bigg|\dfrac{^\star\widehat{x}_{\widetilde{j}, \lambda}  - x_j}{x_j }\bigg|,
\end{equation}
\begin{equation}\label{Sigma6}
^\star\Sigma^6_{\overline{i}, \lambda, \nu}:= \displaystyle{\sum_{j=i-\nu+1}^i }\dfrac{(^\star\widehat{x}_{\overline{j}, \lambda}  - x_j)^2}{|x_j |},   \qquad  ^\star\Sigma^6_{\widetilde{i}, \lambda, \nu}:= \displaystyle{\sum_{j=i-\nu+1}^i }\dfrac{(^\star\widehat{x}_{\widetilde{j}, \lambda}  - x_j)^2}{|x_j |},  
\end{equation}
\begin{equation}\label{Sigma7}
^\star\Sigma^7_{\overline{i}, \lambda, \nu}:=\displaystyle{\sum_{j=i-\nu+1}^i } (^\star\widehat{x}_{\overline{j}, \lambda}  - x_j)^2,   \qquad  ^\star\Sigma^7_{\widetilde{i}, \lambda, \nu}:=\displaystyle{\sum_{j=i-\nu+1}^i } (^\star\widehat{x}_{\widetilde{j}, \lambda}  - x_j)^2,
\end{equation}
\begin{equation}\label{Sigma8}
^\star\Sigma^8_{\overline{i}, \lambda, \nu}:=\displaystyle{\sum_{j=i-\nu+1}^i }|^\star\widehat{x}_{\overline{j}, \lambda}  - x_j|,  \qquad   ^\star\Sigma^8_{\widetilde{i}, \lambda, \nu}:=\displaystyle{\sum_{j=i-\nu+1}^i }|^\star\widehat{x}_{\widetilde{j}, \lambda}  - x_j|.
\end{equation}

For a given optimization criteria $\kappa=1, \dots, 8$, a chosen maximum allowed order $\lambda_{\text{max}}$, and a fixed length $\nu$ such that $\lambda_{\text{max}} + \nu < n$, the least sum of $\star$-mean  (resp., $\star$-median) estimate forecast errors   $^\star\Sigma^{\kappa, \ast}_{\overline{n}, \lambda_{\text{max}}, \nu}$ (resp., $^\star\Sigma^{\kappa, \ast}_{\widetilde{n}, \lambda_{\text{max}}, \nu}$), to forecast $x_n$, of order $\lambda_{\text{max}}$ and length $\nu$,   are defined as follows: 
\begin{equation}\label{optimeanmedSigma}
 ^\star\Sigma^{\kappa, \ast}_{\overline{n}, \lambda_{\text{max}}, \nu} = \min_{\lambda=1, \dots, \lambda_{\text{max}}} {^\star\Sigma^\kappa_{\overline{n}, \lambda, \nu}}  \qquad \text{ and } \qquad  ^\star\Sigma^{\kappa, \ast}_{\widetilde{n}, \lambda_{\text{max}} , \nu} = \min_{\lambda=1, \dots, \lambda_{\text{max}}} {^\star\Sigma^\kappa_{\widetilde{n}, \lambda, \nu}}  
\end{equation}

From the latter, and for a given optimization criteria $\kappa=1, \dots, 8$, a chosen maximum order $\lambda_{\text{max}} = 1, \dots, n-1$ and fixed length $\nu = 1, \dots, n-1$
 such that $\lambda_{\text{max}}  + \nu < n$, we extract the optimal $\star$-mean (resp., $\star$-median) order $^\star\lambda^{k, \ast}_{\overline{\lambda_{\text{max}}}, \nu}$ (resp., $^\star\lambda^{k, \ast}_{\widetilde{\lambda_{\text{max}}}, \nu}$)  from $1$ to $\lambda_{\text{max}}$ to forecast $x_n$ for a  given length $\nu$ as follows: 
 \begin{equation}\label{optilambda}
^\star\lambda^{k, \ast}_{\overline{\lambda_{\text{max}}}, \nu} = \arg\Big( \min _{\lambda = 1, \dots, \lambda_{\text{max}}}  {^\star\Sigma^{\kappa, \ast}_{\overline{n}, \lambda_{\text{max}}, \nu}}\Big) \quad \text{ and } \quad  ^\star\lambda^{k, \ast}_{\widetilde{\lambda_{\text{max}}}, \nu}= \arg\Big( \min _{\lambda = 1, \dots, \lambda_{\text{max}}}  {^\star\Sigma^{\kappa, \ast}_{\widetilde{n}, \lambda_{\text{max}}, \nu}}\Big) 
\end{equation}

These latter four optimal orders $^\star\lambda^{k, \ast}_{\overline{\lambda_{\text{max}}}, \nu} $ and $^\star\lambda^{k, \ast}_{\widetilde{\lambda_{\text{max}}}, \nu}$, where $\star$ is either $\sharp$ or $\flat$,   are then used as preferred orders to forecast $x_{n+1}$ from the $n$ observations $x_1, \dots, x_n$ and either rate of interest times series $R(Y)$ or rate of discount time series $D(Y)$. Finally, we propose the following thirty two 2-parameter families of forecasting functions, that is for each of the four type of forecast estimate ($\sharp$-mean, $\sharp$-median, $\flat$-mean, and $\flat$-median) with each one of the eight optimization criteria $\kappa = 1, \dots, 8$. 

\begin{Def} Let $X = (x_1, x_2, \dots, x_n)$ and $Y = (y_1, y_2, \dots, y_n)$ be two time series with $n \geq 2$ observations.  Let $R_Y$ be the rate of interest time series associated to $Y$.  For each optimization criteria $\kappa=1, \dots, 8$, a chosen maximum order $\lambda_{\text{max}} = 1, \dots, n-2$, and a fixed length $\nu$ in  $1, \dots, n-2$ such that $\lambda_{\text{max}}  + \nu < n$, the $\sharp$-mean (resp., $\sharp$-median) forecasting function $^\sharp\Psi^\kappa_{\overline{\lambda_{\text{max}}}, \nu}$  (resp., $^\sharp\Psi^\kappa_{\widetilde{\lambda_{\text{max}}}, \nu}$) of order $\lambda_{\text{max}}$ and length $\nu$  maps the time series $X$ and the rate of interest $R_Y$ to a $\sharp$-mean (resp., $\sharp$-median)  forecast of $x_{n+1}$, denoted $^\sharp\widehat{x}_{\overline{n+1}}^\kappa$ (resp., $^\sharp\widehat{x}_{\widetilde{n+1}}^\kappa$), as follows:
\begin{equation}\label{Phikn+1}
^\sharp\widehat{x}_{\overline{n+1}}^\kappa = {^\sharp\Psi}^\kappa_{\overline{\lambda_{\text{max}}}, \nu}(X, R_Y) := x_{n} \cdot \Big(1+ \overline{r}_{n-1, ^\sharp\lambda^{\kappa, \ast}_{\overline{\lambda_{\text{max}}}, \nu}}\Big)
\end{equation} 
resp.,
\begin{equation}\label{Phikn+1}
^\sharp\widehat{x}_{\widetilde{n+1}}^\kappa = {^\sharp\Psi}^\kappa_{\widetilde{\lambda_{\text{max}}}, \nu}(X, R_Y) := x_{n} \cdot \Big(1+ \widetilde{r}_{n-1, ^\sharp\lambda^{\kappa, \ast}_{\widetilde{\lambda_{\text{max}}}, \nu}}\Big)
\end{equation} 
where the $\sharp$-mean (resp., $\sharp$-median) optimal order $^\sharp\lambda^{k, \ast}_{\overline{\lambda_{\text{max}}}, \nu}$ (resp., $^\sharp\lambda^{k, \ast}_{\widetilde{\lambda_{\text{max}}}, \nu}$) is as in (\ref{optilambda}). \end{Def}

\begin{Def} Let $X = (x_1, x_2, \dots, x_n)$ and $Y = (y_1, y_2, \dots, y_n)$ be two time series with $n \geq 2$ observations.  Let $D_Y$ be the rate of discount time series associated to $Y$.  For each optimization criteria $\kappa=1, \dots, 8$, a chosen maximum order $\lambda_{\text{max}} = 1, \dots, n-2$,  and a fixed length $\nu$ in  $1, \dots, n-2$ such that $\lambda_{\text{max}}  + \nu < n$, the $\flat$-mean (resp., $\flat$-median) forecasting function $^\flat\Psi^\kappa_{\overline{\lambda_{\text{max}}}, \nu}$  (resp., $^\flat\Psi^\kappa_{\widetilde{\lambda_{\text{max}}}, \nu}$) of order $\lambda_{\text{max}}$ and length $\nu$  maps the time series $X$ and the rate of discount $D_Y$ to a $\flat$-mean  (resp., $\flat$-median) forecast  of $x_{n+1}$, denoted $^\flat\widehat{x}_{\overline{n+1}}^\kappa$ (resp., $^\flat\widehat{x}_{\widetilde{n+1}}^\kappa$), as follows:
\begin{equation}\label{Phikn+1}
^\flat\widehat{x}_{\overline{n+1}}^\kappa = {^\flat\Psi}^\kappa_{\overline{\lambda_{\text{max}}}, \nu}(X, D_Y) := x_{n} \div \Big(1- \overline{d}_{n-1, ^\flat\lambda^{\kappa, \ast}_{\overline{\lambda_{\text{max}}}, \nu}}\Big)
\end{equation} 
resp.,
\begin{equation}\label{Phikn+1}
^\flat\widehat{x}_{\widetilde{n+1}}^\kappa = {^\flat\Psi}^\kappa_{\widetilde{\lambda_{\text{max}}}, \nu}(X, D_Y) := x_{n} \div\Big(1+ \widetilde{d}_{n-1, ^\flat\lambda^{\kappa, \ast}_{\widetilde{\lambda_{\text{max}}}, \nu}}\Big)
\end{equation} 
where the $\flat$-mean (resp., $\flat$-median) optimal order $^\flat\lambda^{k, \ast}_{\overline{\lambda_{\text{max}}}, \nu}$ (resp., $^\flat\lambda^{k, \ast}_{\widetilde{\lambda_{\text{max}}}, \nu}$) is as in (\ref{optilambda}). \end{Def}

One can optimize within a pair of forecast estimates, that is, choosing two forecast estimate types among the four. There are six possible pairs, and thus one can define another forty eight 2-parameter families of forecasting functions. We propose in the following  only two pairs (among the six) leading to a $\sharp$- (resp., $\flat$-) forecasting function, which combines the $\sharp$-mean (resp., $\flat$-mean) and  $\sharp$-median (resp., $\flat$-median) forecasting functions.

\begin{Def} Let $X = (x_1, x_2, \dots, x_n)$ and $Y = (y_1, y_2, \dots, y_n)$ be two time series with $n \geq 2$ observations.  Let $R_Y$ (resp., $D_Y$) be the rate of interest (resp., discount) time series associated to $Y$.  For each optimization criteria $\kappa=1, \dots, 8$, a chosen maximum order $\lambda_{\text{max}} = 1, \dots, n-2$, and a fixed length $\nu$  in $1, \dots, n-2$ such that $\lambda_{\text{max}}  + \nu < n$, the $\sharp$- (resp., $\flat$-) forecasting function $^\sharp\Psi^\kappa_{\lambda_{\text{max}}, \nu}$ (resp.,$^\flat\Psi^\kappa_{\lambda_{\text{max}}, \nu}$) of order $\lambda_{\text{max}}$ and length $\nu$  maps the time series $X$ and the rate of interest (resp., discount) $R_Y$ (resp., $D_Y)$ to a $\sharp$- (resp., $\flat$-) forecast of $x_{n+1}$, denoted $^\sharp\widehat{x}_{n+1}^\kappa$ (resp., $^\flat\widehat{x}_{n+1}^\kappa$), as follows:
\begin{equation}\label{Phikn+1}
^\sharp\widehat{x}_{n+1}^\kappa = {^\sharp\Psi}^\kappa_{\lambda_{\text{max}}, \nu}(X, R_Y) := \left\{\begin{array}{ccc}^\sharp\widehat{x}_{\overline{n+1}}^\kappa  &  \text{ if } & ^\sharp\Sigma^{\kappa, \ast}_{\overline{n}, \lambda_{\text{max}}, \nu} \leq {^\sharp\Sigma^{\kappa, \ast}_{\widetilde{n}, \lambda_{\text{max}}, \nu}}  \\^\sharp\widehat{x}_{\widetilde{n+1}}^\kappa & \text{ if } & ^\sharp\Sigma^{\kappa, \ast}_{\overline{n}, \lambda_{\text{max}}, \nu} > {^\sharp\Sigma^{\kappa, \ast}_{\widetilde{n}, \lambda_{\text{max}}, \nu}}\end{array}\right.
\end{equation} 
resp., 
\begin{equation}
^\flat\widehat{x}_{n+1}^\kappa = {^\flat\Psi}^\kappa_{\lambda_{\text{max}}, \nu}(X, D_Y) := \left\{\begin{array}{ccc}^\flat\widehat{x}_{\overline{n+1}}^\kappa  &  \text{ if } & ^\flat\Sigma^{\kappa, \ast}_{\overline{n}, \lambda_{\text{max}}, \nu} \leq {^\flat\Sigma^{\kappa, \ast}_{\widetilde{n}, \lambda_{\text{max}}, \nu}}  \\^\flat\widehat{x}_{\widetilde{n+1}}^\kappa & \text{ if } & ^\flat\Sigma^{\kappa, \ast}_{\overline{n}, \lambda_{\text{max}}, \nu} > {^\flat\Sigma^{\kappa, \ast}_{\widetilde{n}, \lambda_{\text{max}}, \nu}}\end{array}\right.
\end{equation} 
where the least sum of $\sharp$- (resp., $\flat$-) mean estimate forecast error $^\sharp\Sigma^{\kappa, \ast}_{\overline{n}, \lambda_{\text{max}}, \nu}$ (resp., $^\flat\Sigma^{\kappa, \ast}_{\overline{n}, \lambda_{\text{max}}, \nu}$) and the least sum of $\sharp$- (resp., $\flat$) median estimate forecast error ${^\sharp\Sigma^{\kappa, \ast}_{\widetilde{n}, \lambda_{\text{max}}, \nu}}$ (resp., ${^\flat\Sigma^{\kappa, \ast}_{\widetilde{n}, \lambda_{\text{max}}, \nu}}$) are as in (\ref{optimeanmedSigma}). 
\end{Def}

Similarly, one can optimize within a triple of forecast estimates. There are four possible triples, and thus one can add another thirty two 2-parameter families of forecasting functions. Lastly, one can optimize across the four forecast estimates, and thus add another eight 2-parameter families of forecasting functions, as shown in the following proposition. 

\begin{Def} Let $X = (x_1, x_2, \dots, x_n)$ and $Y = (y_1, y_2, \dots, y_n)$ be two time series with $n \geq 2$ observations.  Let $R_Y$ (resp., $D_Y$) be the rate of interest (resp., discount) times series associated to $Y$.  For each optimization criteria $\kappa=1, \dots, 8$, a chosen maximum order $\lambda_{\text{max}} = 1, \dots, n-2$,  and fixed length $\nu$ in  $1, \dots, n-2$ such that $\lambda_{\text{max}}  + \nu < n$, the forecasting function ${^\natural}\Psi^\kappa_{\lambda_{\text{max}}, \nu}$ of order $\lambda_{\text{max}}$ and length $\nu$  maps the time series $X$, the rate of interest $R_Y$, and the rate of discount $D_Y$ to a forecast of $x_{n+1}$, denoted ${^\natural}\widehat{x}_{n+1}^\kappa$, as follows:
\begin{equation}\label{Phikn+1}
{^\natural}\widehat{x}_{n+1}^\kappa = {{^\natural}\Psi}^\kappa_{\lambda_{\text{max}}, \nu}(X, R_Y, D_Y) := \left\{\begin{array}{ccc}^\sharp\widehat{x}_{n+1}^\kappa  &  \text{ if } & ^\sharp\Sigma^{\kappa, \ast}_{n, \lambda_{\text{max}}, \nu} \leq  {^\flat\Sigma^{\kappa, \ast}_{n, \lambda_{\text{max}}, \nu} }\\^\flat\widehat{x}_{n+1}^\kappa & \text{ if } &^\sharp\Sigma^{\kappa, \ast}_{n, \lambda_{\text{max}}, \nu}  > {^\flat\Sigma^{\kappa, \ast}_{n, \lambda_{\text{max}}, \nu}}\end{array}\right.
\end{equation}  
where the least $\sharp$-estimate forecast error $^\sharp\Sigma^{\kappa, \ast}_{n, \lambda_{\text{max}}, \nu}$ is 
\begin{eqnarray}
^\sharp\Sigma^{\kappa, \ast}_{n, \lambda_{\text{max}}, \nu} := \min(^\sharp\Sigma^{\kappa, \ast}_{\overline{n}, \lambda_{\text{max}}, \nu},  {^\sharp\Sigma^{\kappa, \ast}_{\widetilde{n}, \lambda_{\text{max}}, \nu}}), 
\end{eqnarray}
and the least $\flat$-estimate forecast error $^\flat\Sigma^{\kappa, \ast}_{n, \lambda_{\text{max}}, \nu}$ is 
\begin{eqnarray}
^\flat\Sigma^{\kappa, \ast}_{n, \lambda_{\text{max}}, \nu} := \min(^\flat\Sigma^{\kappa, \ast}_{\overline{n}, \lambda_{\text{max}}, \nu},  {^\flat\Sigma^{\kappa, \ast}_{\widetilde{n}, \lambda_{\text{max}}, \nu}})
\end{eqnarray}\end{Def}

Note that given a times series $X$, one can use its associated rate of interest $R_X$ and/or rate of discount $D_X$, i.e., set $Y = X$ in for forecasting functions. Note also that one can define a total of one hundred twenty forecasting functions. However, we focus our attention on only twenty four of them, i.e., ${^\sharp}\Psi^{\kappa}, {^\flat}\Psi^{\kappa}$ and ${^\natural}\Psi^{\kappa}$ for $\kappa = 1, \dots, 8$.

\begin{Ex}  Let $X = (19.17, 18.92, 18.87, 18.98, 18.60, 18.82, 16.36, 16.17, 15.72, 16.01)$ be a time series with $10$ observations. We illustrate in the following the steps to compute $^\sharp\Psi^{\kappa}_{\overline{5},4}(X, R_X)$, $^\sharp\Psi^{\kappa}_{\widetilde{5},4}(X, R_X)$, $^\flat\Psi^{\kappa}_{\overline{5},4}(X, D_X)$, $^\flat\Psi^{\kappa}_{\widetilde{5},4}(X, D_X)$, $^\sharp\Psi^{\kappa}_{5,4}(X, R_X)$, $^\flat\Psi^{\kappa}_{5,4}(X, D_X)$ and ${^\natural}\Psi^{\kappa}_{5,4}(X, R_X, D_X)$  for each one of the optimization criteria $\kappa=1, \dots, 8$. Note that $x_{11} =15.64$, and  $X$ together with $x_{11}$ correspond to 11 consecutive daily closing prices of a given stock.  

Using (\ref{meanforest}), the $\sharp$-mean estimate forecasts $^\sharp\widehat{x}_{\overline{i}, \lambda }$ for $i=7 , \dots, 10$ and order $\lambda = 1, \dots, 5$ are 
\begin{center}
\begin{tabular}{l|c|c|c|c|c}
$^\sharp\widehat{x}_{\overline{i}, \lambda }$    &   $\lambda =1$ &       $\lambda =2$  &       $\lambda =3$  &       $\lambda =4$  &       $\lambda =5$ \\ \hline
$i=10$ & 15.28252 &15.40998 &14.82839 &15.09777& 15.15927\\
$i=9$ &15.98221 &15.01930 &15.46662 &15.56153 &15.70207\\
$i=8$ &14.22155 &15.38753 &15.60250& 15.81572 &15.91593\\
$i=7$ &19.04260 &18.74290 &18.80517 &18.79644 &18.75207
\end{tabular}
\end{center}

Using (\ref{medianforest}), the $\sharp$-median estimate forecasts $^\sharp\widehat{x}_{\widetilde{i}, \lambda }$ for $i=7, \dots, 10$ and order $\lambda = 1, \dots, 5$ are 

\begin{center}
\begin{tabular}{l|c|c|c|c|c}
    $^\sharp\widehat{x}_{\widetilde{i}, \lambda}$    &   $\lambda =1$ &       $\lambda =2$  &       $\lambda =3$  &       $\lambda =4$  &       $\lambda =5$\\ \hline
$i=10$ & 15.28252& 15.40998 &15.28252 &15.40998 &15.40527\\
$i=9$ &   15.98221 &15.01930& 15.98221 &15.91423 &15.98221\\
$i=8$ &  14.22155 &15.38753 &16.03246 &16.24391 &16.31677\\
$i=7$ & 19.04260 &18.74290 &18.92971 &18.84999 &18.77026\end{tabular}
\end{center}

Using (\ref{meanDforest}), the $\flat$-mean estimate forecasts $^\flat\widehat{x}_{\overline{i}, \lambda }$ for $i=7, \dots, 10$ and order $\lambda = 1, \dots, 5$ are 
\begin{center}
\begin{tabular}{l|c|c|c|c|c}
$^\flat\widehat{x}_{\overline{i}, \lambda }$    &   $\lambda =1$ &       $\lambda =2$  &       $\lambda =3$  &       $\lambda =4$  &       $\lambda =5$ \\ \hline
$i=10$ &15.28252 &15.40892 &14.78026 &15.04647 &15.11689\\
$i=9$ & 15.98221 &14.95756 &15.39791 &15.50760 &15.65325\\
$i=8$ & 14.22155 &15.29918 &15.53604 &15.75610 &15.86513\\
$i=7$ &19.04260 &18.73811 &18.80154 &18.79371 &18.74947
\end{tabular}
\end{center}

Using (\ref{medianDforest}), the $\flat$-median estimate forecasts $^\flat\widehat{x}_{\widetilde{i}, \lambda }$ for $i=7, \dots, 10$ and order $\lambda = 1, \dots, 5$ are

\begin{center}
\begin{tabular}{l|c|c|c|c|c}
    $^\flat\widehat{x}_{\widetilde{i}, \lambda}$    &   $\lambda =1$ &       $\lambda =2$  &       $\lambda =3$  &       $\lambda =4$  &       $\lambda =5$\\ \hline
$i=10$& 15.28252 &15.40892& 15.28252& 15.40892 &15.40527\\
$i=9$&  15.98221 &14.95756 &15.98221& 15.91394 &15.98221\\
$i=8$&  14.22155 &15.29918 &16.03246 &16.24116 &16.31677\\
$i=7$&  19.04260 &18.73811 &18.92971 &18.84965 &18.77026
\end{tabular}
\end{center}

Using (\ref{Sigma1} -- \ref{Sigma8}), the sum of $\sharp$-mean estimate forecast errors to forecast $x_{10}$ for each optimization criteria $\kappa=1, \dots, 8$, order $\lambda = 1, \dots, 5$ and length $\nu=4$ are

\begin{center}
\begin{tabular}{l|c|c|c|c|c}
    $^\sharp\Sigma^\kappa_{\overline{10}, \lambda, 4}$     & $\lambda =1$ &       $\lambda =2$  &       $\lambda =3$  &       $\lambda =4$  &       $\lambda =5$\\ \hline
$\kappa=1$ & 0.04115135 &0.02244212 &0.02484805 &0.02105823 &0.01967779\\
$\kappa=2$ & 0.34188850 &0.26357807 &0.26246703 &0.22262790 &0.20078705\\
$\kappa=3$ & 0.68378978 &0.39879634 &0.43688700& 0.38048595 &0.35695762\\
$\kappa=4$ &  0.04374984 &0.02694821& 0.02927703 &0.02600744 &0.02445041\\
$\kappa=5$ &  0.34658969 &0.27609648& 0.27447910 &0.23789604 &0.21620445\\
$\kappa=6$ &  0.71208790 &0.43866460 &0.47666540 &0.42418913 &0.39897276\\
$\kappa=7$ & 11.59078051 &7.14149881 &7.76132502 &6.91904497 &6.51060090\\
$\kappa=8$ & 5.62073402 &4.46609961 &4.44766242 &3.86142213 &3.51479163
\end{tabular}
\end{center}
Using (\ref{Sigma1} -- \ref{Sigma8}), the sum of $\sharp$-median estimate forecast errors to forecast $x_{10}$ for each optimization criteria $\kappa=1, \dots, 8$, order $\lambda = 1, \dots, 5$ and length $\nu=4$ are
\begin{center}
\begin{tabular}{l|c|c|c|c|c}
    $^\sharp\Sigma^k_{\widetilde{10}, \lambda, 4}$     & $\lambda =1$ &       $\lambda =2$  &       $\lambda =3$  &       $\lambda =4$  &       $\lambda =5$\\ \hline
$\kappa=1$ & 0.04115135 &0.02244212 &0.02103677 &0.01913483 &0.01837979\\
$\kappa=2$ &  0.34188850 &0.26357807 &0.20833721 &0.18778721 &0.19306440\\
$\kappa=3$ & 0.68378978 &0.39879634 &0.38894910 &0.35498453 &0.33885933\\
$\kappa=4$ & 0.04374984 &0.02694821 &0.02708708 &0.02474289 &0.02349248\\
$\kappa=5$ &  0.34658969 &0.27609648 &0.22769753 &0.20660431 &0.21085497\\
$\kappa=6$ &  0.71208790& 0.43866460 &0.44223021 &0.40420042 &0.38364383\\
$\kappa=7$ & 11.59078051 &7.14149881 &7.22029532 &6.60324808 &6.26536616\\
$\kappa=8$ & 5.62073402 &4.46609961 &3.69693673 &3.35815310 &3.42396750
  
    \end{tabular}
\end{center}
Using (\ref{Sigma1} -- \ref{Sigma8}), the sum of $\flat$-mean estimate forecast errors to forecast $x_{10}$ for each optimization criteria $\kappa=1, \dots, 8$, order $\lambda = 1, \dots, 5$ and length $\nu=4$ are

\begin{center}
\begin{tabular}{l|c|c|c|c|c}
    $^\flat\Sigma^k_{\overline{10}, \lambda, 4}$     & $\lambda =1$ &       $\lambda =2$  &       $\lambda =3$  &       $\lambda =4$  &       $\lambda =5$\\ \hline
$\kappa=1$ & 0.04115135 &0.02346669& 0.02588849 &0.02174764 &0.02011944\\
$\kappa=2$ &0.34188850& 0.27381425 &0.27478417 &0.23349849 &0.21000321\\
$\kappa=3$  &0.68378978 &0.41369099 &0.45197944 &0.39063998 &0.36342793\\
$\kappa=4$ & 0.04374984 &0.02779207& 0.03012902 &0.02600744 &0.02481771\\
$\kappa=5$ &0.34658969 &0.28526036 &0.28574487 &0.24805096 &0.22494066\\
$\kappa=6$& 0.71208790& 0.45212834 &0.49028507 &0.43349172 &0.40484953\\
$\kappa=7$& 11.59078051 &7.35634421& 7.76132502& 7.06777918 &6.60462705\\
$\kappa=8$ & 5.62073402& 4.61244611& 4.44766242 &4.02353907 &3.65420470
\end{tabular}
\end{center}

Using (\ref{Sigma1} -- \ref{Sigma8}), the sum of $\flat$-median estimate forecast errors to forecast $x_{10}$ for each optimization criteria $\kappa=1, \dots, 8$, order $\lambda = 1, \dots, 5$ and length $\nu=4$ are

\begin{center}
\begin{tabular}{l|c|c|c|c|c}
    $^\flat\Sigma^k_{\widetilde{10}, \lambda, 4}$     & $\lambda =1$ &       $\lambda =2$  &       $\lambda =3$  &       $\lambda =4$  &       $\lambda =5$\\ \hline
$\kappa=1$ & 0.04115135 &0.02346669 &0.02103677 &0.01913432 &0.01837979\\
$\kappa=2$ & 0.34188850 &0.27381425 &0.20833721 &0.18765602 &0.19306440\\
$\kappa=3$ &0.68378978 &0.41369099 &0.38894910 &0.35495355 &0.33885933\\
$\kappa=4$ & 0.04374984 &0.02779207& 0.02708708 &0.02473957 &0.02349248\\
$\kappa=5$ &0.34658969 &0.28526036 &0.22769753 &0.20646085 &0.21085497\\
$\kappa=6$ & 0.71208790 &0.45212834 &0.44223021 &0.40414501 &0.38364383\\
$\kappa=7$ &11.59078051 &7.35634421& 7.22029532& 6.60232323 &6.26536616\\
$\kappa=8$ &5.62073402 &4.61244611&3.69693673 &3.35582711 &3.42396750
\end{tabular}
\end{center}
From the latter four tables, one can notice  that the least sum of $\sharp$-mean estimate forecast errors $^\sharp\Sigma^{\kappa, \ast}_{\overline{10}, \lambda_{\text{max}}, 4}$ and the least sum of $\flat$-mean estimate forecast errors $^\flat\Sigma^{\kappa, \ast}_{\overline{10}, \lambda_{\text{max}}, 4}$,  for every $\kappa=1, \dots, 8$, $\lambda_{\text{max}} = 5$ and length $\nu = 4$,  are achieved for $\lambda = 5$, i.e., ${^\sharp\lambda^{\kappa, \ast}_{\overline{5}, 4}} = {^\flat\lambda^{\kappa, \ast}_{\overline{5}, 4}}  = 5$. Therefore, for every $\kappa=1, \dots, 8$,
\begin{eqnarray*}
 ^\sharp\widehat{x}_{\overline{11}}^\kappa&=& {^\sharp\Psi}^\kappa_{\overline{5}, 4}(X) = x_{10}\cdot \Big(1+ \overline{r}_{9, {^\sharp\lambda^{\kappa, \ast}_{\overline{5}, 4}}} \Big)=  x_{10}\cdot \Big(1 + \mean(r_5, r_6, r_7, r_8, r_9) \Big)= 15.56211,\\
 ^\flat\widehat{x}_{\overline{11}}^\kappa &=& {^\flat\Psi}^\kappa_{\overline{5}, 4}(X) = x_{10}/ \Big(1- \overline{d}_{9, {^\flat\lambda^{\kappa, \ast}_{\overline{5}, 4}}} \Big)=  x_{10} / \Big(1 - \mean(d_5, d_6, d_7, d_8, d_9) \Big)= 15.51074.
 \end{eqnarray*}
However, the least sum of $\sharp$-median estimate forecast errors $^\sharp\Sigma^{\kappa, \ast}_{\widetilde{10}, \lambda_{\text{max}}, 4}$ and   least sum of $\flat$-median estimate forecast errors $^\flat\Sigma^{\kappa, \ast}_{\widetilde{10}, \lambda_{\text{max}}, 4}$ are achieved for $\lambda = 5$ when $\kappa=1, 3, 4, 6,7$,  while they  are achieved for $\lambda = 4$ when $\kappa=2, 5, 8$. Therefore, for $\kappa=1, 3, 4, 6,7$, we have ${^\sharp\lambda^{\kappa, \ast}_{\widetilde{5}, 4}} = {^\flat\lambda^{\kappa, \ast}_{\widetilde{5}, 4}} =  5$, and thus 
\begin{eqnarray*}
^\sharp\widehat{x}_{\widetilde{11}}^\kappa &=& {^\sharp\Psi}^\kappa_{\widetilde{5}, 4}(X) = x_{10}\cdot\Big(1+ \widetilde{r}_{9, {^\sharp\lambda^{\kappa, \ast}_{\widetilde{5}, 4}}} \Big)=  x_{10}\cdot \Big(1 + \med(r_5, r_6, r_7, r_8, r_9) \Big)= 15.82406,\\
 ^\flat\widehat{x}_{\widetilde{11}}^\kappa &=& {^\flat\Psi}^\kappa_{\widetilde{5}, 4}(X) = x_{10}/ \Big(1- \widetilde{d}_{9, {^\flat\lambda^{\kappa, \ast}_{\widetilde{5}, 4}}} \Big)=  x_{10}/\Big(1 - \med(d_5, d_6, d_7, d_8, d_9) \Big)= 15.82406.
 \end{eqnarray*}
while for $\kappa=2, 5, 8$, we have ${^\sharp\lambda^{\kappa, \ast}_{\widetilde{5}, 4}} = 4$, and thus
\begin{eqnarray*}
^\sharp\widehat{x}_{\widetilde{11}}^\kappa &=& {^\sharp\Psi}^\kappa_{\widetilde{5}, 4}(X) = x_{10}\cdot\Big(1+ \widetilde{r}_{9, {^\sharp\lambda^{\kappa, \ast}_{\widetilde{5}, 4}}} \Big)=  x_{10}\cdot \Big(1 + \med(r_6, r_7, r_8, r_9) \Big)= 15.69426, \\
^\flat\widehat{x}_{\widetilde{11}}^\kappa &=& {^\flat\Psi}^\kappa_{\widetilde{5}, 4}(X) = x_{10}/\Big(1-\widetilde{d}_{9, {^\flat\lambda^{\kappa, \ast}_{\widetilde{5}, 4}}} \Big)=  x_{10}/ \Big(1 - \med(d_6, d_7, d_8, d_9) \Big)= 15.69319
\end{eqnarray*}
Note that for every $\kappa=1, \dots, 8$, the least sum of $\sharp$-median forecast errors are smaller than the least sum of $\sharp$-mean forecast errors. Therefore, the $\sharp$-forecasts are:
\begin{equation*}
^\sharp\widehat{x}_{11}^\kappa = \left\{\begin{array}{ccl}15.82406 & \text{ for } & \kappa = 1, 3, 4, 6, 7 \\15.69426 & \text{ for } & \kappa = 2, 5, 8\end{array}\right.
\end{equation*}
Similarly, for every $\kappa=1, \dots, 8$, the least sum of $\flat$-median forecast errors are smaller than the least sum of $\flat$-mean forecast errors. Therefore, the $\flat$-forecasts are:
\begin{equation*}
^\flat\widehat{x}_{11}^\kappa = \left\{\begin{array}{ccl}15.82406 & \text{ for } & \kappa = 1, 3, 4, 6, 7 \\15.69319 & \text{ for } & \kappa = 2, 5, 8\end{array}\right.
\end{equation*}
Finally, the least sum $\flat$-median forecast errors are the smallest amount the four types, therefore the ${^\natural}$-forecasts are:
\begin{equation*}
{^\natural}\widehat{x}_{11}^\kappa = \left\{\begin{array}{ccl}15.82406 & \text{ for } & \kappa = 1, 3, 4, 6, 7 \\15.69319 & \text{ for } & \kappa = 2, 5, 8\end{array}\right.
\end{equation*}
Recall that $x_{11} =15.64$. The optimization criteria $k=2, 5, 8$ led to a better forecast ex-post. In Figure \ref{Ex1Fig}, we plotted the time series $X$ (10 observations) in solid lines, connected the tenth observation to the eleventh observation by a dashed line, and plotted in blue dots the different forecasted values of the eleventh observation. 
\begin{figure}[htbp]
\begin{center}
\includegraphics[scale=0.3]{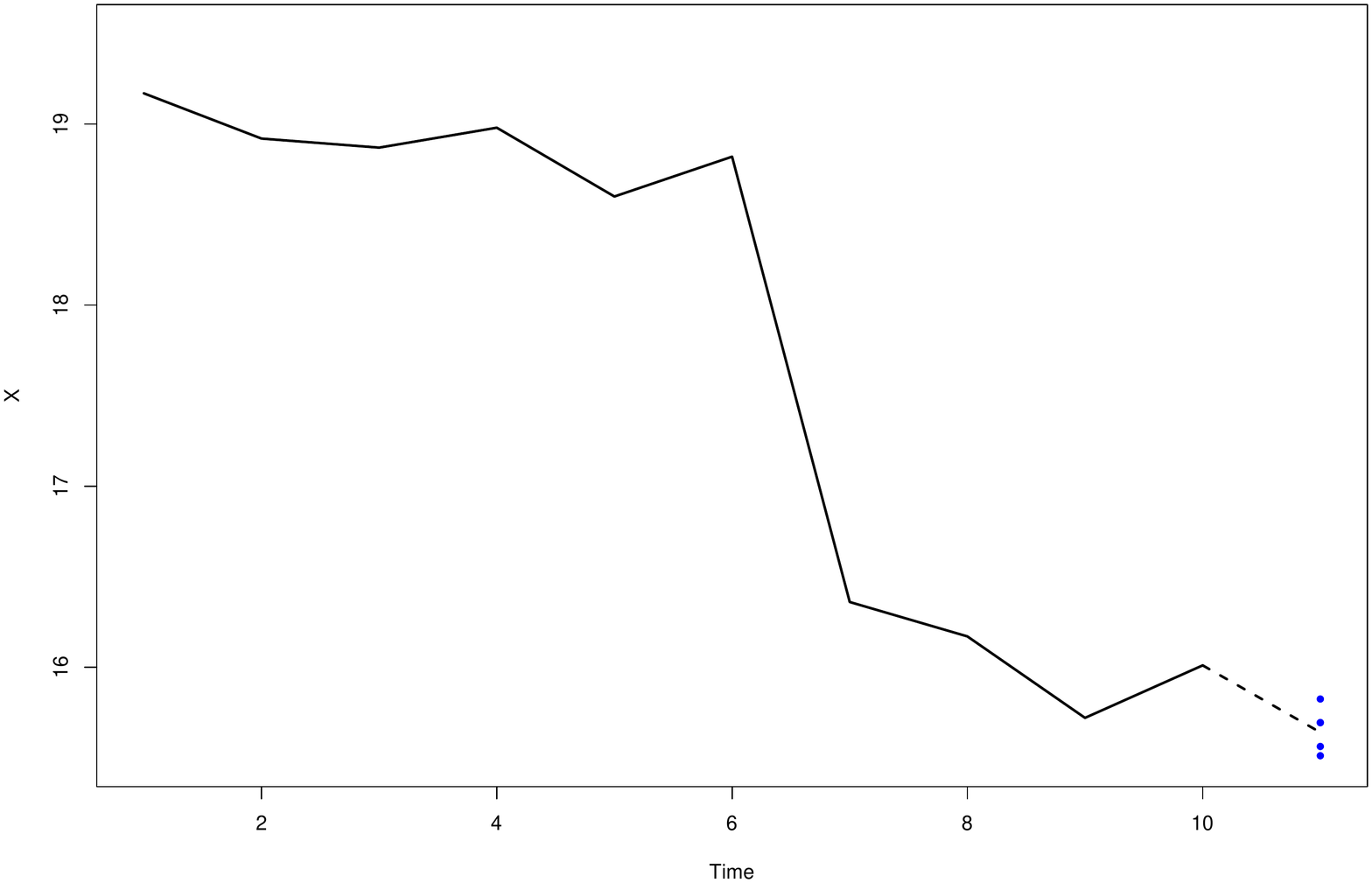}
\caption{Time series $X$ and forecasted values of  $x_{11}$ (in blue)}
\label{Ex1Fig}
\end{center}
\end{figure}
\end{Ex}


\section{Non-Seasonal  Balanced Forecasting Models and Powering}

The floor function is the function that maps a real number $x$ to the greatest integer less than or equal to $x$, denoted  $\lfloor x\rfloor$. For instance, $\lfloor 2.5\rfloor = 2$, $\lfloor 2\rfloor = 2$ and $\lfloor -2.5\rfloor = -3$. Similarly, the ceiling function maps $x$ to the least integer greater than or equal to $x$, denoted $\lceil x \rceil$. For instance, $\lceil 2.5\rceil = 3$, $\lceil 2\rceil= 2$ and $\lceil -2.5\rceil = -2$. We defined in the previous section one hundred twenty $2$-parameter families of forecasting functions with parameters $\lambda_{\text{max}} = 1, \dots, n-2$ (order) and $\nu = 1, \dots, n-2$ (length) such that $\lambda_{\text{max}} + \nu < n$, where $n$ is the size of the time series $X$. Thus, one can choose $n(n-1)/2$ pairs of parameters $(\lambda_{\text{max}}, \nu)$ to forecast $x_{n+1}$ for each optimization criteria $\kappa = 1, \dots, 8$. However, many of these pairs won't lead to the best forecast. While the optimal pair depends on the time series $X$, we propose the following choice of a pair that we found to be often optimal (or near optimal) across several time series.  In what follows, we will focus only on the $^\sharp \Psi$, $^\flat \Psi$ and ${^\natural}\Psi$ forecasting functions.

\begin{Def}[Non-seasonal balanced forecasting models]Let $X = (x_1, \dots, n)$ and $Y= (y_1, \dots, y_n)$ be two time series with $n \geq 2$ observations.   Let $R_Y$ (resp., $D_Y$) be the rate of interest (resp., discount) time series associated to $Y$.  For each optimization criteria $\kappa=1, \dots, 8$, the $\sharp$- (resp., $\flat$-, $\natural$-) balanced forecasting function ${^\sharp}\Psi^\kappa$ (resp.,${^\flat}\Psi^\kappa$, ${^\natural}\Psi^\kappa$) maps the time series $X$ and the rate of interest (resp., discount) time series $R_Y$ (resp., $D_Y$) to a $\sharp$- (resp., $\flat$-, $\natural$-) balanced forecast of $x_{n+1}$, denoted $^\sharp\widehat{x}^\kappa_{n+1}$ (resp., $^\flat\widehat{x}^\kappa_{n+1}$, $^\natural\widehat{x}^\kappa_{n+1}$) , as follows:
\begin{eqnarray}
^\sharp\widehat{x}^\kappa_{n+1}&:=& {^\sharp}\Psi^\kappa_{\lceil \frac{n-1}{2} \rceil,  \lfloor \frac{n-1}{2} \rfloor}(X, R_Y)\\
^\flat\widehat{x}^\kappa_{n+1}&:=& {^\flat}\Psi^\kappa_{\lceil \frac{n-1}{2} \rceil,  \lfloor \frac{n-1}{2} \rfloor}(X, D_Y)\\
{^\natural}\widehat{x}^\kappa_{n+1}&:=& {^\natural}\Psi^\kappa_{\lceil \frac{n-1}{2} \rceil,  \lfloor \frac{n-1}{2} \rfloor}(X, R_Y, D_Y)\end{eqnarray}
\end{Def}
In Example 1., $X$ has $n=10$ observations,  and we used a maximum order $\lambda_{\text{max}} = 5 = \lceil (10-1)/2\rceil $ and length $\nu = 4 = \lfloor (10-1)/2\rfloor$. Thus $^\sharp\widehat{x}^\kappa_{11}$, $^\flat\widehat{x}^\kappa_{11}$ and ${^\natural}\widehat{x}^\kappa_{11}$  were actually balanced forecasts. \\

In order to compare these balanced forecasting functions for a given time series $X = (x_1, \dots, x_n)$, where $n\geq 2$, let's compute the sequence of forecasts of $x_3$, $x_4$, \dots, $x_n$. First, let's adopt the following notation. Given a time series $X = (x_1, \dots, x_n)$, the $(i,j)$-truncated time series of $X$, denoted $X_{[i:j]}$ is the time series 
\begin{equation}
X_{[i:j]} := (x_i, x_{i+1}, \dots, x_{j-1}, x_j)
\end{equation}
In particular, $X = X_{[1:n]}$. The sequence of one-step-ahead forecasts for $X$ are then 
\begin{eqnarray}
\Big({^\sharp}\widehat{x}^\kappa_{i+1} \Big)_{i=2, \dots, n-1}&:=& \Big({^\sharp}\Psi^\kappa_{\lceil \frac{i-1}{2} \rceil,  \lfloor \frac{i-1}{2} \rfloor}(X_{[1:i]}, R_{Y_{[1:i]}}) \Big)_{i=2, \dots, n-1}\\
\Big({^\flat}\widehat{x}^\kappa_{i+1} \Big)_{i=2, \dots, n-1}&:=& \Big({^\flat}\Psi^\kappa_{\lceil \frac{i-1}{2} \rceil,  \lfloor \frac{i-1}{2} \rfloor}(X_{[1:i]}, D_{Y_{[1:i]}}) \Big)_{i=2, \dots, n-1}\\
\Big({^\natural}\widehat{x}^\kappa_{i+1} \Big)_{i=2, \dots, n-1}&:=& \Big({^\natural}\Psi^\kappa_{\lceil \frac{i-1}{2} \rceil,  \lfloor \frac{i-1}{2} \rfloor}(X_{[1:i]}, R_{Y_{[1:i]}}, D_{Y_{[1:i]}}) \Big)_{i=2, \dots, n-1}
\end{eqnarray}

To compare these balanced forecasting functions, one may use the sum of absolute errors, or the sum of square errors. Indeed, for each optimization criteria $\kappa = 1,\dots, 8$, the $\sharp$- (resp., $\flat$-, $\natural$-) sum of absolute errors and the $\sharp$- (resp., $\flat$,  $\natural$-) sum of square errors are defined as follows: 
\begin{eqnarray}
^\sharp \text{SAE}^\kappa &:= \displaystyle \sum_{i=3}^n \Big|{^\sharp}\widehat{x}^\kappa_{i}  - x_{i}\Big| \qquad \text{ and } \qquad ^\sharp \text{SSE}^\kappa &:= \sum_{i=3}^n \Big({^\sharp}\widehat{x}^\kappa_{i}  - x_{i}\Big)^2 \\
^\flat \text{SAE}^\kappa &:=\displaystyle \sum_{i=3}^n \Big|{^\flat}\widehat{x}^\kappa_{i}  - x_{i}\Big| \qquad \text{ and } \qquad ^\flat \text{SSE}^\kappa &:= \sum_{i=3}^n \Big({^\flat}\widehat{x}^\kappa_{i}  - x_{i}\Big)^2 \\
 {^\natural}\text{SAE}^\kappa &:= \displaystyle \sum_{i=3}^n \Big|{^\natural}\widehat{x}^\kappa_{i}  - x_{i}\Big| \qquad \text{ and } \qquad  {^\natural}\text{SSE}^\kappa &:= \sum_{i=3}^n \Big({^\natural}\widehat{x}^\kappa_{i}  - x_{i}\Big)^2 
\end{eqnarray}
The model that minimizes either the sum of absolute errors or the sum of square errors (depending on the forecaster's interests) is said to be optimal for $X$. We illustrate the latter in the following example with nine time series.

\begin{Ex}\label{9TSEx}Let's consider nine time series of total fall enrollment in degree-granting postsecondary institutions, by control and level of institution: 1970 through 2019 (Table 303.25, Digest of Education Statistics, National Center for Education Statistics): all institutions (total, 4-year and 2-year), all public institutions (total, 4-year and 2-year), and all private institutions (total, 4-year and 2-year). For each one of these nine time series, we compute the successive $\sharp$, $\flat$ and $\natural$ balanced forecasted values using each one of the eight optimization criteria, that is, compute 
$$\Big({^\sharp}\widehat{x}^\kappa_{i} \Big)_{i=3, \dots, 50}, \Big({^\flat}\widehat{x}^\kappa_{i} \Big)_{i=3, \dots, 50} \text{ and } \Big({^\natural}\widehat{x}^\kappa_{i} \Big)_{i=3, \dots, 50} \text{ for } \kappa = 1, \dots, 8$$

For each one of the nine time series, we compute the $\sharp$- (resp., $\flat$,  $\natural$-) sum of absolute errors $^\sharp \text{SAE}^\kappa$, where  $\kappa = 1, \dots, 8$. We have then

\begin{table}[H]
\begin{center}
\begin{tabular}{c|r|r|r|r|r|r|r|r}
$\kappa$ & 1 & 2 & 3 & 4 & 5 & 6 & 7 & 8 \\ \hline
$\sharp $  & 12083817 &12551669 &11708995 &11874435 &12410154 &11718068 &11881138& 12088965\\

$\flat$& 12080266 &12539211 &11701715 &11920601 &12471746 &11723804 &11883470 &12524919\\
$\natural$& 12070194 &12537797&11695372 &11910528& 12470331 &11720681 &11877127& 12523519\\ 
\end{tabular}
\end{center}
\caption{ $\sharp$-, $\flat$- and $\natural$-sum of absolute errors for All Institutions Time Series}
\label{default}
\end{table}

\begin{table}[H]
\begin{center}
\begin{tabular}{c|r|r|r|r|r|r|r|r}
$\kappa$ & 1 & 2 & 3 & 4 & 5 & 6 & 7 & 8 \\ \hline
$\sharp$& 4907270 &4972938 &4907270 &4907270& 4909500 &4907270 &4907270 &4909500\\

$\flat $& 4891219 &4968685 &4891219 &4891219& 4904695 &4891219 &4886620 &4904695\\

$\natural$ & 4891138 &4968605 &4891138& 4891138 &4904614& 4891138& 4886540 &4904614
\\
\end{tabular}
\end{center}
\caption{ $\sharp$-, $\flat$- and $\natural$-sum of absolute errors for All 4-year Institutions Time Series}
\label{default}
\end{table}

\begin{table}[H]
\begin{center}
\begin{tabular}{c|r|r|r|r|r|r|r|r}
$\kappa$ & 1 & 2 & 3 & 4 & 5 & 6 & 7 & 8 \\ \hline
$\sharp $ &8903246 &8922155& 8944882& 9241077 &9008852 &8795744& 8717161 &8849391
\\
$\flat$ & 8881884 &8710132 &8820023 &9171905& 8815584& 8692556 &8739160 &8962604\\

$\natural$ & 8886170 &8711491& 8818060 &9187113 &8816916 &8714935 &8743327 &8962247
\\ 
\end{tabular}
\end{center}
\caption{ $\sharp$-, $\flat$- and $\natural$-sum of absolute errors for All 2-year Institutions Time Series}
\label{default}
\end{table}

\begin{table}[H]
\begin{center}
\begin{tabular}{c|r|r|r|r|r|r|r|r} 
$\kappa$ & 1 & 2 & 3 & 4 & 5 & 6 & 7 & 8 \\ \hline
$\sharp$ & 9880555 & 10414381 & 9880555 & 9983663 &10307811 &10119261  &9913959 &10414381
\\
$\flat $ & 9852694 &10148433  &9547249 &9852694 &10340392 & 9742631  &9801691 &10148433
\\
$\natural $ & 9842326 &10143138  &9842326 & 9915269& 10339465& 10054378  &9872083 &10143558
\\ 
\end{tabular}
\end{center}
\caption{ $\sharp$-, $\flat$- and $\natural$-sum of absolute errors for Public Institutions Time Series}
\label{default}
\end{table}

\begin{table}[H]
\begin{center}
\begin{tabular}{c|r|r|r|r|r|r|r|r}
$\kappa$ & 1 & 2 & 3 & 4 & 5 & 6 & 7 & 8 \\ \hline
$\sharp$ & 4026867 &4089837 &4190572 &4026867 &4089837& 4190572 &4122036 &3962909
\\
$\flat$ & 4037233 &4068829 &4130286 &4023680 &4068829 &4179911 &3949545 &3944460
\\
$\natural$ & 4023173& 4068898 &4185809& 4022107 &4068898 &4184744 &3950688 &3944529
\\ 
\end{tabular}
\end{center}
\caption{ $\sharp$-, $\flat$- and $\natural$-sum of absolute errors for 4-year Public Institutions Time Series}
\label{default}
\end{table}

\begin{table}[H]
\begin{center}
\begin{tabular}{c|r|r|r|r|r|r|r|r}
$\kappa$ & 1 & 2 & 3 & 4 & 5 & 6 & 7 & 8 \\ \hline
$\sharp $ & 8621104 &8459579 &8635789 &8503495 &8618638 &8254545 &8509111 &8422546
\\
$\flat $ & 8567851 &8554066& 8444806 &8482210 &8604265 &8327451& 8424247 &8274292
\\
$\natural $ & 8591681 &8520928 &8462315 &8478871& 8624201& 8316568 &8436471 &8272759
\\ \end{tabular}
\end{center}
\caption{ $\sharp$-, $\flat$- and $\natural$-sum of absolute errors for 2-year Public Institutions Time Series}
\label{default}
\end{table}

\begin{table}[H]
\begin{center}
\begin{tabular}{c|r|r|r|r|r|r|r|r}
$\kappa$ & 1 & 2 & 3 & 4 & 5 & 6 & 7 & 8 \\ \hline
$\sharp $ & 3089306 &2568205&3074759 &3089306 &2605606 &3112370 &3092287 &2605606
\\
$\flat $ & 3091305 &2569594 &3071050& 3085748 &2578103 &3071050 &3080076 &2606371
\\
$\natural $ & 3094605 &2570380 &3070191 &3085219& 2608376 &3070191 &3079560 &2608376
\\ 
\end{tabular}
\end{center}
\caption{ $\sharp$-, $\flat$- and $\natural$-sum of absolute errors for Private Institutions Time Series}
\label{default}
\end{table}

\begin{table}[H]
\begin{center}
\begin{tabular}{c|r|r|r|r|r|r|r|r} 
$\kappa$ & 1 & 2 & 3 & 4 & 5 & 6 & 7 & 8 \\ \hline
$\sharp $ & 2004801 &2178096& 2304175 &2200788& 2178096& 2291032& 2364460 &2175763
\\
$\flat $ & 1974659 &2175029 &2284100& 2217746& 2206345 &2269743 &2343571 &2208269
\\
$\natural $ & 1975003 &2183106 &2286858 &2218248 &2183106 &2269836 &2343489 &2177646
\\ 
\end{tabular}
\end{center}
\caption{ $\sharp$-, $\flat$- and $\natural$-sum of absolute errors for Private 4-year Institutions Time Series}
\label{default}
\end{table}

\begin{table}[H]
\begin{center}
\begin{tabular}{c|r|r|r|r|r|r|r|r}
$\kappa$ & 1 & 2 & 3 & 4 & 5 & 6 & 7 & 8 \\ \hline
$\sharp $ & 893238.4 &947318.4 &873704.5 &892711.5 &923941.5 &888010.8 &895441.6 &917444.6
\\
$\flat $ & 885683.5 &915072.1 &869553.6 &889317.2 &922349.9& 888324.6 &941594.4 &904691.7
\\
$\natural $ & 893304.0 &931738.8 &876702.7 &892202.8 &930250.5 &889701.6 &893442.3 &916869.8
\\
\end{tabular}
\end{center}
\caption{ $\sharp$-, $\flat$- and $\natural$-sum of absolute errors for Private 2-year  Institutions Time Series}
\label{default}
\end{table}

For each one of the nine  time series, we extract the best model among the 24 models, i.e., determine the best of the $\sharp$, $\flat$ and $\natural$ along with the best optimization criteria $\kappa = 1, \dots, 8$. This lead to the following summary Table \ref{Best9TS} and plots in Figure \ref{NineTSFig}.,  where the time series are plotted in solid bold black line, and the sequence of forecasts are plotted in a solid red line.

\begin{table}[H]

\begin{center}
\begin{tabular}{l|c|c| r}
 Time Series & $\sharp, \flat, \natural $ & $\kappa $ &  $ \text{SAE}^\kappa$ \\ \hline
 All Institutions & $\natural $ & 3 &   11,695,372\\ 
 All 4-year Institutions & $\natural $ & 7 & 4,886,540 \\
 All 2-year Institutions & $\flat$ & 6 & 8,692,556\\ \hline 
 Public Institutions & $\flat$ & 3 & 9,547,249 \\
 Public 4-year Institutions & $\flat$ & 8 & 3,944,460\\
 Public 2-year Institutions &$\sharp $ & 6 &  8,254,545 \\ \hline
 Private Institutions &  $\sharp$ & 2 &  2,568,205\\
 Private 4-year Institutions & $\flat$ &1 & 1,974,659 \\
 Private 2-year Institutions & $\natural$ & 3 &  876,703 
\end{tabular}
\end{center}
\caption{Optimal Sum of Absolute  Errors}
\label{Best9TS}
\end{table}%

\begin{figure}[H]
\begin{center}
\includegraphics[scale=0.5]{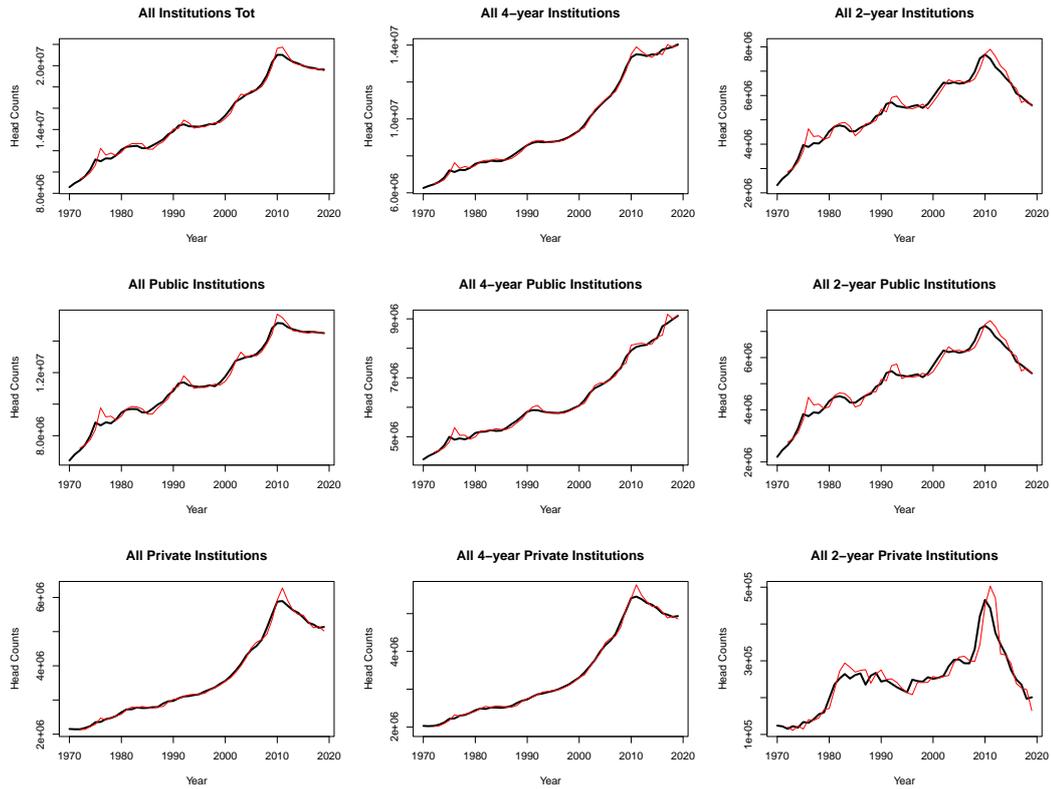}
\caption{Total fall enrollment in degree-granting postsecondary institutions, 1970 through 2019 (in black) and balanced forecasts  $\Big(\widehat{x}^\kappa_{i} \Big)$ (in red)}
\label{NineTSFig}
\end{center}
\end{figure}
\end{Ex}

\begin{Rem}Given a time series, one may compare the three balanced forecasting methods for all optimization criteria, using either the sum of absolute or square errors, to determine  the optimal method and the optimal optimization criteria. The optimal model can then be used to forecast the next (unknown) observation.  \end{Rem}

One can vary the outcomes of the forecasts by considering a function of the time series $Y$ used to generate both the rate of interest time series $R_Y$ and the rate of discount time series $D_Y$.  For instance, one may consider a power of the time series $Y^\alpha$ where $\alpha$ is a real number between 0 and 1. We call this process powering.

\begin{Def}[Non-seasonal $\alpha$-power balanced forecasting models]Let $X = (x_1, \dots, n)$ and $Y= (y_1, \dots, y_n)$ be two time series with $n \geq 2$ observations.  Let $\alpha$ be a real number in $[0,1]$. Let $R_{Y^\alpha}$ (resp., $D_{Y^\alpha}$) be the rate of interest (resp., discount) time series associated to $Y^\alpha$.  For each optimization criteria $\kappa=1, \dots, 8$, the $\alpha$-power $\sharp$- (resp., $\flat$-, $\natural$-) balanced forecast of $x_{n+1}$, denoted $^\sharp\widehat{x}^{\kappa, \alpha}_{n+1}$ (resp., $^\flat\widehat{x}^{\kappa\alpha}_{n+1}$, $^\natural\widehat{x}^{\kappa,\alpha}_{n+1}$) , is defined as follows: 
\begin{eqnarray} \label{sharppowerforecast}
^\sharp\widehat{x}^{\kappa,\alpha}_{n+1}&:=& {^\sharp}\Psi^\kappa_{\lceil \frac{n-1}{2} \rceil,  \lfloor \frac{n-1}{2} \rfloor}(X, R_{Y^\alpha})\\\label{flatpowerforecast}
^\flat\widehat{x}^{\kappa,\alpha}_{n+1}&:=& {^\flat}\Psi^\kappa_{\lceil \frac{n-1}{2} \rceil,  \lfloor \frac{n-1}{2} \rfloor}(X, D_{Y^\alpha})\\\label{naturalpowerforecast}
{^\natural}\widehat{x}^{\kappa, \alpha}_{n+1}&:=& {^\natural}\Psi^\kappa_{\lceil \frac{n-1}{2} \rceil,  \lfloor \frac{n-1}{2} \rfloor}(X, R_{Y^\alpha}, D_{Y^\alpha})\end{eqnarray}
\end{Def}

In order to compare these $\alpha$-power balanced forecasts for a given time series $X = (x_1, \dots, x_n)$, where $n\geq 2$, let's compute the sequence of $\alpha$-power forecasts of $x_3$, $x_4$, \dots, $x_n$ for a given real number $\alpha$ in $[0,1]$. 

\begin{eqnarray}
\Big({^\sharp}\widehat{x}^{\kappa,\alpha}_{i+1} \Big)_{i=2, \dots, n-1}&:=& \Big({^\sharp}\Psi^\kappa_{\lceil \frac{i-1}{2} \rceil,  \lfloor \frac{i-1}{2} \rfloor}(X_{[1:i]}, R_{Y^\alpha_{[1:i]}}) \Big)_{i=2, \dots, n-1}\\
\Big({^\flat}\widehat{x}^{\kappa, \alpha}_{i+1} \Big)_{i=2, \dots, n-1}&:=& \Big({^\flat}\Psi^\kappa_{\lceil \frac{i-1}{2} \rceil,  \lfloor \frac{i-1}{2} \rfloor}(X_{[1:i]}, D_{Y^\alpha_{[1:i]}}) \Big)_{i=2, \dots, n-1}\\
\Big({^\natural}\widehat{x}^{\kappa, \alpha}_{i+1} \Big)_{i=2, \dots, n-1}&:=& \Big({^\natural}\Psi^\kappa_{\lceil \frac{i-1}{2} \rceil,  \lfloor \frac{i-1}{2} \rfloor}(X_{[1:i]}, R_{Y^\alpha_{[1:i]}}, D_{Y^\alpha_{[1:i]}}) \Big)_{i=2, \dots, n-1}
\end{eqnarray}

To compare these $\alpha$-power balanced forecasts, one may use either the sum of absolute errors, or the sum of square errors. Indeed, for each optimization criteria $\kappa = 1,\dots, 8$,  the $\alpha$-power $\sharp$- (resp., $\flat$-, $\natural$-) sum of absolute errors and the $\alpha$-power $\sharp$- (resp., $\flat$,  $\natural$-) sum of square errors are defined as follows: 
\begin{eqnarray}
^\sharp \text{SAE}^{\kappa, \alpha} &:= \displaystyle \sum_{i=3}^n \Big|{^\sharp}\widehat{x}^{\kappa, \alpha}_{i}  - x_{i}\Big| \qquad \text{ and } \qquad ^\sharp \text{SSE}^{\kappa, \alpha} &:= \sum_{i=3}^n \Big({^\sharp}\widehat{x}^{\kappa, \alpha}_{i}  - x_{i}\Big)^2 \\
^\flat \text{SAE}^{\kappa, \alpha} &:=\displaystyle \sum_{i=3}^n \Big|{^\flat}\widehat{x}^{\kappa, \alpha}_{i}  - x_{i}\Big| \qquad \text{ and } \qquad ^\flat \text{SSE}^{\kappa, \alpha} &:= \sum_{i=3}^n \Big({^\flat}\widehat{x}^{\kappa, \alpha}_{i}  - x_{i}\Big)^2 \\
 {^\natural}\text{SAE}^{\kappa, \alpha} &:= \displaystyle \sum_{i=3}^n \Big|{^\natural}\widehat{x}^{\kappa, \alpha}_{i}  - x_{i}\Big| \qquad \text{ and } \qquad  {^\natural}\text{SSE}^{\kappa, \alpha} &:= \sum_{i=3}^n \Big({^\natural}\widehat{x}^{\kappa, \alpha}_{i}  - x_{i}\Big)^2 
\end{eqnarray}

\begin{Ex} Let's reconsider the nine times series of total fall enrollment in degree-granting postsecondary institutions in Example \ref{9TSEx}. For each one of these nine time series,  and for each optimization criteria $\kappa = 1, \dots, 8$, we compute the successive $\sharp$, $\flat$ and $\natural$ balanced forecasted values using  a power $\alpha = 0, 0.1, 0.2, \dots, 0.8, 0.9$ and $1$, and determine the power $\alpha^\ast$ that minimizes the sum of absolute errors. 
 
 \begin{table}[H]
\begin{center}
\begin{tabular}{c|c|r||c|c||c|c}
$\kappa$& $\alpha^\ast$ &  $^\sharp \text{SAE}^{\kappa, \alpha^\ast}$ & $\alpha^\ast$ &  $^\flat \text{SAE}^{\kappa, \alpha^\ast}$  & $\alpha^\ast$ &  $^\natural \text{SAE}^{\kappa, \alpha^\ast}$ \\ \hline 
1 & 0.6 & 10,664,793 &  0.6 & 10,664,258 &0.6 &10,662,800  \\
2 &0.6 &10,951,124 & 0.6 & 10,820,357 & 0.6 & 10,944,929  \\
3& 0.6 & 10,664,793  & 0.6 & 10,664,258 & 0.6 & 10,006,662 \\
4& 0.6 & 10,664,793 & 0.6 & 10,664,258 & 0.6 & 10,662,800 \\
5&  0.6 & 11,037,315 & 0.6 &  10,820,357 & 0.6 & 11,031,120 \\
6& 0.6 & 10,703,072 & 0.6 & 10,704,325  & 0.6 & 10,701,079\\
7& 0.6 & 10,664,793 & 0.6 & 10,508,277 & 0.6 & 10,662,800  \\
8& 0.6 &  10,885,776 & 0.6 & 10,891,924 &  0.6 & 10,879,581 \\
\end{tabular}
\end{center}
\label{default}
\caption{Best $\alpha$-power  $\sharp$-, $\flat$- and $\natural$-SAE for All Institutions Time Series}

\end{table}%

The smallest sum of absolute error for the All Institutions time series is achieved for a power $\alpha =0.6$ with the $\natural$-balanced forecasts and optimization criteria $\kappa = 3$. We have ${^\natural}\text{SAE}^{3, 0.6} = 10,006,662$,  which is a $-14.44\%$ from  ${^\natural}\text{SAE}^{3, 1} = 11,695,372$.    Optimizing ${^\natural}\text{SAE}^{3, \alpha}$ for $\alpha \in [0.5, 0.7]$ leads to $\alpha^\ast = 0.64550$ for which the sum of absolute error is  ${^\natural}\text{SAE}^{3, 0.64550} = 9,976,492$, which is a $-14.70\%$ from ${^\natural}\text{SAE}^{3, 1} = 11,695,372$.

\begin{table}[H]
\begin{center}
\begin{tabular}{c|c|r||c|c||c|c}
$\kappa$& $\alpha^\ast$ &  $^\sharp \text{SAE}^{\kappa, \alpha^\ast}$ & $\alpha^\ast$ &  $^\flat \text{SAE}^{\kappa, \alpha^\ast}$  & $\alpha^\ast$ &  $^\natural \text{SAE}^{\kappa, \alpha^\ast}$ \\ \hline 
1 &  0.8 & 4,613,854 & 0.8 & 4,613,508 & 0.8 & 4,613,232 \\
2 & 0.8 & 4,733,663 &0.8 & 4,731,937& 0.8 & 4,731,395\\
3& 0.8 & 4,613,854 & 0.8 & 4,613,508 & 0.8 & 4,503,882\\
4& 0.8 & 4,640,079 & 0.8 & 4,639,523& 0.8 &4,638,840 \\
5&0.8. &4,682,985 & 0.8 & 4,680,907  & 0.8 & 4,680,364 \\
6& 0.8 & 4,639,463 & 0.8 & 4,613,508  & 0.8 & 4,638,840\\
7& 0.8 & 4,613,854 & 0.8 & 4,613,508 & 0.8 & 4,613,232 \\
8& 0.9 &4,647,772 & 0.9 & 4,644,540 & 0.9 & 4,644,474\\
\end{tabular}
\end{center}
\label{default}
\caption{Best $\alpha$-power  $\sharp$-, $\flat$- and $\natural$-SAE for All 4-year Institutions Time Series}
\end{table}%
 
The smallest sum of absolute error for the All 4-year  Institutions time series is achieved for a power $\alpha =0.8$ with the $\natural$-balanced forecasts and optimization criteria $\kappa = 3$. We have ${^\natural}\text{SAE}^{3, 0.8} = 4,503,882$,  which is a $-7.83\%$ from  ${^\natural}\text{SAE}^{7, 1} = 4,886,540$.    Optimizing ${^\natural}\text{SAE}^{3, \alpha}$ for $\alpha \in [0.7, 0.9]$ leads to $\alpha^\ast = 0.86063$ for which the sum of absolute error is  ${^\natural}\text{SAE}^{3, 0.86063} = 4,463,779$, which is a $-8.65\%$ from ${^\natural}\text{SAE}^{7, 1} =  4,886,540$.

 \begin{table}[H]
\begin{center}
\begin{tabular}{c|c|r||c|c||c|c}
$\kappa$& $\alpha^\ast$ &  $^\sharp \text{SAE}^{\kappa, \alpha^\ast}$ & $\alpha^\ast$ &  $^\flat \text{SAE}^{\kappa, \alpha^\ast}$  & $\alpha^\ast$ &  $^\natural \text{SAE}^{\kappa, \alpha^\ast}$ \\ \hline 
1 & 0.5 & 7,409,595 & 0.5 &7,497,790 & 0.5 & 7,411,853 \\
2 & 0.4 &7,543,155 & 0.4 &7,522,901 & 0.4 &  7,541,927\\
3& 0.5 & 7,482,474 & 0.5 & 7,451,336 & 0.6 &7,403,820 \\
4& 0.5 & 7,451,048 & 0.5 & 7,451,336 & 0.4 &  7,453,700\\
5& 0.4 & 7,488,020 & 0.4 & 7,506,161 & 0.4 & 7,485,853 \\
6& 0.5 & 7,451,048 & 0.5 & 7,476,940 & 0.5 & 7,461,854 \\
7& 0.4 & 7,436,378 & 0.5 & 7,425,560 & 0.4 & 7,437,952\\
8& 0.4 & 7,606,405 & 0.3 &7,584,707 & 0.4 & 7,604,927 \\
\end{tabular}
\end{center}
\label{default}
\caption{Best $\alpha$-power  $\sharp$-, $\flat$- and $\natural$-SAE for All 2-year Institutions Time Series}
\end{table}%

The smallest sum of absolute error for the All 2-year  Institutions time series is achieved for a power $\alpha =0.6$ with the $\natural$-balanced forecasts and optimization criteria $\kappa = 3$. We have ${^\natural}\text{SAE}^{3, 0.6} = 7,403,380$,  which is a $-14.83\%$ from  ${^\flat}\text{SAE}^{6, 1} = 8,692,556$.    Optimizing ${^\natural}\text{SAE}^{3, \alpha}$ for $\alpha \in [0.5, 0.7]$ leads to $\alpha^\ast = 0.51155$ for which the sum of absolute error is  ${^\natural}\text{SAE}^{3, 0.51155} = 7,309,500$, which is a $-15.91\%$ from ${^\flat}\text{SAE}^{6, 1} =  8,692,556$.

\begin{table}[H]
\begin{center}
\begin{tabular}{c|c|r||c|c||c|c}
$\kappa$& $\alpha^\ast$ &  $^\sharp \text{SAE}^{\kappa, \alpha^\ast}$ & $\alpha^\ast$ &  $^\flat \text{SAE}^{\kappa, \alpha^\ast}$  & $\alpha^\ast$ &  $^\natural \text{SAE}^{\kappa, \alpha^\ast}$ \\ \hline 
1 & 0.5 &8,567,196 & 0.5 & 8,556,403 &0.5 & 8,569,512  \\
2 & 0.6 &8,590,567 & 0.6 & 8,571,975 & 0.6 & 8,585,951 \\
3& 0.5 & 8,567,196 & 0.5 &  8,556,403 & 0.5 & 8,328,066 \\
4&0.4 &8,598,591 & 0.4 & 8,600,214 & 0.4 & 8,599,964  \\
5& 0.6 & 8,590,567 &  0.6 & 8,571,975  &  0.6 & 8,586,863\\
6& 0.4 & 8,598,591 & 0.4 & 8,600,214 &  0.4 &  8,599,964 \\
7& 0.4 & 8,574,403 & 0.4 & 8,600,214 & 0.4 &  8,575,776\\
8& 0.6 & 8,590,567& 0.6 & 8,571,975 & 0.6 & 8,586,863\\
\end{tabular}
\end{center}
\label{default}
\caption{Best $\alpha$-power  $\sharp$-, $\flat$- and $\natural$-SAE for Public Institutions Time Series}
\end{table}%

The smallest sum of absolute error for the Public  Institutions time series is achieved for a power $\alpha =0.5$ with the $\natural$-balanced forecasts and optimization criteria $\kappa = 3$. We have ${^\natural}\text{SAE}^{3, 0.5} = 8,328,066$,  which is a $-12.77\%$ from  ${^\flat}\text{SAE}^{3, 1} = 9,547,249$.    Optimizing ${^\natural}\text{SAE}^{3, \alpha}$ for $\alpha \in [0.4, 0.6]$ leads to $\alpha^\ast = 0.56124$ for which the sum of absolute error is  ${^\natural}\text{SAE}^{3, 0.56124} = 8,202,540$, which is a $-8.65\%$ from ${^\flat}\text{SAE}^{3, 1} = 9,547,249$.

\begin{table}[H]
\begin{center}
\begin{tabular}{c|c|r||c|c||c|c}
$\kappa$& $\alpha^\ast$ &  $^\sharp \text{SAE}^{\kappa, \alpha^\ast}$ & $\alpha^\ast$ &  $^\flat \text{SAE}^{\kappa, \alpha^\ast}$  & $\alpha^\ast$ &  $^\natural \text{SAE}^{\kappa, \alpha^\ast}$ \\ \hline 
1 & 0.6 & 3,392,015 & 0.6 & 3,395,082 & 0.6 &  3,394,902 \\
2 & 0.7& 3,510,392 & 0.7 &3,509,022& 0.7 & 3,508,203\\
3& 0.6 & 3,364,669 & 0.6 & 3,366,828 & 0.7 & 3,210,471  \\
4& 0.6 &3,392,015& 0.6 & 3,395,082 & 0.6 & 3,394,900 \\
5&0.7 & 3,510,392 & 0.7 &3,509,022& 0.7 &  3,508,454  \\
6& 0.6 &3,364,669& 0.6 & 3,366,828& 0.6 &  3,367,556 \\
7&0.6 &3,364,669 & 0.6 & 3,366,828& 0.6 & 3,367,558 \\
8& 0.7 & 3,510,392& 0.7 & 3,509,022 & 0.7 &  3,508,454 \\
\end{tabular}
\end{center}
\label{default}
\caption{Best $\alpha$-power  $\sharp$-, $\flat$- and $\natural$-SAE for Public 4-year Institutions Time Series}
\end{table}%

The smallest sum of absolute error for the Public 4-year Institutions time series is achieved for a power $\alpha =0.7$ with the $\natural$-balanced forecasts and optimization criteria $\kappa = 3$. We have ${^\natural}\text{SAE}^{3, 0.7} = 3,210,471$,  which is a $-18.61\%$ from  ${^\flat}\text{SAE}^{8,1} =  3,944,460$.    Optimizing ${^\natural}\text{SAE}^{3, \alpha}$ for $\alpha \in [0.6, 0.8]$ leads to $\alpha^\ast = 0.66485$ for which the sum of absolute error is  ${^\natural}\text{SAE}^{3, 0.66485} = 3,193,186$, which is a $-19.05\%$ from ${^\flat}\text{SAE}^{8, 1} = 3,944,460$.

\begin{table}[H]
\begin{center}
\begin{tabular}{c|c|r||c|c||c|c}
$\kappa$& $\alpha^\ast$ &  $^\sharp \text{SAE}^{\kappa, \alpha^\ast}$ & $\alpha^\ast$ &  $^\flat \text{SAE}^{\kappa, \alpha^\ast}$  & $\alpha^\ast$ &  $^\natural \text{SAE}^{\kappa, \alpha^\ast}$ \\ \hline 
1 &  0.6 &7,062,567 & 0.6 & 7,149,477& 0.6 & 7,061,976 \\
2 &0.4 & 7,050,119& 0.4 & 7,062,291& 0.4 & 7,049,241\\
3& 0.4 & 7,078,305& 0.4 & 7,082,528 & 0.6 & 6,887,148  \\
4&0.6 &6,932,983& 0.4 & 6,973,883& 0.6 & 6,940,619  \\
5& 0.4 & 7,159,756 & 0.4 &7,137,362 & 0.4 &  7,159,639 \\
6& 0.6 &7,016,628 & 0.6 & 7,027,167 & 0.6&  7,024,265 \\
7& 0.6 & 6,990,405& 0.4 & 6,971,065& 0.4 & 6,968,992 \\
8&0.4 &7,202,190& 0.4 &7,060,734 & 0.4&  7,202,073\\
\end{tabular}
\end{center}
\label{default}
\caption{Best $\alpha$-power  $\sharp$-, $\flat$- and $\natural$-SAE for Public 2-year  Institutions Time Series}
\end{table}%

The smallest sum of absolute error for the Public 2-year Institutions time series is achieved for a power $\alpha =0.6$ with the $\natural$-balanced forecasts and optimization criteria $\kappa = 3$. We have ${^\natural}\text{SAE}^{3, 0.6} = 6,887,148$,  which is a $-16.57\%$ from  ${^\sharp}\text{SAE}^{6,1} =  3,944,460$.    Optimizing ${^\natural}\text{SAE}^{3, \alpha}$ for $\alpha \in [0.5, 0.7]$ leads to $\alpha^\ast = 0.62457$ for which the sum of absolute error is  ${^\natural}\text{SAE}^{3, 0.62457} = 6,832,340$, which is a $-17.23\%$ from ${^\sharp}\text{SAE}^{6, 1} = 8,254,545$.

\begin{table}[H]
\begin{center}
\begin{tabular}{c|c|r||c|c||c|c}
$\kappa$& $\alpha^\ast$ &  $^\sharp \text{SAE}^{\kappa, \alpha^\ast}$ & $\alpha^\ast$ &  $^\flat \text{SAE}^{\kappa, \alpha^\ast}$  & $\alpha^\ast$ &  $^\natural \text{SAE}^{\kappa, \alpha^\ast}$ \\ \hline 
1 & 0.9 &2,698,686& 0.9 & 2,703,641& 0.9&2,703,741  \\
2 & 1.0 & 2,568,205& 1.0 & 2,569,594& 1.0 & 2,570,380\\
3& 0.8 &2,693,536& 0.8 & 2,694,572& 0.8 &  2,698,871\\
4& 0.9 &2,719,235& 0.9 & 2,737,380& 0.9& 2,737,284 \\
5& 0.8 &2,601,835& 1.0 & 2,578,103& 0.8 & 2,601,733  \\
6& 0.9 &2,701,260& 0.9 &  2,701,367& 0.9& 2,701,514 \\
7&0.8 & 2,692,108& 0.9 &2,694,137& 0.8&  2,692,502\\
8&0.8 &2,561,153& 0.8 &2,560,631& 0.8 &  2,561,051\\
\end{tabular}
\end{center}
\label{default}
\caption{Best $\alpha$-power  $\sharp$-, $\flat$- and $\natural$-SAE for Private Institutions Time Series}
\end{table}%

The smallest sum of absolute error for the Private Institutions time series is achieved for a power $\alpha =0.8$ with the $\flat$-balanced forecasts and optimization criteria $\kappa = 8$. We have ${^\flat}\text{SAE}^{8, 0.8} = 2,560,631$,  which is a $-0.29\%$ from  ${^\sharp}\text{SAE}^{2,1} =  2,568,205$.    Optimizing ${^\flat}\text{SAE}^{8, \alpha}$ for $\alpha \in [0.7, 0.9]$ leads to $\alpha^\ast = 0.80631$ for which the sum of absolute error is  ${^\flat}\text{SAE}^{8, 0.80631} = 2,554,254$, which is a $-0.54\%$ from ${^\sharp}\text{SAE}^{2, 1} = 2,568,205$.

\begin{table}[H]
\begin{center}
\begin{tabular}{c|c|r||c|c||c|c}
$\kappa$& $\alpha^\ast$ &  $^\sharp \text{SAE}^{\kappa, \alpha^\ast}$ & $\alpha^\ast$ &  $^\flat \text{SAE}^{\kappa, \alpha^\ast}$  & $\alpha^\ast$ &  $^\natural \text{SAE}^{\kappa, \alpha^\ast}$ \\ \hline 
1 & 0.9 &1,938,181& 0.9 &1,938,817& 0.9 & 1,938,797  \\
2 & 0.9 &1,991,630& 0.9 & 1,991,851& 0.9 &  1,991,516\\
3& 0.9 &1,948,956& 0.9 &1,938,817& 0.8 & 2,042,833 \\
4& 0.9 & 1,938,181& 0.9 &1,938,817& 0.9 & 1,938,797 \\
5& 0.9 & 1,991,630& 0.9 & 1,991,851& 0.9 & 1,991,730\\
6& 0.9 &1,948,956& 0.9 & 1,938,817& 0.9 & 1,938,797  \\
7& 0.8 &2,027,714& 0.8 &2,028,053& 0.8 &  2,028,074\\
8& 0.9 &1,982,756& 0.9 &2,011,278& 0.9 & 1,982,642 \\
\end{tabular}
\end{center}
\label{default}
\caption{Best $\alpha$-power  $\sharp$-, $\flat$- and $\natural$-SAE for Private 4-year Institutions Time Series}
\end{table}%

The smallest sum of absolute error for the Private 4-year Institutions time series is achieved for a power $\alpha =0.9$ with the $\sharp$-balanced forecasts and optimization criteria $\kappa = 1$. We have ${^\sharp}\text{SAE}^{1, 0.9} = 1,938,181$,  which is a $-1.85\%$ from  ${^\flat}\text{SAE}^{1,1} =  1,974,659$.    Optimizing ${^\sharp}\text{SAE}^{1, \alpha}$ for $\alpha \in [0.8, 1]$ leads to $\alpha^\ast = 0.90287$ for which the sum of absolute error is  ${^\sharp}\text{SAE}^{1, 0.90287} = 1,937,398$, which is a $-1.89\%$ from ${^\flat}\text{SAE}^{1, 1} = 1,974,659$.

\begin{table}[H]
\begin{center}
\begin{tabular}{c|c|r||c|c||c|c}
$\kappa$& $\alpha^\ast$ &  $^\sharp \text{SAE}^{\kappa, \alpha^\ast}$ & $\alpha^\ast$ &  $^\flat \text{SAE}^{\kappa, \alpha^\ast}$  & $\alpha^\ast$ &  $^\natural \text{SAE}^{\kappa, \alpha^\ast}$ \\ \hline 
1 & 0.4 &752,841& 0.4 &752,968& 0.4 & 752,866  \\
2 & 0.6 & 794,828 & 0.6 & 792,245& 0.6 & 796,085 \\
3& 0.4 &750,137& 0.4 &749,646& 0.5 & 747173 \\
4&0.3 &781,220& 0.3 &781,270& 0.3 & 781,410 \\
5&0.6 & 795,014& 0.5 & 806,706& 0.5 &   809,264\\
6& 0.4 & 745,038& 0.4 & 745,637& 0.4 & 745,894   \\
7& 0.5 &746,759& 0.5 & 747,093& 0.5 & 747,304  \\
8& 0.4 &786,882& 0.4 & 787,248& 0.4 &  787,432\\
\end{tabular}
\end{center}
\label{default}
\caption{Best $\alpha$-power  $\sharp$-, $\flat$- and $\natural$-SAE for Private 2-year Institutions Time Series}
\end{table}%
The smallest sum of absolute error for the Private 2-year Institutions time series is achieved for a power $\alpha =0.4$ with the $\sharp$-balanced forecasts and optimization criteria $\kappa = 6$. We have ${^\sharp}\text{SAE}^{6, 0.4} = 745,038$,  which is a $-15.02\%$ from  ${^\natural}\text{SAE}^{3,1} =  876,703$.    Optimizing ${^\sharp}\text{SAE}^{6, \alpha}$ for $\alpha \in [0.3, 0.5]$ leads to $\alpha^\ast = 0.41801$ for which the sum of absolute error is  ${^\sharp}\text{SAE}^{6, 0.41801} = 741,095$, which is a $-15.47\%$ from $^\natural\text{SAE}^{3, 1} = 876,703$.\\

The following Table \ref{Best9TSsuite} is the continuation of Table \ref{Best9TS} containing the smallest sum of square error among 24 different balanced forecasts ($\sharp$, $\flat$ and $\natural$ for $\kappa = 1, \dots, 8$) to showcase the effect of our powering process.  

\begin{table}[H]

\begin{center}
\begin{tabular}{l|r|r|| r| r| r| r| r| r| r}
 Time Series & $\bullet\kappa$ &  $ \text{SAE}^\kappa$ & $\bullet \kappa$ &  $\alpha$  &  $\text{SAE}^{\kappa, \alpha}$  & $\%$  & $\alpha$  &  $\text{SAE}^{\kappa, \alpha}$  & $\%$  \\ \hline
 All& $\natural 3$ &   11,695,372 & $\natural 3$ & 0.6 & 10,006,622 &-14.44 & 0.64550 & 9,976,492 & -14.70   \\ 
 All 4-year  & $\natural  7$ & 4,886,540 & $\natural 3$ & 0.8 & 4,503,882 & -7.83 &0.86063 & 4,463,779 & -8.65 \\
All 2-year  & $\flat 6$ & 8,692,556  & $\natural 3$ & 0.6 & 7,403,380 & -14.83 & 0.51155 & 7,309,500 & -15.91\\ \hline 
 Public & $\flat 3$ & 9,547,249  & $\natural 3$ & 0.5 & 8,328,066 & -12.77 & 0.56124 & 8,202,540 & -8.65\\
 Public 4-year & $\flat 8$ & 3,944,460  & $\natural 3$ & 0.7 & 3,210,471 & -18.61 & 0.66485 & 3,193,186 &   -19.05\\
 Public 2-year  &$\sharp  6$ &  8,254,545 & $\natural 3$  & 0.6 & 6,887,148 & -16.57 & 0.62457 & 6,832,340 & -17.23\\ \hline
 Private  &  $\sharp 2$ &  2,568,205 & $\flat 8$ & 0.8 & 2,560,631 & -0.29 & 0.80631 & 2,554,254 & -0.54\\
 Private 4-year & $\flat 1$ & 1,974,659 & $\sharp 1$ & 0.9 & 1,193,181 & -1.85 & 0.90287 & 1,937,398 & -1.89 \\
 Private 2-year & $\natural  3$ &  876,703 & $\sharp 6$ & 0.4 & 745,038 & -15.02 & 0.41801 & 741,095 & -15.47  
\end{tabular}
\caption{Optimal Sum of Absolute  Errors}
\label{Best9TSsuite}
\end{center}
\end{table}%

One can notice that while the best in 24 models using balanced forecasts were achieved for various optimization criteria and methods ($\sharp$, $\flat$ and $\natural$), the best $\alpha$-power balanced forecast were achieved by $\natural$-balanced forecasts with the optimization criteria $\kappa = 3$ (goodnes-of-fit type) for six of the nine time series. Moreover, the reduction in sum of absolute errors for these six times series is between 8\% and 19\%.

\begin{figure}
\begin{center}
\includegraphics[scale=0.5]{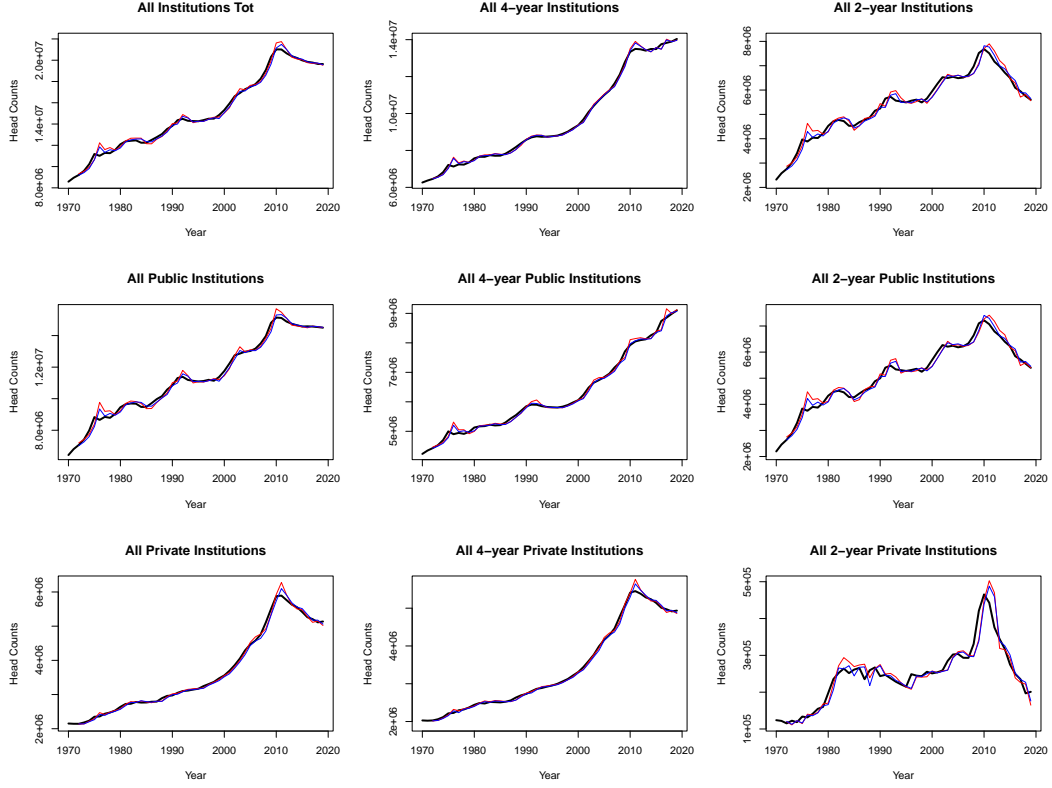}
\caption{Total fall enrollment in degree-granting postsecondary institutions, 1970 through 2019 (in black) and balanced forecasts  $\Big(\widehat{x}^\kappa_{i} \Big)$ (in red) and $\Big(\widehat{x}^{\kappa,\alpha^\ast}_{i} \Big)$ (in blue)}
\label{9TSSuite}
\end{center}
\end{figure}

Figure \ref{9TSSuite} is a continuation of Figure \ref{NineTSFig} where we added the best $\alpha$-powered balanced forecasts (plotted in solid blue lines).

\end{Ex}

In light of the results in the latter example pertaining to the $\alpha$-power balanced forecasts, i.e., to powering, we propose two $\alpha$-optimized balanced forecasting models.

\begin{Def}[Optimal SAE $\alpha$-power] Let $X = (x_1, \dots, n)$ and $Y= (y_1, \dots, y_n)$ be two time series with $n \geq 2$ observations. Let $\alpha$ be a real number in $[0,1]$.  For each optimization criteria $\kappa=1, \dots, 8$, the $i$th-optimal sum of absolute error $\alpha$-power for the $\sharp$- (resp., $\flat$-, $\natural$-) balanced forecast, denoted ${^\sharp \alpha^{\kappa}_i}$ (resp., ${^\flat\alpha^{\kappa}_i}$, ${^\natural\alpha^{\kappa}_i}$), where $i > 2$, is defined as 

\begin{eqnarray}\label{sharpalphai}
{^\sharp \alpha^{\kappa}_3}  = 1, \quad  \text{ and }\quad   {^\sharp \alpha^{\kappa}_i} = \arg\Bigg(  \min_{\alpha \in [0,1]} \sum_{j=3}^{j} \Big({^\sharp \widehat{x}}_j^{\kappa, \alpha} - x_j\Big)\Bigg), \quad i=4, \dots, n
\end{eqnarray}
resp., 
\begin{eqnarray}\label{flatalphai}
{^\flat \alpha^{\kappa}_3} = 1, \quad  \text{ and }\quad   {^\flat \alpha^{\kappa}_i} = \arg\Bigg(  \min_{\alpha \in [0,1]} \sum_{j=3}^{j} \Big({^\flat \widehat{x}}_j^{\kappa, \alpha} - x_j\Big)\Bigg),\qquad  i=4, \dots, n\\\label{naturalalphai}
{^\natural \alpha^{\kappa}_3} = 1, \quad  \text{ and }\quad   {^\natural \alpha^{\kappa}_i} = \arg\Bigg(  \min_{\alpha \in [0,1]} \sum_{j=3}^{j} \Big({^\natural \widehat{x}}_j^{\kappa, \alpha} - x_j\Big)\Bigg), \quad \quad i=4, \dots, n
\end{eqnarray}
where ${^\sharp \widehat{x}}_j^{\kappa, \alpha}$ (resp., ${^\flat \widehat{x}}_j^{\kappa, \alpha}$, ${^\natural \widehat{x}}_j^{\kappa, \alpha}$) is the $\alpha$-power $\sharp$- (resp., $\flat$, $\natural$) balanced forecast of $x_i$ as  defined in (\ref{sharppowerforecast}) (resp., (\ref{flatpowerforecast}), (\ref{naturalpowerforecast})). 
\end{Def}

\begin{Def}[Optimal SAE $\alpha$-power time series] Let $X = (x_1, \dots, n)$ and $Y= (y_1, \dots, y_n)$ be two time series with $n \geq 2$ observations.  For each optimization criteria $\kappa=1, \dots, 8$, the optimal sum of absolute error $\alpha$-power time series for the $\sharp$- (resp., $\flat$-, $\natural$-) balanced forecasts, denoted ${^\sharp \alpha^{\kappa, \ast}}$ (resp., ${^\flat\alpha^{\kappa, \ast}}$, ${^\natural\alpha^{\kappa, \ast}}$), is the time series of size $(n-2)$ defined as 

\begin{eqnarray}
{^\sharp \alpha^\ast} = ({^\sharp \alpha^{k}_3}, {^\sharp \alpha^{k}_4}, {^\sharp \alpha^{k}_5}, \dots, {^\sharp \alpha^{k}_n} )\end{eqnarray}
resp., 
\begin{eqnarray}
{^\flat \alpha^\ast} = ({^\flat \alpha^{k}_3}, {^\flat \alpha^{k}_4}, {^\sharp \alpha^{k}_5}, \dots, {^\sharp \alpha^{k}_n} )\\
{^\natural \alpha^\ast} = ({^\natural \alpha^{k}_5}, {^\natural \alpha^{k}_4}, {^\sharp \alpha^{k}_5}, \dots, {^\sharp \alpha^{k}_n} )
\end{eqnarray}
where ${^\sharp \alpha^{k}_i}$ (resp., ${^\sharp \alpha^{k}_i}$, ${^\sharp \alpha^{k}_i}$) is as in (\ref{sharpalphai}) (resp., (\ref{flatalphai}), (\ref{naturalalphai})). 
\end{Def}

Note that optimal SSE $\alpha$-power times series is defined similarly. Given a time series $X$ of size $n$, together with a time series $Y$ (or $X$ itself) of size $n$ to determine either the rate of interest  or discount time series, or both, one can use the optimal SAE $\alpha$-power time series to forecast $x_{n+1}$. This can be done in several ways. We propose in the following two possible methods. The first uses the mean of the optimal SAE $\alpha$-power time series as a preferred $\alpha$-power for $Y$. The second method uses the latest optimal SAE $\alpha$-power, that is, ${^\sharp \alpha^{k}_n}$, ${^\flat\alpha^{k}_n}$ or ${^\natural\alpha^{k}_n}$, as a preferred $\alpha$-power.

\begin{Def}[Non-seasonal mean-optimized SAE $\alpha$-power balanced forecasts]Let $X = (x_1, \dots, n)$ and $Y= (y_1, \dots, y_n)$ be two time series with $n \geq 2$ observations. Let ${^\sharp \alpha^{\kappa, \ast}}$ (resp., ${^\flat\alpha^{\kappa, \ast}}$, ${^\natural\alpha^{\kappa, \ast}}$) be their corresponding optimal SAE $\alpha$-power time series for the $\sharp$- (resp., $\flat$-, $\natural$-) balanced forecasts.   For each optimization criteria $\kappa=1, \dots, 8$, the mean-optimized sum of absolute error $\alpha$-power $\sharp$- (resp., $\flat$-, $\natural$-) balanced forecast of $x_{n+1}$, denoted $^\sharp\widehat{x}^{\kappa,\overline{\alpha}^\ast}_{n+1}$ (resp., $^\flat\widehat{x}^{\kappa,\overline{\alpha}^\ast}_{n+1}$, $^\natural\widehat{x}^{\kappa,\overline{\alpha}^\ast}_{n+1}$) is defined as follows
\begin{eqnarray}
^\sharp\widehat{x}^{\kappa,\overline{\alpha^\ast}}_{n+1}&:=& {^\sharp}\Psi^\kappa_{\lceil \frac{n-1}{2} \rceil,  \lfloor \frac{n-1}{2} \rfloor}(X, R_{Y^{\overline{^\sharp \alpha^{\kappa, \ast}}}})\\
^\flat\widehat{x}^{\kappa,\overline{\alpha^\ast}}_{n+1}&:=& {^\flat}\Psi^\kappa_{\lceil \frac{n-1}{2} \rceil,  \lfloor \frac{n-1}{2} \rfloor}(X, D_{Y^{\overline{^\flat \alpha^{\kappa, \ast}}}})\\
{^\natural}\widehat{x}^{\kappa, \overline{\alpha^\ast}}_{n+1}&:=& {^\natural}\Psi^\kappa_{\lceil \frac{n-1}{2} \rceil,  \lfloor \frac{n-1}{2} \rfloor}(X, R_{Y^{\overline{^\natural \alpha^{\kappa, \ast}}}}, D_{Y^{\overline{^\natural \alpha^{\kappa, \ast}}}})\end{eqnarray}
where ${\overline{^\sharp \alpha^{\kappa, \ast}}}$ (resp., ${\overline{^\sharp \alpha^{\kappa, \ast}}}$, ${\overline{^\sharp \alpha^{\kappa, \ast}}}$) is the mean of ${^\sharp \alpha^{\kappa, \ast}}$ (resp., ${^\flat\alpha^{\kappa, \ast}}$, ${^\natural\alpha^{\kappa, \ast}}$). 
\end{Def}

\begin{Def}[Non-seasonal latest-optimized SAE $\alpha$-power balanced forecasts]Let $X = (x_1, \dots, n)$ and $Y= (y_1, \dots, y_n)$ be two time series with $n \geq 2$ observations. Let ${^\sharp \alpha^{\kappa}_n}$ (resp., ${^\flat\alpha^{\kappa}_n}$, ${^\natural\alpha^{\kappa}_n}$ be the latest optimal sum of absolute error $\alpha$-power for the $\sharp$- (resp., $\flat$-, $\natural$-) balanced forecasts. For each optimization criteria $\kappa=1, \dots, 8$, the latest-optimized sum of absolute error $\alpha$-power $\sharp$- (resp., $\flat$-, $\natural$-) balanced forecast of $x_{n+1}$, denoted $^\sharp\widehat{x}^{\kappa,{^\sharp \alpha^{\kappa}_n} }_{n+1}$ (resp., $^\flat\widehat{x}^{\kappa,{^\sharp \alpha^{\kappa}_n}}_{n+1}$, $^\natural\widehat{x}^{\kappa,{^\sharp \alpha^{\kappa}_n}}_{n+1}$) is defined as follows
\begin{eqnarray}
^\sharp\widehat{x}^{\kappa,{^\sharp \alpha^{\kappa}_n}}_{n+1}&:=& {^\sharp}\Psi^\kappa_{\lceil \frac{n-1}{2} \rceil,  \lfloor \frac{n-1}{2} \rfloor}(X, R_{Y^{{^\sharp}\alpha^{\kappa}_n}}) 
\\^\flat\widehat{x}^{\kappa,{^\flat \alpha^{\kappa}_n}}_{n+1}&:=& {^\flat}\Psi^\kappa_{\lceil \frac{n-1}{2} \rceil,  \lfloor \frac{n-1}{2} \rfloor}(X, D_{Y^{{^\flat}\alpha^{\kappa}_n}})\\
{^\natural}\widehat{x}^{\kappa, {^\natural \alpha^{\kappa}_n}}_{n+1}&:=& {^\natural}\Psi^\kappa_{\lceil \frac{n-1}{2} \rceil,  \lfloor \frac{n-1}{2} \rfloor}(X, R_{Y^{^\sharp \alpha^{\kappa}_n}}, D_{Y^{^\sharp \alpha^{\kappa}_n}})
\end{eqnarray}
\end{Def}

In the following example, we compute the successive mean- and latest-optimized SAE $\alpha$-power balanced forecasts and compare the sum of absolute error of all twenty four models.

\begin{Ex} Let's reconsider the nine times series of total fall enrollment in degree-granting postsecondary institutions in Example \ref{9TSEx}. For each one of these nine time series, and for each optimization criteria $\kappa = 1, \dots, $ we compute the successive mean- and latest-optimal  SAE $\alpha$-power balanced forecasts for $x_3, \dots, x_{50}$ and compare the sum of absolute error of each sequence of forecasts. Note that we performed a discrete optimization with $\alpha = 0, 0.1, \dots, 0.9, 1$.

 \begin{table}[H]
\begin{center}
\begin{tabular}{ c| r|r||r|r||r|r }
$\kappa$ &  $^\sharp\text{SAE}^{\kappa,\overline{\alpha}^\ast}$& $^\sharp\text{SAE}^{\kappa,{^\sharp \alpha^{\kappa}_n}}$  &  $^\flat\text{SAE}^{\kappa,\overline{\alpha}^\ast}$& $^\flat\text{SAE}^{\kappa,{^\sharp \alpha^{\kappa}_n}}$ &  $^\natural\text{SAE}^{\kappa,\overline{\alpha}^\ast}$& $^\natural\text{SAE}^{\kappa,{^\sharp \alpha^{\kappa}_n}}$\\ \hline
1 &11,077,321 & 11,325,008& 11,074,981&11,320,308&11,073,927&11,323,300 \\
2 & 11,428,905 & 11,574,248&11,296,162&11,362,929&11,420,459& 11,572,920 \\
3 & 11,088,531& 11,443,206&11,086,081&11,279,091&10,349,358& 10,782,300 \\
4 &11,086,684& 11,313,372&11,084,169&11,268,381&11,083,290& 11,311,664 \\
5 &11,428,751& 11,684,148&11,293,405&11,470,667&11,420,305&11,682,820 \\
6 &11,146,686&11,822,899&11,002,907&11,820,798&11,145,118&11,821,192 \\
7 & 10,982,006& 11,147,209&10,980,506&11,105,650&10,978,612& 11,145,501\\
8 &11,366,425& 11,279,755&11,370,166&11,289,392&11,357,979& 11,278,427  \\
\end{tabular}
\end{center}
\caption{SAE of the  mean- and latest-optimal  SAE $\alpha$-power balanced forecasts All Institutions}
\label{default}
\end{table}%

The smallest sum of absolute error for the All Institutions time series is achieved by the mean-optimal SAE $\alpha$-power for the $\natural$-balanced forecasts and optimization criteria $\kappa = 3$. We have $^\natural\text{SAE}^{3,\overline{\alpha}^\ast} = 10,349,358$, which is a $-11.51\%$ from  ${^\natural}\text{SAE}^{3, 1} = 11,695,372$.

 \begin{table}[H]
\begin{center}
\begin{tabular}{ c| r|r||r|r||r|r }
$\kappa$ &  $^\sharp\text{SAE}^{\kappa,\overline{\alpha}^\ast}$& $^\sharp\text{SAE}^{\kappa,{^\sharp \alpha^{\kappa}_n}}$  &  $^\flat\text{SAE}^{\kappa,\overline{\alpha}^\ast}$& $^\flat\text{SAE}^{\kappa,{^\sharp \alpha^{\kappa}_n}}$ &  $^\natural\text{SAE}^{\kappa,\overline{\alpha}^\ast}$& $^\natural\text{SAE}^{\kappa,{^\sharp \alpha^{\kappa}_n}}$\\ \hline
1 & 4,822,683& 4,772,390& 4,822,183& 4,770,618& 4,821,936& 4,770,487 \\
2 & 5,117,012 & 5,029,440& 5,113,847& 5,028,952& 5,113,722& 5,028,994 \\
3 &4,986,866 & 4,826,222& 4,986,353& 4,824,412& 4,882,337& 4,724,232 \\
4 &4,879,855& 4,794,188& 4,879,341& 4,793,812& 4,879,108& 4,793,721 \\
5 &4,841,964& 4,847,401& 4,838,285& 4,848,072&4,837,745& 4,847,519 \\
6 &4,986,210& 4,844,686& 4,985,697& 4,844,311&4,985,464& 4,844,219 \\
7 & 4,976,369& 4,822,798& 4,975,856 & 4,822,422& 4,975,622& 4,822,331\\
8 & 4,824,965& 4,785,079& 4,821,898& 4,785,751&4,821,358& 4,785,197  \\
\end{tabular}
\end{center}
\caption{SAE of the  mean- and latest-optimal  SAE $\alpha$-power balanced forecasts All 4-year Institutions}
\label{default}
\end{table}%

The smallest sum of absolute error for the All 4-year Institutions time series is achieved by the latest-optimal SAE $\alpha$-power for the $\natural$-balanced forecasts and optimization criteria $\kappa = 3$. We have $^\natural\text{SAE}^{3,{^\sharp \alpha^{3}_n}} = 4,724,232 $, which is a $-3.32\%$ from  ${^\natural}\text{SAE}^{7, 1} = 4,886,540$.

 \begin{table}[H]
\begin{center}
\begin{tabular}{ c| r|r||r|r||r|r }
$\kappa$ &  $^\sharp\text{SAE}^{\kappa,\overline{\alpha}^\ast}$& $^\sharp\text{SAE}^{\kappa,{^\sharp \alpha^{\kappa}_n}}$  &  $^\flat\text{SAE}^{\kappa,\overline{\alpha}^\ast}$& $^\flat\text{SAE}^{\kappa,{^\sharp \alpha^{\kappa}_n}}$ &  $^\natural\text{SAE}^{\kappa,\overline{\alpha}^\ast}$& $^\natural\text{SAE}^{\kappa,{^\sharp \alpha^{\kappa}_n}}$\\ \hline
1 & 7,912,331& 7,783,649& 7,923,341& 7,867,412& 7,911,178& 7,783,385 \\
2 & 8,036,105 & 7,948,069&7,993,746& 7,920,251& 8,023,657& 7,940,811 \\
3 &7,930,768 & 7,872,715& 7,916,945& 7,860,548& 7,781,393& 7,574,632 \\
4 & 7,893,691& 7,935,665& 7,919,471& 7,935,562& 7,906,932& 7,942,080 \\
5 & 8,173,617& 7,846,102&8,172,647& 7,849,318& 8,162,395& 7,837,984 \\
6 & 7,968,614& 8,284,838& 7,985,359& 8,290,639& 7,983,197& 8,291,253 \\
7 & 7,859,731& 7,848,066& 7,875,505& 7,854,013& 7,866,660& 7,852,898\\
8 & 8,142,389& 8,054,366& 8,134,907& 8,004,044& 8,130,181& 8,047,108  \\
\end{tabular}
\end{center}
\caption{SAE of the  mean- and latest-optimal  SAE $\alpha$-power balanced forecasts All 2-year Institutions}
\label{default}
\end{table}%

The smallest sum of absolute error for the All 2-year Institutions time series is achieved by the latest-optimal SAE $\alpha$-power for the $\natural$-balanced forecasts and optimization criteria $\kappa = 3$. We have $^\natural\text{SAE}^{3,{^\sharp \alpha^{3}_n}}= 7,574,632$, which is a $-12.86\%$ from  ${^\flat\text{SAE}^{6, 1}} = 8,692,556$.  

 \begin{table}[H]
\begin{center}
\begin{tabular}{ c| r|r||r|r||r|r }
$\kappa$ &  $^\sharp\text{SAE}^{\kappa,\overline{\alpha}^\ast}$& $^\sharp\text{SAE}^{\kappa,{^\sharp \alpha^{\kappa}_n}}$  &  $^\flat\text{SAE}^{\kappa,\overline{\alpha}^\ast}$& $^\flat\text{SAE}^{\kappa,{^\sharp \alpha^{\kappa}_n}}$ &  $^\natural\text{SAE}^{\kappa,\overline{\alpha}^\ast}$& $^\natural\text{SAE}^{\kappa,{^\sharp \alpha^{\kappa}_n}}$\\ \hline
1 & 9,009,883& 9,009,486& 8,958,947& 9,012,350& 9,073,048& 9,011,836 \\
2 &  8,991,093& 8,974,013& 8,975,961& 8,974,994& 8,985,627& 8,972,314 \\
3 & 9,004,076& 9,156,666&8,993,326& 9,158,058& 8,697,732& 8,997,022 \\
4 &9,064,496& 9,091,883& 9,060,250& 9,093,268& 9,057,832& 9,092,759 \\
5 &9,099,509& 8,974,013&9,100,137& 8,974,994& 9,089,997& 8,972,314 \\
6 & 9,069,072& 9,145,986&9,064,842& 9,147,378& 9,062,470& 9,146,862 \\
7 & 9,069,072& 9,145,989&9,033,369&9,147,378& 9,033,130& 9,146,862\\
8 & 8,991,093& 8,974,013& 9,100,137& 8,974,994& 8,980,949& 8,972,314  \\
\end{tabular}
\end{center}
\caption{SAE of the  mean- and latest-optimal  SAE $\alpha$-power balanced forecasts Public Institutions}
\label{default}
\end{table}%

The smallest sum of absolute error for the Public Institutions time series is achieved by the mean-optimal SAE $\alpha$-power for the $\natural$-balanced forecasts and optimization criteria $\kappa = 3$. We have $^\natural\text{SAE}^{3,\overline{\alpha}^\ast} = 8,697,732$, which is a $-9.90\%$ from  ${^\flat}\text{SAE}^{3, 1} = 9,547,249$.

 \begin{table}[H]
\begin{center}
\begin{tabular}{ c| r|r||r|r||r|r }
$\kappa$ &  $^\sharp\text{SAE}^{\kappa,\overline{\alpha}^\ast}$& $^\sharp\text{SAE}^{\kappa,{^\sharp \alpha^{\kappa}_n}}$  &  $^\flat\text{SAE}^{\kappa,\overline{\alpha}^\ast}$& $^\flat\text{SAE}^{\kappa,{^\sharp \alpha^{\kappa}_n}}$ &  $^\natural\text{SAE}^{\kappa,\overline{\alpha}^\ast}$& $^\natural\text{SAE}^{\kappa,{^\sharp \alpha^{\kappa}_n}}$\\ \hline
1 &3,546,458 & 3,572,023& 3,546,101& 3,572,522& 3,545,677& 3,572,127 \\
2 & 3,771,443 & 3,799,141& 3,769,605& 3,799,829& 3,769,026& 3,799,203 \\
3 & 3,503,607& 3,532,882&3,502,393& 3,532,033& 3,357,567& 3,403,461 \\
4 & 3,535,165& 3,560,228& 3,534,825& 3,560,288& 3,553,372& 3,559,763 \\
5 & 3,771,443& 3,799,141& 3,769,605& 3,799,829& 3,769,134& 3,799,203 \\
6 &3,509,739& 3,532,882&3,502,393& 3,532,033& 3,502,814& 3,532,419 \\
7 & 3,503,607& 3,532,882&3,502,393& 3,532,033& 3,502,817& 3,532,652\\
8 & 3,776,382& 3,773,416&3,774,543& 3,774,105&3,774,073& 3,773,478  \\
\end{tabular}
\end{center}
\caption{SAE of the  mean- and latest-optimal  SAE $\alpha$-power balanced forecasts Public 4-year Institutions}
\label{default}
\end{table}%

The smallest sum of absolute error for the Public 4-year Institutions time series is achieved by the mean-optimal SAE $\alpha$-power for the $\natural$-balanced forecasts and optimization criteria $\kappa = 3$. We have $^\natural\text{SAE}^{3,\overline{\alpha}^\ast} = 3,357,567$, which is a $-14.88\%$ from  ${^\flat}\text{SAE}^{8, 1} = 3,944,460$.

 \begin{table}[H]
\begin{center}
\begin{tabular}{ c| r|r||r|r||r|r }
$\kappa$ &  $^\sharp\text{SAE}^{\kappa,\overline{\alpha}^\ast}$& $^\sharp\text{SAE}^{\kappa,{^\sharp \alpha^{\kappa}_n}}$  &  $^\flat\text{SAE}^{\kappa,\overline{\alpha}^\ast}$& $^\flat\text{SAE}^{\kappa,{^\sharp \alpha^{\kappa}_n}}$ &  $^\natural\text{SAE}^{\kappa,\overline{\alpha}^\ast}$& $^\natural\text{SAE}^{\kappa,{^\sharp \alpha^{\kappa}_n}}$\\ \hline
1 & 7,565,375 & 7,571,224& 7,578,884& 7,571,916& 7,563,457& 7,563,858 \\
2 &  7,727,757& 7,380,502&7,671,310& 7,375,044& 7,717,267& 7,373,777 \\
3 & 7,512,239& 7,411,642& 7,514,452& 7,411,454& 7,262,486& 7,222,771 \\
4 & 7,356,661& 7,372,029& 7,368,975& 7,380,444& 7,367,289& 7,373,815 \\
5 & 7,627,587& 7,546,622& 7,623,355& 7,506,599& 7,616,968& 7,540,658 \\
6 & 7,448,254& 7,562,936& 7,452,411& 7,569,345& 7,460,760& 7,564,712 \\
7 & 7,358,151& 7,625,157&7,441,827& 7,589,108& 7,441,500& 7,586,295\\
8 & 7,871,764& 7,576,451& 7,810,599& 7,537,341& 7,441,500& 7,586,295  \\
\end{tabular}
\end{center}
\caption{SAE of the  mean- and latest-optimal  SAE $\alpha$-power balanced forecasts Public 2-year Institutions}
\label{default}
\end{table}%

The smallest sum of absolute error for the Public 2-year Institutions time series is achieved by the latest-optimal SAE $\alpha$-power for the $\natural$-balanced forecasts and optimization criteria $\kappa = 3$. We have $^\natural\text{SAE}^{3,{^\sharp \alpha^{3}_n}}= 7,076,983$, which is a $-14.27\%$ from  ${^\sharp}\text{SAE}^{6, 1} = 8,254,545$.

 \begin{table}[H]
\begin{center}
\begin{tabular}{ c| r|r||r|r||r|r }
$\kappa$ &  $^\sharp\text{SAE}^{\kappa,\overline{\alpha}^\ast}$& $^\sharp\text{SAE}^{\kappa,{^\sharp \alpha^{\kappa}_n}}$  &  $^\flat\text{SAE}^{\kappa,\overline{\alpha}^\ast}$& $^\flat\text{SAE}^{\kappa,{^\sharp \alpha^{\kappa}_n}}$ &  $^\natural\text{SAE}^{\kappa,\overline{\alpha}^\ast}$& $^\natural\text{SAE}^{\kappa,{^\sharp \alpha^{\kappa}_n}}$\\ \hline
1 & 2,902,758 & 2,931,159& 2,903,261& 2,933,221& 2,902,725& 2,931,631 \\
2 & 2,723,120 & 2,663,448& 2,705,166& 2,665,216& 2,723,346& 2,665,628 \\
3 & 2,862,387& 2,971,373& 2,863,492& 2,973,250& 2,940,614& 2,952,808 \\
4 &2,899,131& 2,942,460& 2,884,016& 2,945,180& 2,899,405& 2,943,492 \\
5 & 2,758,598& 2,760,733& 2,761,207& 2,734,004& 2,758,961& 2,763,804 \\
6 & 2,887,218& 2,969,280& 2,888,450& 2,971,597& 2,887,492& 2,968,671 \\
7 & 2,857,226& 2,965,479& 2,870,162& 2,976,320& 2,857,696& 2,964,970\\
8 &2,741,639& 2,699,844& 2,744,335& 2,700,988& 2,742,004& 2,702,619  \\
\end{tabular}
\end{center}
\caption{SAE of the  mean- and latest-optimal  SAE $\alpha$-power balanced forecasts Private Institutions}
\label{default}
\end{table}%

The smallest sum of absolute error for the Private Institutions time series is achieved by the latest-optimal SAE $\alpha$-power for the $\sharp$-balanced forecasts and optimization criteria $\kappa = 2$. We have $^\sharp\text{SAE}^{2,{^\sharp \alpha^{2}_n}}=  2,663,448$, which is a $+3.71\%$ from  ${^\sharp\text{SAE}^{2, 1}} = 2,568,205$.

 \begin{table}[H]
\begin{center}
\begin{tabular}{ c| r|r||r|r||r|r }
$\kappa$ &  $^\sharp\text{SAE}^{\kappa,\overline{\alpha}^\ast}$& $^\sharp\text{SAE}^{\kappa,{^\sharp \alpha^{\kappa}_n}}$  &  $^\flat\text{SAE}^{\kappa,\overline{\alpha}^\ast}$& $^\flat\text{SAE}^{\kappa,{^\sharp \alpha^{\kappa}_n}}$ &  $^\natural\text{SAE}^{\kappa,\overline{\alpha}^\ast}$& $^\natural\text{SAE}^{\kappa,{^\sharp \alpha^{\kappa}_n}}$\\ \hline
1 & 2,053,192 & 2,036,917& 2,053,383& 2,024,944& 2,052,185& 2,025,090 \\
2 & 2,197,592 & 2,090,988& 2,198,237& 2,090,641&2,197,671& 2,091,190 \\
3 & 2,105,968& 2,068,163& 2,106,383& 2,056,225& 2,125,007& 2,258,121 \\
4 &2,079,899& 2,125,798& 2,079,861& 2,140,450& 2,079,668& 2,140,613 \\
5 &2,200,423& 2,097,624&2,220,958& 2,097,277& 2,200,501& 2,098,039 \\
6 & 2,109,281& 2,095,676& 2,109,688& 2,085,902& 2,109,518& 2,083,941 \\
7 & 2,111,902 & 2,277,338& 2,112,304& 2,274,958& 2,112,135& 2,277,019\\
8 & 2,217,394& 2,085,155& 2,274,958& 2,216,075& 2,217,473& 2,085,357  \\
\end{tabular}
\end{center}
\caption{SAE of the  mean- and latest-optimal  SAE $\alpha$-power balanced forecasts Private 4-year Institutions}
\label{default}
\end{table}%

The smallest sum of absolute error for the Private 4-year Institutions time series is achieved by the latest-optimal SAE $\alpha$-power for the $\flat$-balanced forecasts and optimization criteria $\kappa = 1$. We have $^\flat\text{SAE}^{1,{^\flat \alpha^{1}_n}}=  2,024,944$, which is a $+2.55\%$ from  ${^\flat}\text{SAE}^{1, 1} =  1,937,398$.

 \begin{table}[H]
\begin{center}
\begin{tabular}{ c| r|r||r|r||r|r }
$\kappa$ &  $^\sharp\text{SAE}^{\kappa,\overline{\alpha}^\ast}$& $^\sharp\text{SAE}^{\kappa,{^\sharp \alpha^{\kappa}_n}}$  &  $^\flat\text{SAE}^{\kappa,\overline{\alpha}^\ast}$& $^\flat\text{SAE}^{\kappa,{^\sharp \alpha^{\kappa}_n}}$ &  $^\natural\text{SAE}^{\kappa,\overline{\alpha}^\ast}$& $^\natural\text{SAE}^{\kappa,{^\sharp \alpha^{\kappa}_n}}$\\ \hline
1 &771,560 & 783,568& 771,636& 781,805& 771,385& 783,492 \\
2 &803,341  & 827,799&803,536& 825,368& 803,668& 828,137 \\
3 & 768,938& 791,335&769,038& 788,632& 765,508& 793,123 \\
4 &800,142& 833,480& 800,670& 832,264& 800,371& 833,962 \\
5 & 820,787& 833,957& 820,627& 832,328& 820,759& 833,962 \\
6 &766,046& 770,581&766,553& 769,648& 766,760& 774,335 \\
7 & 771,641& 774,893& 771,657& 773,442&771,754& 775,000\\
8 &803,900& 818,336& 804,090& 817,044& 804,278& 818,673  \\
\end{tabular}
\end{center}
\caption{SAE of the  mean- and latest-optimal  SAE $\alpha$-power balanced forecasts Private 2-year  Institutions}
\label{default}
\end{table}%

The smallest sum of absolute error for the Public Institutions time series is achieved by the mean-optimal SAE $\alpha$-power for the $\natural$-balanced forecasts and optimization criteria $\kappa = 3$. We have $^\natural\text{SAE}^{3,\overline{\alpha}^\ast} = 765,507.9 $, which is a $-12.68\%$ from  ${^\natural}\text{SAE}^{3, 1} = 876,702.7$. \\

The following Table \ref{Best9TSsuite2} is a second continuation of Table \ref{Best9TS} containing the smallest sum of square error among 24 mean-optimal SAE $\alpha$-power and 24 latest-optimal SAE $\alpha$-power balanced forecasts ($\sharp$, $\flat$ and $\natural$ for $\kappa = 1, \dots, 8$) to showcase the effect of these two $\alpha$-powering models.  

\begin{table}[h]
\begin{center}
\begin{tabular}{l|r|r|| r| l| r| r}
 Time Series & $\bullet\kappa$ &  $ \text{SAE}^\kappa$ & $\bullet \kappa$ &  $\text{SAE}$ type  &  $\text{SAE}$ value  & $\%$  \\ \hline
 All& $\natural 3$ &   11,695,372 & $\natural 3$ & $^\natural\text{SAE}^{3,\overline{\alpha}^\ast}$  & 10,349,358 & -11.51  \\ 
 All 4-year  & $\natural  7$ & 4,886,540 & $\natural 3$ &  $^\natural\text{SAE}^{3,{^\sharp \alpha^{3}_n}}$  & 4,724,232 &-3.32 \\
All 2-year  & $\flat 6$ & 8,692,556  & $\natural 3$ & $^\natural\text{SAE}^{3,{^\sharp \alpha^{3}_n}}$ & 7,574,632&-12.86 \\ \hline 
 Public & $\flat 3$ & 9,547,249  & $\natural 3$ & $^\natural\text{SAE}^{3,\overline{\alpha}^\ast}$ &8,697,732&-9.90 \\
 Public 4-year & $\flat 8$ & 3,944,460  & $\natural 3$ & $^\natural\text{SAE}^{3,\overline{\alpha}^\ast}$ &3,357,567&-14.88 \\
 Public 2-year  &$\sharp  6$ &  8,254,545 & $\natural 3$ & $^\natural\text{SAE}^{3,{^\sharp \alpha^{3}_n}}$& 7,076,983&-14.27 \\ \hline
 Private  &  $\sharp 2$ &  2,568,205 & $\sharp 2$ & $^\sharp\text{SAE}^{2,{^\sharp \alpha^{2}_n}}$ &  2,663,448&+3.71\\
 Private 4-year & $\flat 1$ & 1,974,659 & $\flat 1$ &  $^\flat\text{SAE}^{1,{^\flat \alpha^{1}_n}}$ &  2,024,944&+2.55 \\
 Private 2-year & $\natural  3$ &  876,703 & $\natural 3$ & $^\natural\text{SAE}^{3,\overline{\alpha}^\ast}$& 765,507.9 & -12.68
\end{tabular}
\caption{Optimal mean and latest SAE $\alpha$-power Sum of Absolute  Errors }
\label{Best9TSsuite2}
\end{center}
\end{table}%

\end{Ex}

\section{Non-Seasonal balanced $\delta$-Forecasting Models }

For large time series (for instance, $n \geq 120$), one may want to use only a portion of the observations rather than the entire time series. We propose in the following a model that uses the latest $\delta$ observations to forecast the next unknown observation. 

 \begin{Def}[Non-seasonal $\delta$-forecasting models]Let $X = (x_1, \dots, x_n)$ and $Y= (y_1, \dots, y_n)$ be two time series with $n \geq 2$ observations.  Let $\delta$ be a fixed integer in $1, \dots, n$.  Let $R_{Y_{[n-\delta+1:n]}}$ (resp., $D_{Y_{[n-\delta+1:n]}}$) be the rate of interest (resp., discount) time series associated to $Y_{[n-\delta+1:n]}$.  For each optimization criteria $\kappa=1, \dots, 8$, the $\sharp$- (resp., $\flat$-, $\natural$-) balanced $\delta$-forecasting function ${^\sharp_\delta}\Psi^\kappa$ (resp., ${^\flat_\delta}\Psi^\kappa$, ${^\natural_\delta}\Psi^\kappa$) maps the time series $X$ and the rate of interest (resp., discount) time series $R_Y$ (resp., $D_Y$) to a $\sharp$- (resp., $\flat$-, $\natural$-) balanced $\delta$-forecast of $x_{n+1}$, denoted ${^\sharp_\delta}\widehat{x}^\kappa_{n+1}$ (resp., $({^\flat_\delta}\widehat{x}^\kappa_{n+1}$, ${^\natural_\delta}\widehat{x}^\kappa_{n+1}$) , as follows:
\begin{eqnarray}
{^\sharp_\delta}\widehat{x}^\kappa_{n+1}=&  {^\sharp_\delta}\Psi^\kappa(X, R_Y) &:= {^\sharp}\Psi^\kappa_{\lceil \frac{\delta-1}{2} \rceil,  \lfloor \frac{\delta-1}{2} \rfloor}(X_{[n-\delta+1:n]}, R_{Y_{[n-\delta+1:n]}})\\
{^\flat_\delta}\widehat{x}^\kappa_{n+1}=& {^\flat_\delta}\Psi^\kappa(X, D_Y) &:= {^\flat}\Psi^\kappa_{\lceil \frac{\delta-1}{2} \rceil,  \lfloor \frac{\delta-1}{2} \rfloor}(X_{[n-\delta+1:n]}, D_{Y_{[n-\delta+1:n]}})\\
{^\natural_\delta}\widehat{x}^\kappa_{n+1}=& {^\natural_\delta}\Psi^\kappa(X, R_Y, D_Y)&:= {^\natural}\Psi^\kappa_{\lceil \frac{\delta-1}{2} \rceil,  \lfloor \frac{\delta-1}{2} \rfloor}(X_{[n-\delta+1:n]}, R_{Y_{[n-\delta+1:n]}}, D_{Y_{[n-\delta+1:n]}})\end{eqnarray}
\end{Def}

In particular,  the balanced $n$-forecast is simply the balanced forecast. In order to compare these balanced forecasting functions for a given time series $X = (x_1, \dots, x_n)$, where $n\geq 2$, let's compute the sequence of $\delta$-forecasts of $x_3$, $x_4$, \dots, $x_n$ for $\delta$ a fixed integer in $\{2, \dots, n\}$. 
\begin{eqnarray}
\Big({^\sharp_\delta}\widehat{x}^\kappa_{i+1} \Big)_{i=\delta, \dots, n-1}&:=& \Big({^\sharp}\Psi^\kappa_{\lceil \frac{\delta-1}{2} \rceil,  \lfloor \frac{\delta-1}{2} \rfloor}(X_{[i-\delta+1:i]}, R_{Y_{[i-\delta+1:i]}}) \Big)_{i=\delta, \dots, n-1}\\
\Big({^\flat_\delta}\widehat{x}^\kappa_{i+1} \Big)_{i=\delta, \dots, n-1}&:=& \Big({^\flat}\Psi^\kappa_{\lceil \frac{\delta-1}{2} \rceil,  \lfloor \frac{\delta-1}{2} \rfloor}(X_{[i-\delta+1:i]}, D_{Y_{[i-\delta+1:i]}}) \Big)_{i=\delta, \dots, n-1}\\
\Big({^\natural_\delta}\widehat{x}^\kappa_{i+1} \Big)_{i=\delta, \dots, n-1}&:=& \Big({^\natural}\Psi^\kappa_{\lceil \frac{\delta-1}{2} \rceil,  \lfloor \frac{\delta-1}{2} \rfloor}(X_{[i-\delta+1:i]}, R_{Y_{[i-\delta+1:i]}}, D_{Y_{[i-\delta+1:i]}}) \Big)_{i=\delta, \dots, n-1}
\end{eqnarray}
To compare these balanced forecasting functions, one may use the sum of absolute errors, or the sum of square errors. Indeed, for each optimization criteria $\kappa = 1,\dots, 8$, the $\sharp$- (resp., $\flat$-, $\natural$-) sum of absolute errors and the $\sharp$- (resp., $\flat$,  $\natural$-) sum of square errors are defined as follows: 
\begin{eqnarray}
{^\sharp_\delta}\text{SAE}^\kappa &:= \displaystyle \sum_{i=\delta+1}^n \Big|{^\sharp_\delta}\widehat{x}^\kappa_{i}  - x_{i}\Big| \qquad \text{ and } \qquad {^\sharp_\delta}\text{SSE}^\kappa &:= \sum_{i=\delta+1}^n \Big({^\sharp_\delta}\widehat{x}^\kappa_{i}  - x_{i}\Big)^2 \\
{^\flat_\delta}\text{SAE}^\kappa &:=\displaystyle \sum_{i=\delta+1}^n \Big|{^\flat_\delta}\widehat{x}^\kappa_{i}  - x_{i}\Big| \qquad \text{ and } \qquad {^\flat_\delta}\text{SSE}^\kappa &:= \sum_{i=\delta+1}^n \Big({^\flat_\delta}\widehat{x}^\kappa_{i}  - x_{i}\Big)^2 \\
 {^\natural_\delta}\text{SAE}^\kappa &:= \displaystyle \sum_{i=\delta+1}^n \Big|{^\natural_\delta}\widehat{x}^\kappa_{i}  - x_{i}\Big| \qquad \text{ and } \qquad  {^\natural_\delta}\text{SSE}^\kappa &:= \sum_{i=\delta+1}^n \Big({^\natural_\delta}\widehat{x}^\kappa_{i}  - x_{i}\Big)^2 
\end{eqnarray}
The model that minimizes either the sum of absolute errors or the sum of square errors (depending on the forecaster's interests) is said to be optimal for $X$. \\

We illustrate in the next  example the sequence of balanced $\delta$-forecasts, and we compare these models to the two most widely used approaches to time series forecasting, i.e., exponential smoothing (proposed by Holt \cite{H}, Brown \cite{B} and Winters \cite{W}) and ARIMA (Auto Regressive Integrated Moving Average) models  \cite{HA}. 

\begin{Ex}Let's consider the monthly time series of the number of new one family home for sales in the United States (in thousands of units) from August 2001 to July 2021, i.e., 240 months (source: US Department of Housing and Urban Development). 
\begin{figure}[H]
\begin{center}
\includegraphics[scale=0.5]{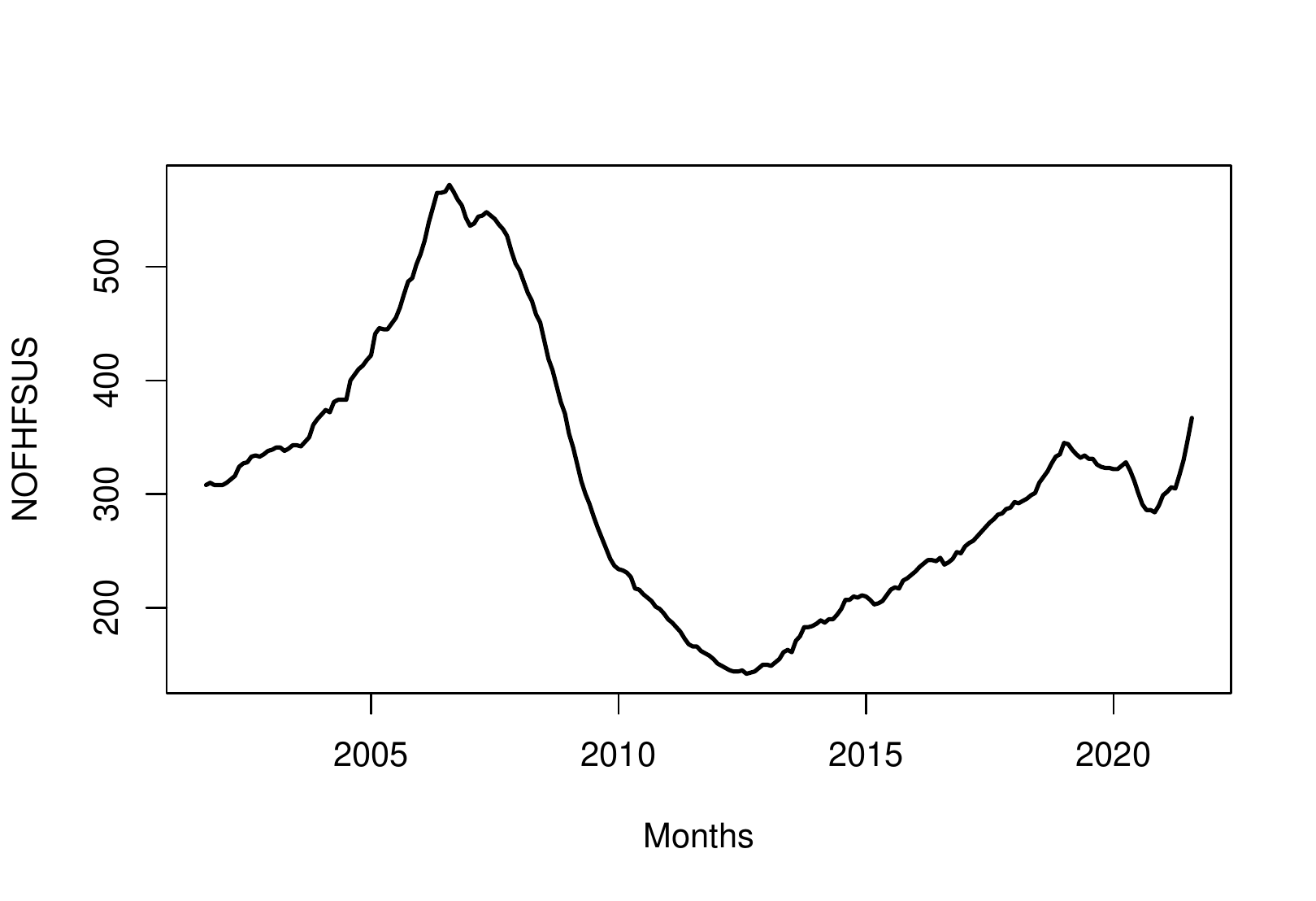}
\caption{Monthly number of new one family home for sales in the United States (in thousands of units) from August 2001 to July 2021}
\end{center}
\end{figure}

Using the Holt--Winters model,  we forecasts the last 100 months, i.e., given time series $X_{[1:m]}$, where $m=140, \dots, 239$, we predict $x_{m+1}$. The sum of absolute and square errors are

\begin{eqnarray}
\text{SAE}^{\text{HW}}  &=&  \displaystyle \sum_{i=141}^{240}\Big|\widehat{x}_{i}^\text{HW}  - x_{i}\Big| = 361.72\\
\text{SSE}^{\text{HW}}  &=&  \displaystyle \sum_{i=141}^{240}\Big(\widehat{x}_{i}^\text{HW}  - x_{i}\Big)^2 = 2139.11\end{eqnarray}

Using the ARIMA model,  we forecasts the last 100 months, i.e., given time series $X_{[1:m]}$, where $m=140, \dots, 239$, we predict $x_{m+1}$. The sum of absolute and square errors are

\begin{eqnarray}
 \text{SAE}^{\text{ARIMA}}  &=&  \displaystyle \sum_{i=141}^{240}\Big|\widehat{x}_{i}^\text{ARIMA}  - x_{i}\Big|  = 313.16\\
 \text{SSE}^{\text{ARIMA}}  &=&  \displaystyle \sum_{i=141}^{240}\Big(\widehat{x}_{i}^\text{ARIMA}  - x_{i}\Big)^2  = 1625.28,
\end{eqnarray}

One can compare two models by counting the number of times a model lead to the closest forecast, i.e., the number of time a model lead to the smallest absolute residual. Comparing Holt--Winters and ARIMA models, we found that out of the 100 forecasts, the Holt--Winters forecasts were the closest 39 times as opposed to 61 times for ARIMA, i/e., $HW / AR = 39 / 61$. Therefore, when taking into account each of  the sum of absolute errors, sum of square errors and the number of smallest absolute residuals, the ARIMA model outperformed the Holt--Winters model. \\

Using the $\sharp$-balanced $\delta$-forecasting function for the optimization criteria $\kappa=1$, we compute the last 100 monthly forecasts for all $\delta = 4, \dots, 120$, and for each $\delta$, we compute the $\sharp$-sum of absolute and square errors 

\begin{eqnarray}
{^\sharp_\delta}\text{SAE}^1 := \displaystyle \sum_{i=141}^{240}\Big|{^\sharp_\delta}\widehat{x}^1_{i}  - x_{i}\Big|  \quad  \text{and} \quad {^\sharp_\delta}\text{SAE}^1 := \displaystyle \sum_{i=141}^{240}\Big({^\sharp_\delta}\widehat{x}^1_{i}  - x_{i}\Big)^2 \quad \text{for }\quad \delta=4, \dots, 120
\end{eqnarray}
In Figure \ref{SAENOFHFSUS}., we plot the graphs of ${^\sharp_\delta}\text{SAE}^1$  and  ${^\sharp_\delta}\text{SSE}^1$ as functions of $\delta$ for $\delta = 4, \dots, 120$. The graphics also contains the sum of absolute errors for both the Holt--Winters and ARIMA models (in dashed lines). One can notice that the sum of balanced absolute errors for the $\delta$-forecasts are low for values of $\delta$ around 24 to 84 months.

\begin{figure}[H]
\begin{center}
\includegraphics[scale=0.42]{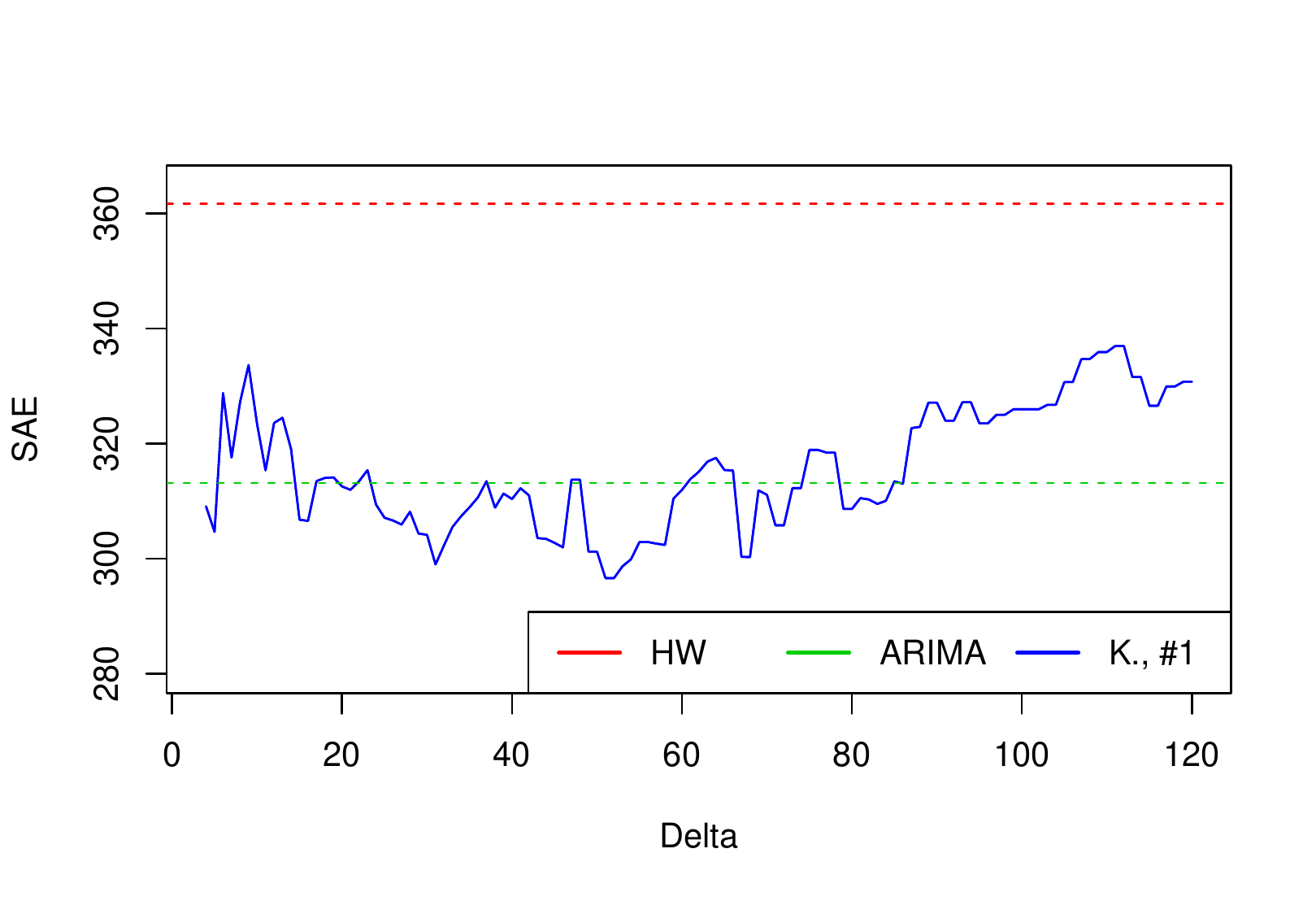} \includegraphics[scale=0.42]{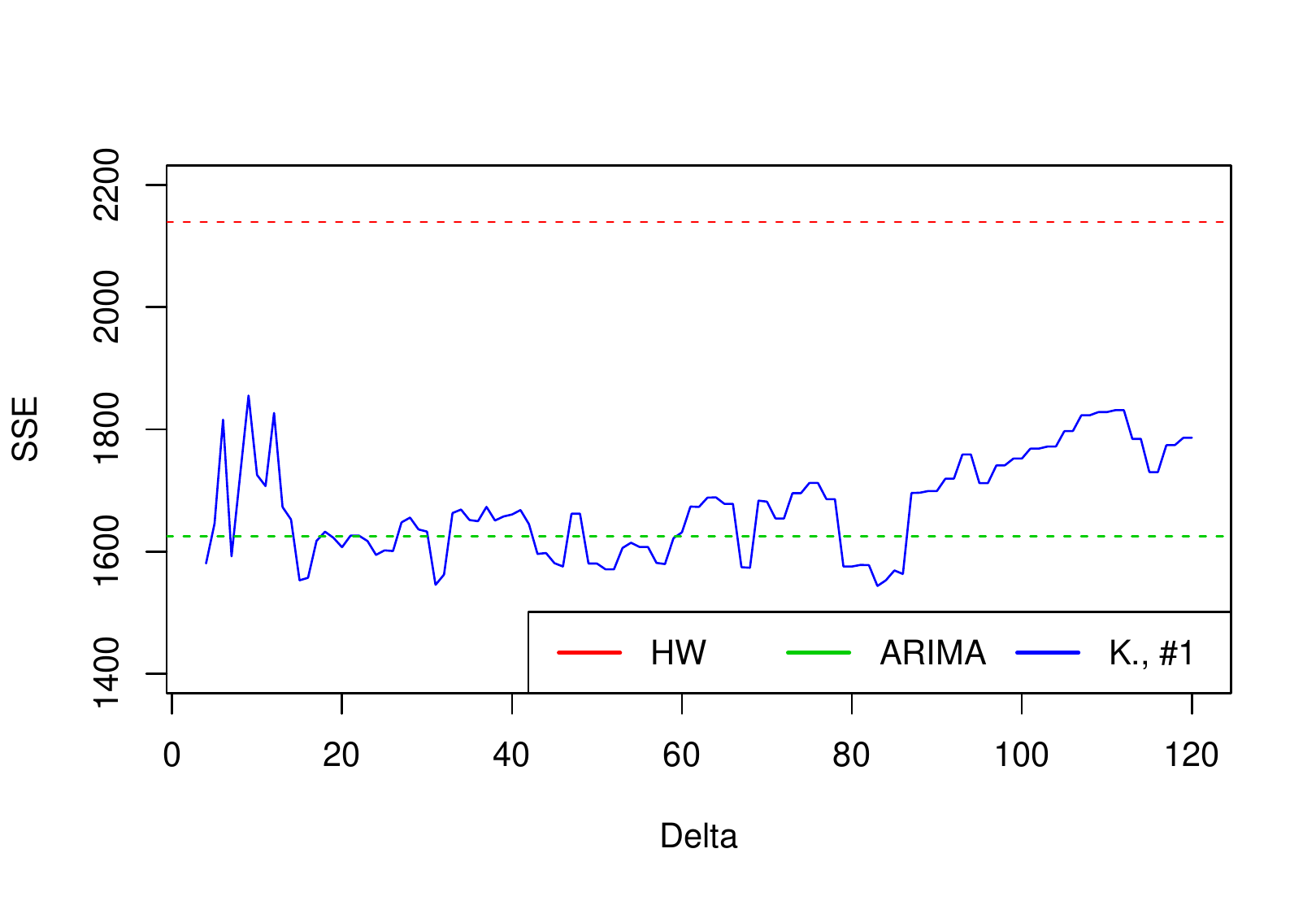}
\caption{The  ${^\sharp_\delta}\text{SAE}^1$  and  ${^\sharp_\delta}\text{SSE}^1$ functions }
\end{center}
\label{SAENOFHFSUS}
\end{figure}

Using $\delta = 24$ (months), we compute the last 100  $\sharp$, $\flat$ and $\natural$ balanced $\delta$-forecasts for each of the eight optimization criteria $\kappa = 1, \dots, 8$. Moreover, we compute the sum of absolute errors ${^\bullet_{24}}\text{SAE}^\kappa$, where $\bullet$ stands for either $\sharp, \flat$ or $\natural$. Finally, we compare these 24 models to both Holt--Winters and ARIMA models by counting the number of time a given model lead to the closest forecast, i.e., the number of smallest absolute residuals. 
\begin{table}[H]
\begin{center}
\begin{tabular}{c|c|c|c|c|c}
$\sharp\kappa$& SAE & SSE & HW / AR /  K. & HW /  K. & AR /  K. \\ \hline
$\sharp 1$ &309.42 & 1594.57 & 33 / 29 / 38 & 44 / 56 & 48 / 52 \\
$\sharp 2$ & 291.05 & 1493.91& 27 / 32 / 41 & 35 / 65 & 49 / 51     \\
$\sharp 3$ & 309.43 &1594.66 & 33 / 29 / 38 & 44 / 56 & 48 /  52\\
$\sharp 4$ & 309.41 & 1594.75  & 33 / 29 / 38 & 44 / 56 & 48 / 56  \\
$\sharp 5$ & 293.65 & 1506.28& 28 / 33 / 39 & 44 / 56 & 50 / 50 \\
$\sharp 6$  & 309.41 & 1594.75 & 33 / 29 / 38 & 44 / 56 & 48 / 52\\
$\sharp 7$ & 309.45 & 1594.82 & 33 / 29 / 38 & 44 / 56 & 48 / 52 \\
$\sharp 8$ & 290.26 & 1490.56& 28 / 31 / 41 & 36 / 64 & 48 / 52\\ 
\end{tabular}
\end{center}
\label{default}
\caption{$\sharp$-Balanced $\delta$-forecasts, Holt--Winters and ARIMA models Comparaison}
\end{table}%

\begin{table}[H]
\begin{center}
\begin{tabular}{c|c|c|c|c|c}
$\flat\kappa$& SAE & SSE & HW / AR /  K. & HW /  K. & AR /  K. \\ \hline
$\flat 1$ & 309.30 & 1595.39& 33 / 29 / 38 & 44 / 56 & 48 / 52 \\
$\flat 2$ &292.84 & 1496.39  &28 / 31 / 41 & 36 / 64 & 48 / 52\\
$\flat 3$& 309.27 & 1595.22& 33 / 29 / 38 & 44 / 56 & 48 / 52 \\
$\flat 4$ & 309.25 & 1595.31& 33 / 29 / 38 & 44 / 56 & 48 / 52\\
$\flat 5$ & 291.94 &1490.08& 29 / 31 / 40 & 37 / 63 & 48 / 52\\
$\flat 6$ & 309.25 & 1595.31 & 33 / 29 / 38 & 44 / 56 & 48 / 52 \\
$\flat 7$ & 309.29 & 1595.38 & 33 / 29 / 30 & 44 / 56 & 48 / 52 \\
$\flat 8$ & 290.80 & 1488.23 & 29 / 30 / 41 & 37 / 63 & 47 / 53 \\ 
\end{tabular}
\end{center}
\label{default}
\caption{$\flat$-Balanced $\delta$-forecasts, Holt--Winters and ARIMA models Comparaison}
\end{table}%

\begin{table}[H]
\begin{center}
\begin{tabular}{c|c|c|c|c|c}
$\natural\kappa$& SAE & SSE & HW / AR /  K. & HW /  K. & AR /  K. \\ \hline
$\natural 1$ & 309.47 & 1594.23 & 33 / 29 / 38 & 44 / 56 & 48 / 52\\
$\natural 2$ & 292.42 & 1495.35  & 28 / 32 / 40 & 36 /  64 & 49 / 51 \\
$\natural 3$ & 309.19 & 1594.23& 33 / 29 / 38 & 44 / 56 & 48 / 52\\
$\natural 4$ & 309.17 &  1594.31& 33 / 29 / 38 & 44 / 56 & 48 / 52 \\
$\natural 5$ & 293.34 &1502.56 & 29 / 32 / 39 & 37 / 63 & 49 / 51 \\
$\natural 6$ & 309.17 & 1594.31& 33 / 29 / 38 & 44 / 56 & 48 / 52\\
$\natural 7$ & 309.21 & 1594.38& 33 / 29 / 38 & 44 / 56 & 48 / 52 \\
$\natural 8$ & 290.72 & 1488.25& 29 / 31 / 40 & 37 / 63 & 48 / 52 \\
\end{tabular}
\end{center}
\label{default}
\caption{$\natural$-Balanced $\delta$-forecasts, Holt--Winters and ARIMA models Comparaison}
\end{table}%
\end{Ex}

\begin{figure}
\begin{center}
\includegraphics[scale=0.55]{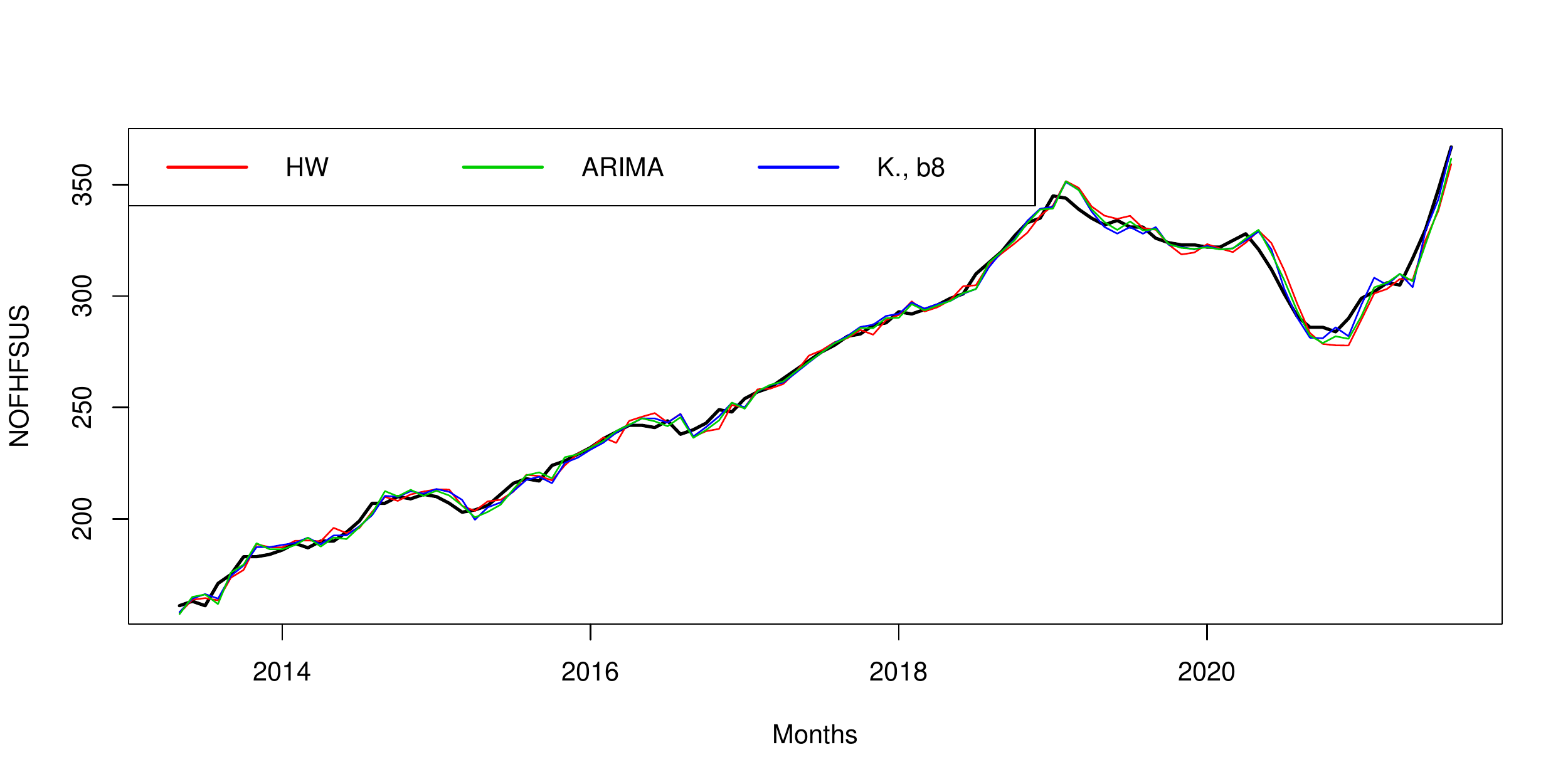}

\caption{The $\flat$-balanced $24$-forecasts ($\kappa = 8$), Holt--Winters and ARIMA models forecasts}
\label{NOFHFSUSTrois}
\end{center}
\end{figure}

In Figure \ref{NOFHFSUSTrois}, we plot the $\flat$-balanced $24$-forecasts ($\kappa = 8$, in blue), Holt--Winters (in red) and ARIMA (in green) models forecasts.

\begin{figure}
\begin{center}

 \includegraphics[scale=0.55]{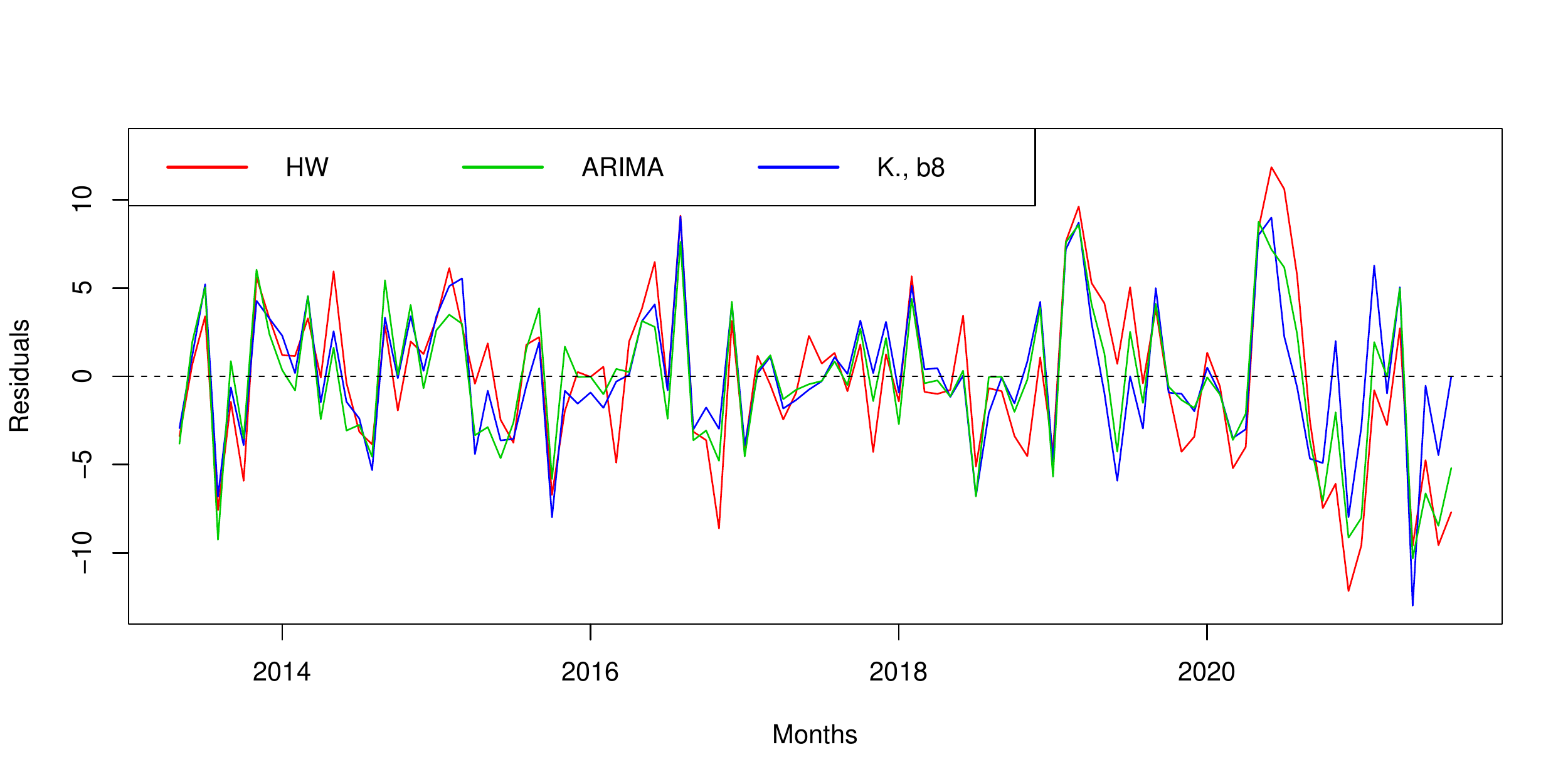} \includegraphics[scale=0.55]{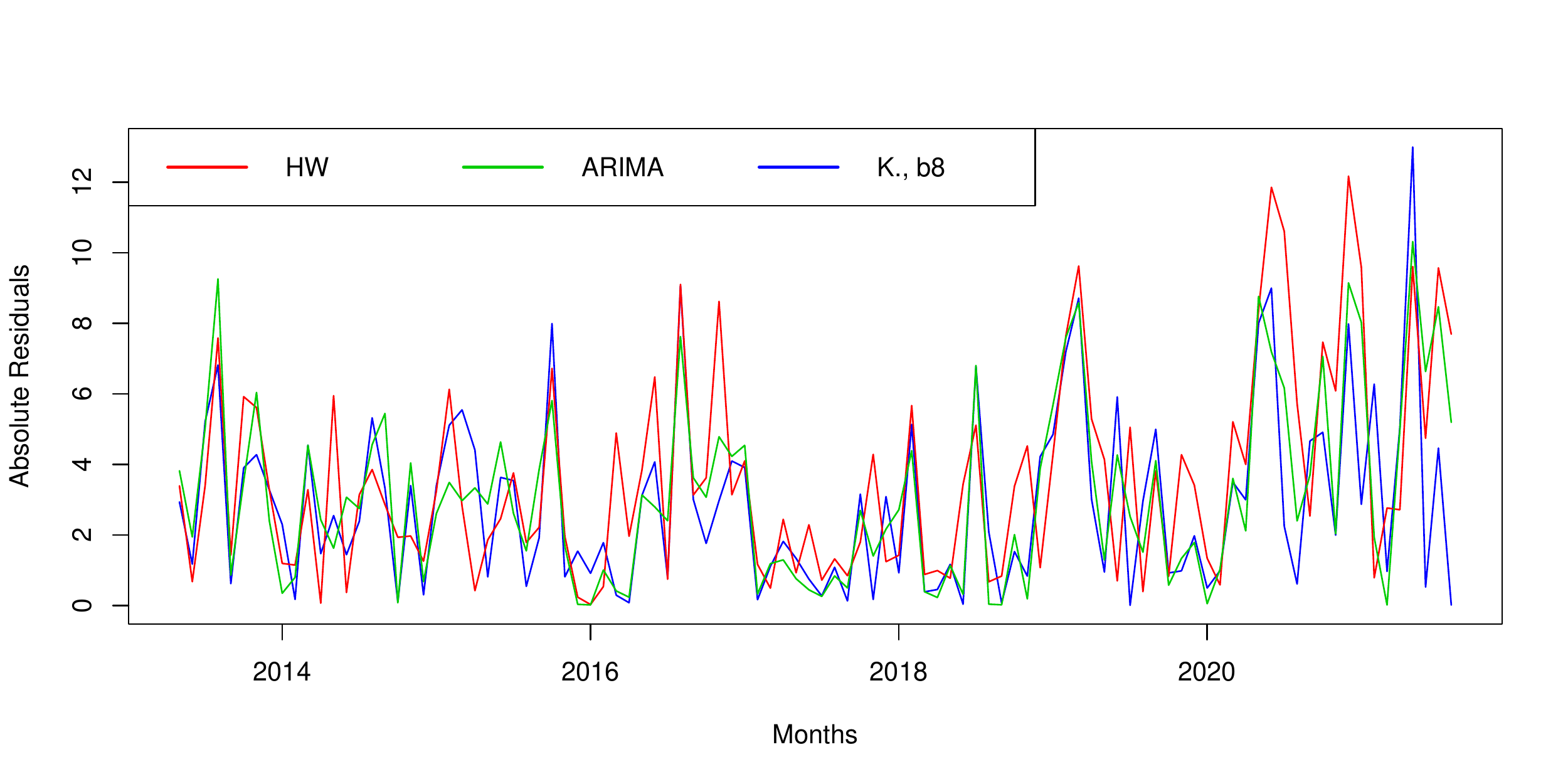}
\caption{The $\flat$-balanced $24$-forecasts ($\kappa = 8$), Holt--Winters and ARIMA models  residuals and absolute residuals}
\label{NOFHFSUSResAbsRes}
\end{center}
\end{figure}

From the above summary Table, all 24 balanced $24$-forecast models outperform both Holt--Winters and ARIMA models with respect to each of the sum of absolute error, sum of square errors, and the number of smallest absolute residuals. Note that the best of the 24 models is the $\flat$-balanced $24$-forecast model for the optimization criteria $\kappa = 8$ (sum of absolute error). Indeed, the sum of absolute error is  ${^\flat_{24}}\text{SAE}^8 = 290.80$ as opposed to 361.72 and 313.16 for Holt--Winters and ARIMA respectively, the sum of square error is  ${^\flat_{24}}\text{SSE}^8 = 1488.23$ as opposed to 2139.11 and 1625.28 for Holt--Winters and ARIMA respectively. In addition, out of 100 forecasts, the $\flat$-balanced $24$-forecast model with $\kappa = 8$ was the closest forecast 41 times, as opposed to 29 and 30 for Holt--Winters and ARIMA respectively, and has 26 more closest forecasts than Holt--Winters models, and 6 more closest forecasts than the ARIMA model. In Figure \ref{NOFHFSUSResAbsRes}., we plot the three models, the residuals and the absolute residuals.

\section{Seasonal  Functionally Balanced Forecasting Models}

Let $X$  be a time series with a frequency $f$ and size $nf$, i.e., 
\begin{eqnarray}
X = \Big(x_1, \dots, x_f, x_{f+1}\dots,  x_{2f}, \dots, x_{(n-1)f+1}, \dots, x_{nf}\Big)
\end{eqnarray}
We denote by $X_{s/f}$ the sub time series of size $n$ defined as follows:
\begin{eqnarray}
X_{s/f} = \Big(x_s, x_{f+s}, x_{2f+s},  \dots, x_{(n-1)f+s}\Big), \quad \text{ where } \quad s = 1, \dots, f
\end{eqnarray}

For instance, if $X$ is a monthly time series starting in January, then the frequency is $f=12$ and $X_{1/12}$ is the January sub time series, $X_{7/12}$ is the July sub time series, and $X_{12/12}$ is the December sub time series. Similarly, if $X$ is a quarterly time series starting in the first quarter of the year (from January $1^\text{st}$ to March $31^\text{st}$), then the frequency is $f=4$, $X_{1/4}$ is the first quarter sub time series, $X_{2/4}$ is the second quarter sub time series, $X_{3/4}$ is the third quarter sub time series and $X_{4/4}$ is the last quarter sub time series.

\begin{Prop}[Seasonal functionally balanced forecasting models]Let $X = (x_1, \dots, x_{nf})$ and $Y= (y_1, \dots, y_{nf})$ be two time series with frequency $f$ and with $nf$ observations, where $n\geq 2$. For each optimization criteria $\kappa=1, \dots, 8$, the $\sharp$- (resp., $\flat$-, $\natural$-) functionally balanced forecasting function ${^\sharp}\Psi^\kappa_f$ (resp.,${^\flat}\Psi^\kappa_f$, ${^\natural}\Psi^\kappa_f$) maps the seasonal time series $X$ and the rate of interest (resp., discount) time series $R_{\Phi(Y)}$ (resp., $D_{\Phi(Y)}$) to a $\sharp$- (resp., $\flat$-, $\natural$-) balanced forecast of $x_{nf+s}$, denoted ${^\sharp}\widehat{x}^\kappa_{nf+s}$ (resp., $({^\flat}\widehat{x}^\kappa_{nf+s}$, ${^\natural}\widehat{x}^\kappa_{nf+s}$), where $s=1, \dots, f$, and $\Phi$ is a function that maps $Y$ to a time series of the same size,  as follows:
\begin{eqnarray}
{^\sharp}\widehat{x}^\kappa_{nf+s}=&  {^\sharp}\Psi^\kappa_f(X, R_{\Phi(Y)}) &:= {^\sharp}\Psi^\kappa_{\lceil \frac{n-1}{2} \rceil,  \lfloor \frac{n-1}{2} \rfloor}(X_{s/f}, R_{{\Phi(Y)_{s/f}}})\\
{^\flat}\widehat{x}^\kappa_{nf+s}=& {^\flat}\Psi^\kappa_f(X, D_{\Phi(Y)}) &:= {^\flat}\Psi^\kappa_{\lceil \frac{n-1}{2} \rceil,  \lfloor \frac{n-1}{2} \rfloor}(X_{s/f}, D_{{\Phi(Y)_{s/f}}})\\
{^\natural}\widehat{x}^\kappa_{nf+s}=& {^\natural}\Psi^\kappa_f(X, R_{\Phi(Y)},  D_{\Phi(Y)})&:= {^\natural}\Psi^\kappa_{\lceil \frac{n-1}{2} \rceil,  \lfloor \frac{n-1}{2} \rfloor}(X_{s/f}, R_{{\Phi(Y)_{s/f}}}, D_{{\Phi(Y)_{s/f}}})\end{eqnarray}
\end{Prop}

Therefore, given a seasonal time series of size $nf$, one can forecast the next $f$ values of the time series, i.e., ${^\bullet}\widehat{x}^\kappa_{nf+1}, {^\bullet}\widehat{x}^\kappa_{nf+2}, \dots, {^\bullet}\widehat{x}^\kappa_{(n+1)f}$, where $\bullet$ is either $\sharp, \flat$ or $\natural$.  In order to compare these seasonal functionally balanced forecasting functions for a given seasonal time series $X = (x_1, \dots, x_{nf})$ with frequency $f$, where $n\geq 2$, let's compute the sequence of forecasts of $x_{2f+1}$, $x_{2f+2}$, \dots, $x_{nf}$. For $s = 1, \dots, f$, and $i=2, \dots, n-1$, the sequence of $f$-step-ahead forecasts for $X$ are then 
\begin{eqnarray}\label{seasxs}
\Big({^\sharp}\widehat{x}^\kappa_{if+s} \Big)_{i, s}&:=& \Big({^\sharp}\Psi^\kappa_{\lceil \frac{i-1}{2} \rceil,  \lfloor \frac{i-1}{2} \rfloor}(X_{[1:if]_{s/f}}, R_{{\Phi(Y_{[1:if]})_{s/f}}}) \Big)_{i, s}\\\label{seasxf}
\Big({^\flat}\widehat{x}^\kappa_{if+s} \Big)_{i, s}&:=& \Big({^\flat}\Psi^\kappa_{\lceil \frac{i-1}{2} \rceil,  \lfloor \frac{i-1}{2} \rfloor}(X_{[1:if]_{s/f}}, D_{{\Phi(Y_{[1:if]})_{s/f}}}) \Big)_{i, s}\\\label{seasxn}
\Big({^\natural}\widehat{x}^\kappa_{if+s} \Big)_{i, s}&:=& \Big({^\natural}\Psi^\kappa_{\lceil \frac{i-1}{2} \rceil,  \lfloor \frac{i-1}{2} \rfloor}(X_{[1:if]_{s/f}}, R_{{\Phi(Y_{[1:if]})_{s/f}}}, D_{{\Phi(Y_{[1:if]})_{j/f}}}) \Big)_{i, s}
\end{eqnarray}

To compare these balanced forecasting functions, one may use two types of  sum of absolute errors, or the sum of square errors: per period and global. Indeed, for each optimization criteria $\kappa = 1,\dots, 8$, the $\sharp$- (resp., $\flat$-, $\natural$-) sum of absolute errors for the $i^\text{th}$ period and the $\sharp$- (resp., $\flat$,  $\natural$-) sum of square errors for the $i^\text{th}$ period, where $i = 3, \dots, n$,  are defined as follows: 
\begin{eqnarray} \label{SAEsi}
^\sharp \text{SAE}^\kappa_i & \hspace{-3mm}:= \displaystyle \sum_{s=1}^f \Big|{^\sharp}\widehat{x}^\kappa_{(i-1)f+s}  - x_{(i-1)f+s}\Big| \quad \text{ and } \quad ^\sharp \text{SSE}^\kappa_i &\hspace{-3mm}:= \sum_{s=3}^f \Big({^\sharp}\widehat{x}^\kappa_{(i-1)f+j}  - x_{(i-1)f+s}\Big)^2 \\  \label{SAEfi}
^\flat \text{SAE}^\kappa_i & \hspace{-3mm}:= \displaystyle \sum_{s=1}^f \Big|{^\flat}\widehat{x}^\kappa_{(i-1)f+s}  - x_{(i-1)f+s}\Big| \quad \text{ and } \quad ^\flat \text{SSE}^\kappa_i &\hspace{-3mm}:= \sum_{s=3}^f \Big({^\flat}\widehat{x}^\kappa_{(i-1)f+j}  - x_{(i-1)f+s}\Big)^2 \\  \label{SAEni}
^\natural \text{SAE}^\kappa_i & \hspace{-3mm}:= \displaystyle \sum_{s=1}^f \Big|{^\natural}\widehat{x}^\kappa_{(i-1)f+s}  - x_{(i-1)f+s}\Big| \quad \text{ and } \quad ^\natural \text{SSE}^\kappa_i &\hspace{-3mm}:= \sum_{s=3}^f \Big({^\natural}\widehat{x}^\kappa_{(i-1)f+j}  - x_{(i-1)f+s}\Big)^2
\end{eqnarray}
and the $\sharp$- (resp., $\flat$-, $\natural$-) sum of absolute errors and the $\sharp$- (resp., $\flat$,  $\natural$-) sum of square errors are defined as follows: 
\begin{eqnarray}
^\sharp \text{SAE}^\kappa & \hspace{-3mm}:= \displaystyle \sum_{i=3}^n  {^\sharp \text{SAE}}^\kappa_i\qquad \text{ and } \qquad ^\sharp \text{SSE}^\kappa&\hspace{-3mm}:= \sum_{i=3}^n {^\sharp \text{SSE}}^\kappa_i\\
^\flat \text{SAE}^\kappa & \hspace{-3mm}:= \displaystyle \sum_{i=3}^n  {^\flat \text{SAE}}^\kappa_i\qquad \text{ and } \qquad ^\flat \text{SSE}^\kappa &\hspace{-3mm}:= \sum_{i=3}^n {^\flat \text{SSE}}^\kappa_i\\
^\natural \text{SAE}^\kappa & \hspace{-3mm}:= \displaystyle \sum_{i=3}^n  {^\natural \text{SAE}}^\kappa_i\qquad \text{ and } \qquad ^\natural \text{SSE}^\kappa &\hspace{-3mm}:= \sum_{i=3}^n {^\natural \text{SSE}}^\kappa_i
\end{eqnarray}

The model that minimizes either the sum of absolute errors or the sum of square errors (depending on the forecaster's interests) is said to be optimal for $X$. One can also count the number of optimal per-period sum of absolute or square errors. We illustrate the latter in the following two examples using one seasonal monthly time series and one seasonal quarterly time series using the following function $\Phi$. 

\begin{Def}[Seasonal power time series mapping] Let  $Y= (y_1, \dots, y_{nf})$ be a time series with frequency $f$ and with $nf$ observations, where $n\geq 2$. Let $\alpha = (\alpha_1, \dots, \alpha_f)$ be an $f$-tuple where $\alpha_s \in [0,1]$ for  $s=1,\dots, f$. The seasonal power times series mapping $\Phi$ maps $Y$ and $\alpha$ to a seasonal time series $\Phi(Y)$ of the same size and frequency such that for each $s=1, \dots, f$, \begin{equation} \label{Phifct}
\Phi(Y)_{s/f} : = Y_{s/f}^{\alpha_{1}} + Y_{(s+1)/f}^{\alpha_{2}} +\dots + Y_{f/f}^{\alpha_{f-s+1}} +  Y_{1/f}^{\alpha_{f-s+2}} +Y_{2/f}^{\alpha_{f-s+3}} + \dots + Y_{(s-1)/f}^{\alpha_{f}}. 
\end{equation}

\end{Def}

\begin{Ex} Let $X$ be the classic  Box and Jenkins airline data \cite{AP}  consisting of the monthly totals of international airline passengers from 1949 to 1960. Thus, $f = 12$ and $X = (x_1, \dots, x_{144})$.  For $s = 1,...,12$ (months), and $i = 2,...,12$ (years), we compute the sequence of twelve-step-ahead forecasts for $X$ using both Holt--Winters and ARIMA models. The sum of absolute errors for each model are
\begin{eqnarray}
 \text{SAE}^{\text{HW}}  = \hspace{-1mm}  \displaystyle \sum_{j=25}^{144}\Big|\widehat{x}_{j}^\text{HW}  - x_{j}\Big|  = 2083 \quad \text{ and } \quad \text{SAE}^{\text{ARIMA}}  = \hspace{-2mm}\displaystyle \sum_{j=25}^{144}\Big|\widehat{x}_{j}^\text{ARIMA}  - x_{j}\Big|  = 2228
\end{eqnarray}
Comparing these $10$ years times $12$ monthly forecasts, i.e., 120 monthly forecasts, the ARIMA model lead to a better monthly forecast (smallest absolute residual)  65 times as opposed to 55 times for the Holt--Winters model, i.e., $HW / AR = 55 / 65$. Moreover, if one compares the sum of absolute error per year, the ARIMA model lead to a better yearly forecast six times as opposed to four times for the Holt--Winters, i.e., $HW / AR = 4 / 6$. \\

For $s = 1,...,12$ (months), and $i = 2,...,12$ (years), we compute the sequence of twelve-step-ahead $\sharp, \flat$ and $\natural$ forecasts for $X$ as in (\ref{seasxs}), (\ref{seasxf}) and (\ref{seasxn}) for each optimization criteria $\kappa = 1, \dots, 8$  using the seasonal power time series mapping $\Phi$ as in $(\ref{Phifct})$ for three examples of $\alpha = (\alpha_1, \dots, \alpha_{12})$. Moreover, we compute the (global) sum of absolute error, and we compare our forecasts to both Holt--Winters and ARIMA models by counting the number of best monthly and yearly forecasts. 

\begin{itemize}
\item When $\alpha = (1/12, 1/11, \dots, 1/2, 1)$, comparing the seasonal power balanced $\sharp$-forecasts lead to 
\end{itemize}
\begin{center}
\begin{tabular}{c||c||c|c|c||c|c|c}
$\sharp \kappa$ & SAE &HW / AR / K. & HW / K. & AR / K. & HW / AR / K. & HW / K. & AR / K. \\ \hline
1 & 1886 & 38 / 37 / 45 & 58 / 62 & 57 / 63 & 3 / 4 / 3 & 5 / 5 & 5 / 5 \\
2 & 1825 & 38 / 38 / 44 & 57 / 63 & 57 / 63 & 3 / 4 / 3 & 5 / 5 & 5 / 5\\
3 &  1886 & 38  /  38  / 44  & 58 / 62  &  58 / 62  & 3 / 4 / 3 & 5 / 5 & 5 / 5\\
4 & 1877 & 38  /  38  / 44  &  58 / 62  &  58 / 62  & 3 / 4 / 3 & 5 / 5 & 5 / 5\\
5 &  1829 &  37 / 38  / 45  &  55 / 65  &  57 / 63   & 3 / 4 / 3 & 5 / 5 & 5 / 5\\
6  &1881  & 38  / 38  / 44  &  57 / 63 &  58 / 62 & 3 / 4 / 3 & 5 / 5 & 5 / 5\\
7  & 1877 &  37 / 38   / 45  &  56 / 64 &  57 / 63  & 3 / 4 / 3 & 5 / 5 & 5 / 5\\
8  &  1834 & 37  / 38  / 45  &  55 / 65 &  57 / 63   & 3 / 4 / 3 & 5 / 5 & 5 / 5
\end{tabular}
\end{center}
The $\flat$ and $\natural$ forecasts lead to equivalent results. 

\begin{itemize}
\item When $\alpha =(1/2^{10}, 1/2^9, \dots, 1/2, 1)$, comparing the seasonal power balanced $\flat$-forecasts lead to 
\end{itemize}
\begin{center}
\begin{tabular}{c||c||c|c|c||c|c|c}
$\flat \kappa$ & SAE &HW / AR / K. & HW / K. & AR / K. & HW / AR / K. & HW / K. & AR / K. \\ \hline
1 & 1906  & 39 / 39 / 42 & 61 / 59 & 59 / 61  & 3 / 4 / 3 & 5 / 5 & 5 / 5 \\
2 & 1834 & 39 / 39 / 42  & 60 / 60 & 58 / 62  & 3 / 4 / 3 & 5 / 5 & 5 / 5 \\
3 &  1906 & 39 / 40 / 41 & 61 / 59 & 60 / 60 & 3 / 4 / 3 & 5 / 5 & 5 / 5 \\
4 & 1888 & 39 / 39 / 42 & 61 / 59 & 59 / 61  & 3 / 4 / 3 & 5 / 5 & 5 / 5 \\
5 & 1823 & 38 / 39 / 43 & 58 / 62 & 58 / 62 &  3 / 4 / 3 & 5 / 5 & 5 / 5 \\
6 &  1899 & 39 / 40 / 41 & 60 / 60 & 60 / 60  & 3 / 4 / 3 & 5 / 5 & 5 / 5 \\
7 & 1894 & 38 / 40 / 42 & 59 / 61 & 59 / 61    & 3 / 4 / 3 & 5 / 5 & 5 / 5 \\
8 & 1847 & 38 / 39 / 42 & 58 / 62 & 58 / 62 &  3 / 4 / 3 & 5 / 5 & 5 / 5

\end{tabular}
\end{center}

The $\sharp$ and $\natural$ forecasts lead to equivalent results.

\begin{itemize}
\item When $\alpha =  (e^{-11}, e^{-10}, \dots, e^{-1}, 1) $, comparing the seasonal power balanced $\natural$-forecasts lead to 
\end{itemize}
\begin{center}
\begin{tabular}{c||c||c|c|c||c|c|c}
$\natural \kappa$ & SAE &HW / AR / K. & HW / K. & AR / K. & HW / AR / K. & HW / K. & AR / K. \\ \hline
1 &  1907 & 38 / 38 / 44 & 59 / 61 & 58 / 62 & 2 / 4 / 4 & 4 / 6 & 5 / 5 \\
2 &  1847 & 37 / 38 / 45 & 59 / 61 & 57 / 63 & 2 / 4 / 4 & 4 / 6 & 5 / 5 \\
3 & 1929 & 40 / 39 / 41 & 59 / 61 & 59 / 61  & 3 / 4 / 3 & 5 / 5 & 5 / 5 \\
4 &1895 & 38 / 39 / 43 & 60 / 60 & 58 / 62  &   2 / 4 / 4 & 4 / 6 & 5 / 5 \\
5 &  1844 & 37 / 38 / 45 & 58 / 62 & 57 / 63 & 2 / 4 / 4 & 4 / 6 & 5 / 5 \\
6 & 1911 & 38 / 40 / 42 & 59 / 61 & 59 / 61  & 2 / 4 / 4 & 4 / 6 & 5 / 5 \\
7 &  1896 & 37 / 40 / 43 & 58 / 62 & 59 / 61  & 2 / 4 / 4 & 4 / 6 & 5 / 5 \\
8 & 1866 & 37 / 38 / 45 & 58 / 62 & 57 / 63 &  2 / 4 / 4 & 4 / 6 & 5 / 5 \\
\end{tabular}
\end{center}
The $\sharp$ and $\flat$ forecasts lead to equivalent results.

\end{Ex}

In this example, we used a (three) fixed $\alpha$ power to compute the seasonally balanced forecasts. While the sum of absolute errors of our $\sharp-, \flat-$ and $\natural$-forecasts were consistently lower than both Holt--Winters and ARIMA models, the number of best monthly and/or yearly forecasts were fairly higher. In the following, we introduce an optimized and dynamic method to select the $\alpha$-powers for the seasonally balanced forecasts. The latter leads to a substantially improved forecasts for seasonal time series.

\begin{Def}[Seasonal  per-period SAE latest-optimized  $\alpha$-power] Let $X = (x_1, \dots, x_{nf})$ and $Y= (y_1, \dots, y_{nf})$ be two time series with frequency $f$ and with $nf$ observations, where $n\geq 2$. For each optimization criteria $\kappa=1, \dots, 8$, the seasonal $\sharp$ (resp., $\flat$, $\natural$)  SAE latest-optimized  $\alpha$-power for the period $i = 3,\dots, n $, denoted $^\sharp{\alpha}^\kappa_{i}$ (resp.,  $^\flat{\alpha}^\kappa_{i}$, $^\natural{\alpha}^\kappa_{i}$), are defined as the $f$-tuples 
\begin{equation}
{^\sharp}{\alpha}^\kappa_{3} =  {^\flat}{\alpha}^\kappa_{3}  ={^\natural}{\alpha}^\kappa_{3}  = (1, 1, \dots, 1)
\end{equation}
and
\begin{eqnarray}
^\sharp{\alpha}^\kappa_{i} =  ({^\sharp}{\alpha}^\kappa_{1,i}, \dots, {^\sharp}{\alpha}^\kappa_{f,i}) &: =& \displaystyle \arg \bigg(  \min_{\beta_1, \dots, \beta_f \in [0,1]} {^\sharp}{SAE}^\kappa_{i-1} \bigg) \quad \text{ for } \quad  i = 4, \dots, n\\
^\flat{\alpha}^\kappa_{i} =  ({^\flat}{\alpha}^\kappa_{1,i}, \dots, {^\flat}{\alpha}^\kappa_{f,i}) &: =& \displaystyle \arg \bigg(  \min_{\beta_1, \dots, \beta_f \in [0,1]} {^\flat}{SAE}^\kappa_{i-1} \bigg) \quad \text{ for } \quad  i = 4, \dots, n\\
^\natural{\alpha}^\kappa_{i} =  ({^\natural}{\alpha}^\kappa_{1,i}, \dots, {^\natural}{\alpha}^\kappa_{f,i}) &: =& \displaystyle \arg \bigg(  \min_{\beta_1, \dots, \beta_f \in [0,1]} {^\natural}{SAE}^\kappa_{i-1} \bigg) \quad \text{ for } \quad  i = 4, \dots, n
\end{eqnarray}
where $\displaystyle {^\sharp}{SAE}^\kappa_i, {^\flat} {SAE}^\kappa_i$ and $\displaystyle {^\natural} {SAE}^\kappa_i$ are as in (\ref{SAEsi}), (\ref{SAEfi}) and  (\ref{SAEni}), and $\Phi(Y)$ as in (\ref{Phifct}).

\end{Def}

We define the sequence of seasonal  per-period SSE latest-optimized  $\alpha$-power in a similar way. While an analytic derivation of the sequence of these seasonal per-period SAE latest-optimized  $\alpha$-powers can be (in practice) tedious, we introduce in the following  stochastic (and more pragmatic) method to obtain an approximate of the sequence of these seasonal per-period SAE latest-optimized  $\alpha$-powers, which we use for the forecasting.

\begin{Def}[Seasonal  per-period SAE stochastic latest-optimized  $\alpha$-power]  Let $X = (x_1, \dots, x_{nf})$ and $Y= (y_1, \dots, y_{nf})$ be two time series with frequency $f$ and with $nf$ observations, where $n\geq 2$. For each optimization criteria $\kappa=1, \dots, 8$,  the random $\sharp$ (resp., $\flat$, $\natural$) SAE $\beta$-power for the period $i = 3,\dots, n $, denoted ${^\sharp_\beta}{\widetilde{SAE}}^\kappa_{i}$ is the sum of absolute error for the $i$th period where the power $\beta = (\beta_1, \dots, \beta_f)$ in $\Phi(Y)$ are such that each of the $\beta_i$  are drawn from a uniform distribution on the interval $[0,1]$. Thus, for each optimization criteria $\kappa =1, \dots, 8$, the seasonal $\sharp$ (resp., $\flat$, $\natural$)  SAE stochastic latest-optimized  $\alpha$-power for the period $i = 3,\dots, n $, denoted $^\sharp{\widetilde{\alpha}}^\kappa_{i}$ (resp.,  $^\flat{\widetilde{\alpha}}^\kappa_{i}$, $^\natural{\widetilde{\alpha}}^\kappa_{i}$), are defined as the $f$-tuples 
\begin{equation}
{^\sharp}{\widetilde{\alpha}}^\kappa_{3} =  {^\flat}{\widetilde{\alpha}}^\kappa_{3}  ={^\natural}{\widetilde{\alpha}}^\kappa_{3}  = (1, 1, \dots, 1)
\end{equation}
and
\begin{eqnarray} \label{salphas}
^\sharp{\widetilde{\alpha}}^\kappa_{i} =  ({^\sharp}{\widetilde{\alpha}}^\kappa_{1,i}, \dots, {^\sharp}{\widetilde{\alpha}}^\kappa_{f,i}) &: =& \displaystyle \arg_{\beta_1, \dots, \beta_f } \bigg(  \min_{j=1, \dots, N} {^\sharp_\beta}{\widetilde{SAE}}^\kappa_{i-1} \bigg) \quad \text{ for } \quad  i = 4, \dots, n\\\label{falphas}
^\flat{\widetilde{\alpha}}^\kappa_{i} =  ({^\flat}{\widetilde{\alpha}}^\kappa_{1,i}, \dots, {^\flat}{\widetilde{\alpha}}^\kappa_{f,i}) &: =& \displaystyle \arg_{\beta_1, \dots, \beta_f } \bigg(  \min_{j=1, \dots, N} {^\flat_\beta}{\widetilde{SAE}}^\kappa_{i-1} \bigg) \quad \text{ for } \quad  i = 4, \dots, n\\ \label{nalphas}
^\natural{\widetilde{\alpha}}^\kappa_{i} =  ({^\natural}{\widetilde{\alpha}}^\kappa_{1,i}, \dots, {^\natural}{\widetilde{\alpha}}^\kappa_{f,i}) &: =& \displaystyle \arg_{\beta_1, \dots, \beta_f } \bigg(  \min_{j=1, \dots, N} {^\natural_\beta}{\widetilde{SAE}}^\kappa_{i-1} \bigg) \quad \text{ for } \quad  i = 4, \dots, n
\end{eqnarray}
where $N$ is a large number (for instance, $N=1000$). 
\end{Def}

\begin{Prop}[Seasonal per-period SAE stochastic latest-optimized  $\widetilde{\alpha}$-power  balanced forecasting models]Let $X = (x_1, \dots, x_{nf})$ and $Y= (y_1, \dots, y_{nf})$ be two time series with frequency $f$ and with $nf$ observations, where $n\geq 2$. For each optimization criteria $\kappa=1, \dots, 8$, the $\sharp$- (resp., $\flat$-, $\natural$-) seasonal per-period SAE stochastic latest-optimized  $\widetilde{\alpha}$-power balanced forecasting function ${^\sharp}\Psi^\kappa_f$ (resp.,${^\flat}\Psi^\kappa_f$, ${^\natural}\Psi^\kappa_f$) maps the seasonal time series $X$ and the rate of interest (resp., discount) time series $R_{\widetilde{\Phi}(Y)}$ (resp., $D_{\widetilde{\Phi}(Y)}$) to a $\sharp$- (resp., $\flat$-, $\natural$-) balanced forecast of $x_{nf+s}$, denoted ${^\sharp}\widehat{x}^\kappa_{nf+s}$ (resp., $({^\flat}\widehat{x}^\kappa_{nf+s}$, ${^\natural}\widehat{x}^\kappa_{nf+s}$), where $s=1, \dots, f$, and $\widetilde{\Phi}$ is as in (\ref{Phifct}) with the $\sharp$-power $ ^\sharp{\widetilde{\alpha}}^\kappa$ (resp., $^\flat{\widetilde{\alpha}}^\kappa$, $^\natural{\widetilde{\alpha}}^\kappa$) is as in (\ref{salphas}) (resp., (\ref{falphas}, (\ref{nalphas})),  as follows:
\begin{eqnarray}\label{}
{^\sharp}\widehat{x}^\kappa_{nf+s}=&  {^\sharp}\Psi^\kappa_f(X, R_{\widetilde{\Phi}(Y)}) &:= {^\sharp}\Psi^\kappa_{\lceil \frac{n-1}{2} \rceil,  \lfloor \frac{n-1}{2} \rfloor}(X_{s/f}, R_{{\widetilde{\Phi}(Y)_{s/f}}})\\
{^\flat}\widehat{x}^\kappa_{nf+s}=& {^\flat}\Psi^\kappa_f(X, D_{\widetilde{\Phi}(Y)}) &:= {^\flat}\Psi^\kappa_{\lceil \frac{n-1}{2} \rceil,  \lfloor \frac{n-1}{2} \rfloor}(X_{s/f}, D_{{\widetilde{\Phi}(Y)_{s/f}}})\\
{^\natural}\widehat{x}^\kappa_{nf+s}=& {^\natural}\Psi^\kappa_f(X, R_{\widetilde{\Phi}(Y)},  D_{\widetilde{\Phi}(Y)})&:= {^\natural}\Psi^\kappa_{\lceil \frac{n-1}{2} \rceil,  \lfloor \frac{n-1}{2} \rfloor}(X_{s/f}, R_{{\widetilde{\Phi}(Y)_{s/f}}}, D_{{\widetilde{\Phi}(Y)_{s/f}}})\end{eqnarray}
\end{Prop}

Thus, for $s = 1, \dots, f$, and $i=2, \dots, n-1$, the sequence of $f$-step-ahead forecasts for $X$ are then 
\begin{eqnarray}\label{seasxss}
\Big({^\sharp}\widehat{x}^\kappa_{if+s} \Big)_{i, s}&:=& \Big({^\sharp}\Psi^\kappa_{\lceil \frac{i-1}{2} \rceil,  \lfloor \frac{i-1}{2} \rfloor}(X_{[1:if]_{s/f}}, R_{{\widetilde{\Phi}(Y_{[1:if]})_{s/f}}}) \Big)_{i, s}\\\label{seasxfs}
\Big({^\flat}\widehat{x}^\kappa_{if+s} \Big)_{i, s}&:=& \Big({^\flat}\Psi^\kappa_{\lceil \frac{i-1}{2} \rceil,  \lfloor \frac{i-1}{2} \rfloor}(X_{[1:if]_{s/f}}, D_{{\widetilde{\Phi}(Y_{[1:if]})_{s/f}}}) \Big)_{i, s}\\\label{seasxns}
\Big({^\natural}\widehat{x}^\kappa_{if+s} \Big)_{i, s}&:=& \Big({^\natural}\Psi^\kappa_{\lceil \frac{i-1}{2} \rceil,  \lfloor \frac{i-1}{2} \rfloor}(X_{[1:if]_{s/f}}, R_{{\widetilde{\Phi}(Y_{[1:if]})_{s/f}}}, D_{{\widetilde{\Phi}(Y_{[1:if]})_{j/f}}}) \Big)_{i, s}
\end{eqnarray}

\begin{Ex}Let's reconsider the classic Box and Jenkins airline data. Recall that the sum of absolute error for the ten twelve-step-forecasts using Holt--Winters and ARIMA models are $2083$ and $2228$ respectively. For $s = 1,...,12$ (months), and $i = 2,...,12$ (years), we compute the sequence of twelve-step-ahead $\sharp, \flat$ and $\natural$ forecasts for $X$ as in (\ref{seasxss}), (\ref{seasxfs}) and (\ref{seasxns}) for each optimization criteria $\kappa = 1, \dots, 8$  using the seasonal  per-period SAE stochastic latest-optimized  $\widetilde{\alpha}$-power  balanced forecasting models (with $N=1000$ to find the optimal latest-optimized alpha power). Moreover, we compute the (global) sum of absolute error, and we compare our forecasts to both Holt--Winters and ARIMA models by counting the number of best monthly and yearly forecasts. The results are summarized in Table \ref{APSL}.

\begin{table}[htp]
\begin{center}
\begin{tabular}{c||c||c|c|c||c|c|c}
$\bullet \kappa$ & SAE &HW / AR / K. & HW / K. & AR / K. & HW / AR / K. & HW / K. & AR / K. \\ \hline
$\sharp 1$ & 1270 & 30 / 28 / 62 & 43 / 77 & 37 / 83 & 2 / 2 / 6 & 2 / 8 & 2 / 8 \\   
$\sharp 2$ & 1300 & 34 / 26 / 60 & 49 / 71 & 34 / 86 & 2 / 1 / 7 & 2 / 8 & 2 / 9 \\
$\sharp 3$ & 1305 & 31 / 25 / 64 & 45 / 75 & 33 / 87 & 2 / 2 / 6 & 2 / 8 & 2 / 8\\ 
$\sharp 4$ & 1273 & 33 / 22 / 65 & 44 / 76 & 31 / 89 & 2 / 2 / 6 & 2 / 8 & 2 / 8 \\ 
$\sharp 5$ & 1290 & 31 / 29 / 60 & 47 / 73 & 38 / 82 & 2 / 2 / 6 & 2 / 8 & 2 / 8 \\ 
$\sharp 6$ & 1304 & 32 / 26 / 62 & 46 / 74 & 35 / 85 & 2 / 2 / 6 & 2 / 8 & 2 / 8 \\ 
$\sharp 7$ & 1284 & 33 / 26 / 61 & 47 / 73 & 36 / 84 & 2 / 2 / 6 & 2 / 8 & 2 / 8 \\
$\sharp 8$ & 1289 & 30 / 28 / 62 & 47 / 73 & 37 / 83 & 2 / 1 / 7 & 2 / 8 & 1 / 9 \\ \hline
$\flat 1$ & 1306 & 34 / 27 / 59 & 47 / 73 & 37 / 83 & 2 / 2 / 6 & 2 / 8 & 2 / 8\\
$\flat 2$ & 1324 & 32 / 30 / 58 & 50 / 70 & 40 / 80 & 2 / 2 / 6 & 2 / 8 & 2 / 8 \\ 
$\flat 3$ & 1306 & 35 / 28 / 57 & 48 / 72 & 38 / 82 & 2 / 2 / 6 & 2 / 8 & 2 / 8\\ 
$\flat 4$ & 1286 & 31 / 28 / 61 & 45 / 75 & 37 / 83 & 2 / 2 / 6 & 2 / 8 & 2 / 8\\
$\flat 5$ & 1273 & 30 / 29 / 61 & 45 / 75 & 38 / 82 & 2 / 1 / 7 & 2 / 8 & 1 / 9 \\
$\flat 6$ & 1255 & 32 / 26 / 62 & 46 / 74 & 36 / 84 & 2 / 2 / 6 & 2 / 8 & 2 / 8\\ 
$\flat 7$ & 1291 & 32 / 28 / 60 & 45 / 75 & 38 / 82 & 2 / 2 / 6 & 2 / 8 & 2 / 8 \\ 
$\flat 8$ & 1305 & 32 / 25 / 63 & 46 / 74 & 33 / 87 & 2 / 1 / 7 & 2 / 8 & 1 / 9 \\ \hline
$\natural 1$ & 1319 & 35 / 25 / 60 & 50 / 70 & 34 / 86 & 2 / 2 / 6 & 2 / 8 & 2 / 8\\
$\natural 2 $ & 1288 & 34 / 24 / 62 & 49 / 71 & 33 / 87 & 2 / 1 / 7 & 2 / 8 & 1 / 9 \\
$\natural 3$ & 1331 & 32 / 30 / 58 & 46 / 74 & 40 / 80 & 2 / 2 / 6 & 2 / 8 & 2 / 8 \\
$\natural 4$ & 1271 & 32 / 27 / 61 & 45 / 75 & 36 / 84 & 2 / 2 / 6 & 2 / 8 & 2 / 8 \\
$\natural 5$ & 1286 & 29 / 27 / 64 & 43 / 77 & 36 / 84 & 2 / 1 / 7 & 2 / 8 & 1 / 9 \\
$\natural 6 $ & 1287 & 32 / 27 / 61 & 46 / 74 & 36 / 84 & 2 / 2 / 6 & 2 / 8 & 2 / 8 \\$\natural 7$ & 1306 & 33 / 28 / 59 & 48 / 72 & 38 / 82 & 2 / 2 / 6 & 2 / 8 & 2 / 8\\
$\natural 8$ & 1260 & 30 / 27 / 63 & 44 / 76 & 36 / 84 & 2 / 1 / 7 & 2 / 8&1 / 9
\end{tabular}
\end{center}
\caption{Comparison of the seasonal per-period SAE stochastic latest-optimized  $\widetilde{\alpha}$-power  balanced forecasting models to the seasonal Holt--Winters and ARIMA models}
\label{APSL}
\end{table}%

One can notice that not only the sum of absolute errors for our models are significantly less than for both Holt--Winters and ARIMA models, but the number of monthly forecasts and year aggregate is substantially    higher for our models.  Figure \ref{APSLGraphs} contains the plots of the three sequences of forecasts, the residuals, the absolute residuals, and the yearly sum of absolute errors.


\begin{figure}[htbp]
\begin{center}
\includegraphics[scale=0.4]{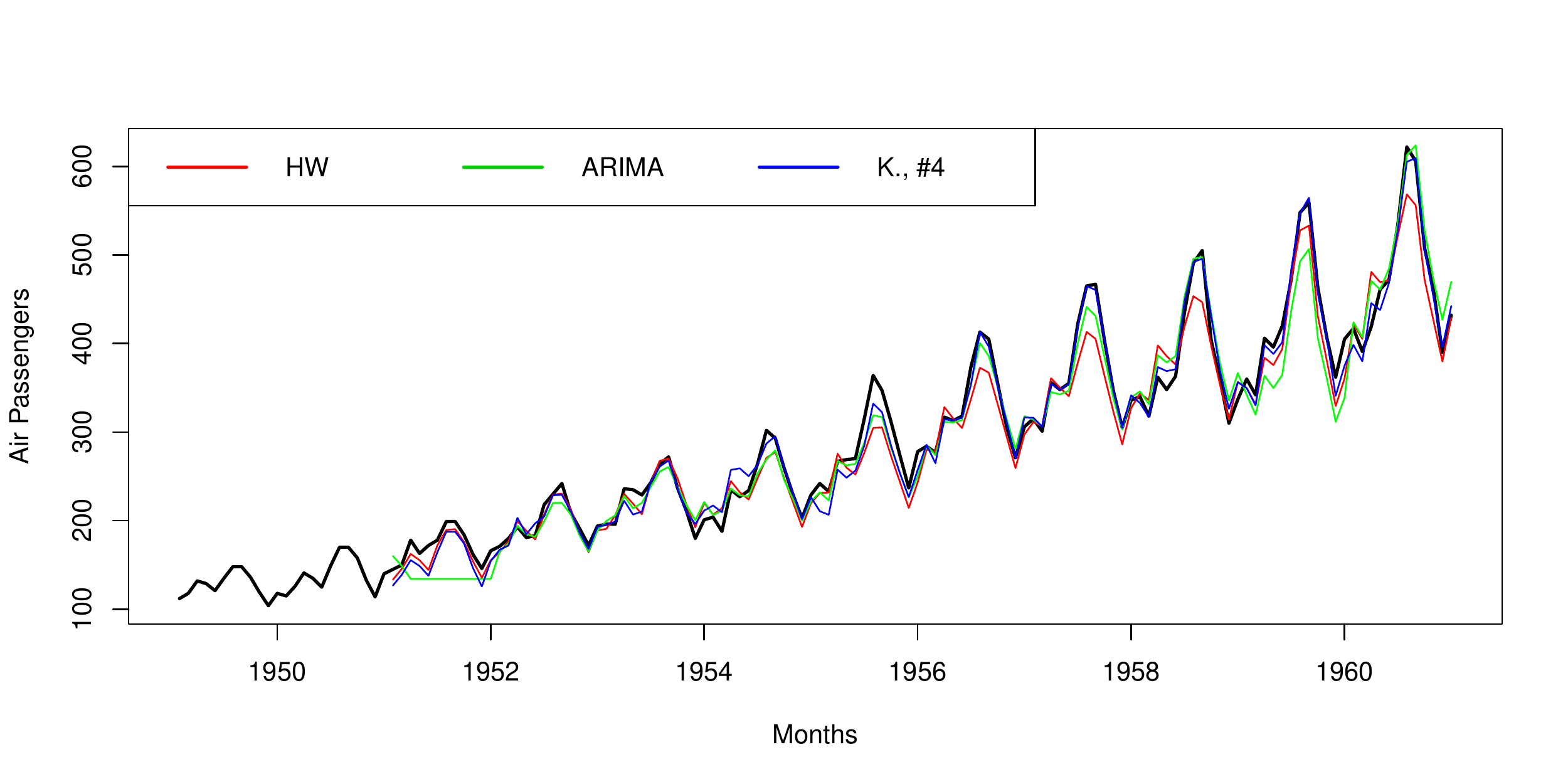} \includegraphics[scale=0.4]{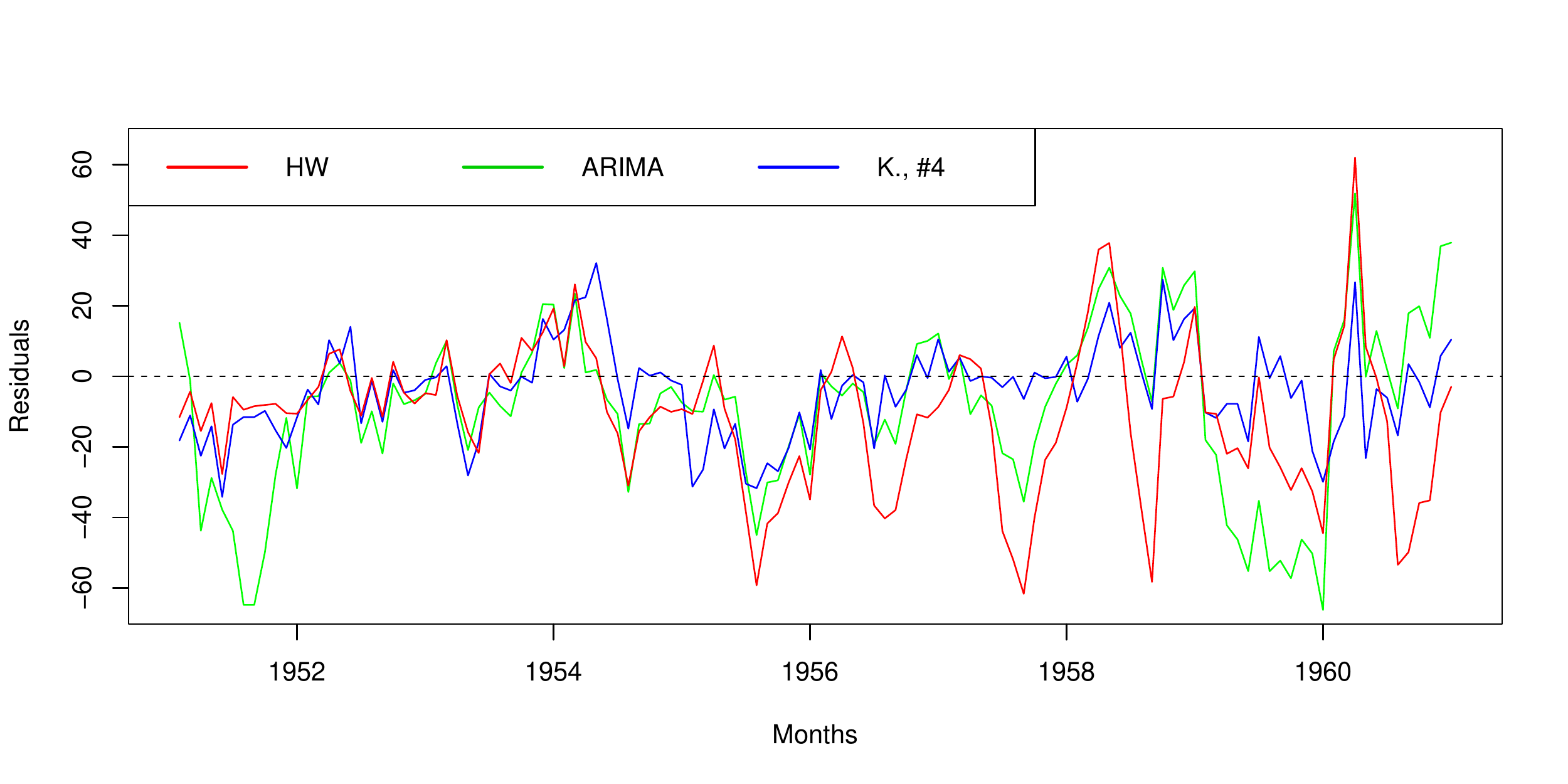}
\includegraphics[scale=0.4]{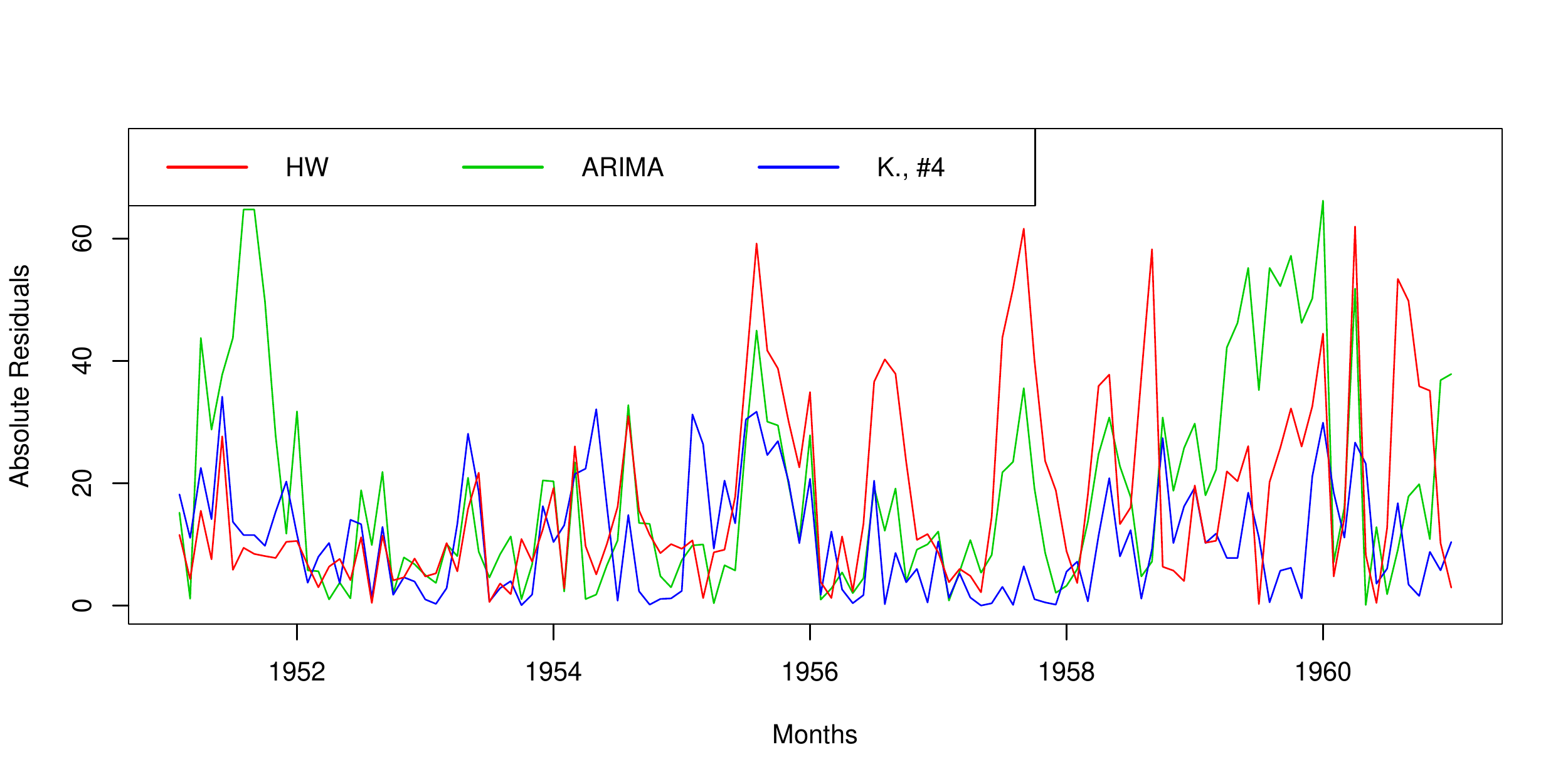} \includegraphics[scale=0.4]{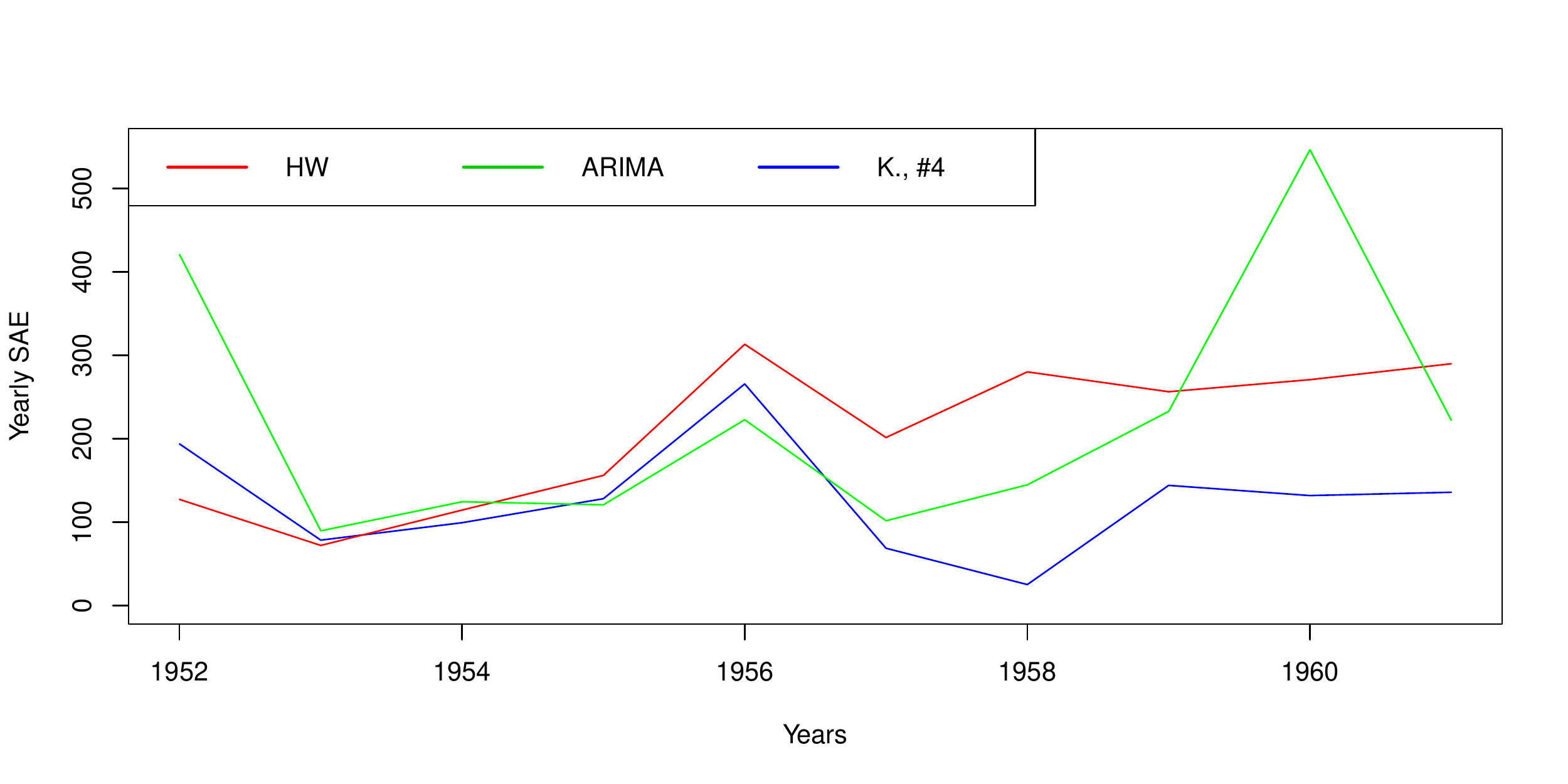}
\caption{Air Passengers: Forecasts, residuals, absolute residuals and yearly sum of absolute errors}
\label{APSLGraphs}
\end{center}
\end{figure}

\end{Ex}

\begin{Ex}Let us consider the non-adjusted quarterly United States total retail sales from  the  first quarter of 2000 to the fourth quarter of 2020 (source: Retail Indicators Branch, U.S. Census Bureau). Thus, $f = 4$ and $X = (x_1, \dots, x_{84})$.  For $s = 1,...,4$ (quarters), and $i = 2,...,19$ (years), we compute the sequence of four-step-ahead forecasts for $X$ using both Holt--Winters and ARIMA models. The sum of absolute errors for each model are
\begin{eqnarray}
 \text{SAE}^{\text{HW}}  = \hspace{-1mm}  \displaystyle \sum_{j=9}^{84}\Big|\widehat{x}_{j}^\text{HW}  - x_{j}\Big|  = 1,716,246 \, \text{ and } \,  \text{SAE}^{\text{ARIMA}}  = \hspace{-2mm}\displaystyle \sum_{j=9}^{84}\Big|\widehat{x}_{j}^\text{ARIMA}  - x_{j}\Big|  = 1,722,353
\end{eqnarray}
Comparing these $19$ years times $4$ quarters forecasts, i.e., 76 quarter forecasts,  the Holt--winters model lead to slightly better quarterly forecast (smallest absolute residual)  39 times as opposed to 37 times for the ARIMA model, i.e., $HW / AR = 39 / 37$. Moreover, if one compares the sum of absolute error per year, the Holt--Winters model lead to a better yearly forecast thirteen times as opposed to six times for ARIMA, i.e., $HW / AR = 13 / 6$. \\

For $s = 1,...,4$ (quarters), and $i = 2,...,21$ (years), we compute the sequence of four-step-ahead $\sharp, \flat$ and $\natural$ forecasts for $X$ as in (\ref{seasxss}), (\ref{seasxfs}) and (\ref{seasxns}) for each optimization criteria $\kappa = 1, \dots, 8$  using the seasonal  per-period SAE stochastic latest-optimized  $\widetilde{\alpha}$-power  balanced forecasting models (with $N=1000$ to find the optimal latest-optimized alpha power). Moreover, we compute the (global) sum of absolute error, and we compare our forecasts to both Holt--Winters and ARIMA models by counting the number of best monthly and yearly forecasts. The results are summarized in Table 34.

\begin{table}[htp]
\label{USQRTSumTab}
\begin{center}
\begin{tabular}{c||c||c|c|c||c|c|c}
$\bullet \kappa$ & SAE &HW / AR / K. & HW / K. & AR / K. & HW / AR / K. & HW / K. & AR / K. \\  \hline
$\sharp 1$ & 1,331,473 & 15 / 15 / 46 & 24 / 52 & 20 / 56 & 2 / 2 / 15 & 3 / 16 & 3 / 16  \\ 
$\sharp 2$ &  1,330,195 & 16 / 15 / 45 & 26 / 50 & 22 / 54 & 2 / 2 / 15 & 3 / 16 & 3 / 16 \\
$\sharp 3$ &1,374,592 & 15 / 13 / 48 & 24 / 52 & 19 / 57 & 2 / 2 / 15 & 3 / 16 & 3 / 16 \\ 
$\sharp 4$ & 1,295,475 & 17 / 11 / 48 & 25 / 51 & 18 / 58 & 2 / 2 / 15 & 3 / 16 & 3 / 16  \\ 
$\sharp 5$ & 1,319,002 & 17 / 15 / 44 & 27 / 49 & 20 / 56 & 2 / 2 / 15 & 3 / 16 & 3 / 16   \\ 
$\sharp 6$ & 1,340,551 & 17 / 13 1 46 & 24 / 52 & 20 / 56 & 2 / 2 / 15 & 3 / 16 & 3 / 16  \\ 
$\sharp 7$ & 1,306,671 & 16 / 15 / 45 & 26 / 50 & 21 / 55 & 2 / 2 / 15 & 3 / 16 & 3 / 16   \\
$\sharp 8$ &  1,319,979 & 18 / 17 / 41 & 28 / 48 & 22 / 54 & 2 / 2 / 15 & 3 / 16 & 3 / 16  \\ \hline
$\flat 1$ & 1,335,252 & 17 / 12 / 47 & 25 / 51 & 17 / 59 & 2 / 2 / 15 & 3 / 16 & 3 / 16 \\
$\flat 2$ & 1,313,326 & 16 / 14 / 46 & 26 / 50 & 19 / 57 & 2 / 2 / 15 & 3 / 16 & 3 / 16  \\ 
$\flat 3$ & 1,347,479 & 16 / 14 / 46 & 24 / 52 & 20 / 56 & 2 / 2 / 15 & 3 / 16 & 3 / 16 \\ 
$\flat 4$ & 1,323,781 & 15 / 15 / 46 & 23 / 53 & 21 / 55 & 2 / 2 / 15 & 3 / 16 & 3 / 16  \\
$\flat 5$ & 1,289,315 & 16 / 13 / 47 & 25 / 51 & 19 / 57 & 2 / 2 / 15 & 3 / 16 & 3 / 16   \\
$\flat 6$ & 1,336,725 & 16 / 13 / 47 & 24 / 52 & 19 / 57 & 2 / 2 / 15 & 3 / 16 & 3 / 16  \\ 
$\flat 7$ & 1,336,048 & 17 / 13 / 46 & 27 / 49 & 19 / 57 & 2 / 2 / 15 & 3 / 16 & 3 / 16 \\
$\flat 8$ &  1,298,426 & 16 / 13 / 47 & 24 / 52 & 19 / 57 & 2 / 2 / 15 & 3 / 16 & 3 / 16 \\ \hline
$\natural 1$ & 1,358,754 & 16 / 13 / 47 & 24 / 52 & 17 / 59 & 2 / 2 / 15 & 3 / 16 & 3 / 16   \\
$\natural 2 $ &1,318,779 & 18 / 14 / 44 & 29 / 47 & 20 / 56 & 2 / 2 / 15 & 3 / 16 & 3 / 16  \\
$\natural 3$ & 1,506,503 & 15 / 21 / 40 & 27 / 49 & 26 / 50 & 3 / 3 / 13 & 5 / 14 & 5 / 14   \\
$\natural 4$ & 1,348,305 & 15 / 16 / 45 & 24 / 52 & 22 / 54 & 2 / 2 / 15 & 3 / 16 & 3 / 16  \\
$\natural 5$ & 1,296,343 & 17 / 14 / 45 & 26 / 50 & 20 / 56 & 2 / 2 / 15 & 3 / 16 & 3 / 16  \\
$\natural 6 $ & 1,339,918 & 16 / 16 / 44 & 25 / 51 & 22 / 54 & 2 / 2 / 15 & 3 / 16 & 3 / 16   \\
$\natural 7$ &1,305,288 & 15 / 15 / 46 & 24 / 52  & 20 / 56 & 2 / 2 / 15 & 3 / 16 & 3 / 16 \\
$\natural 8$ & 1,304,665 & 15 / 12 / 49 & 24 / 52 & 19 / 57& 2 / 2 / 15 & 3 / 16 & 3 / 16 
\end{tabular}

\caption{Comparison of the seasonal per-period SAE stochastic latest-optimized  $\widetilde{\alpha}$-power  balanced forecasting models to the seasonal Holt--Winters and ARIMA models}
\end{center}

\end{table}%

One can again notice that not only the sum of absolute errors for our models are significantly less than both Holt--Winters and ARIMA models, but the number of monthly forecasts and year aggregate is substantially higher for our models.


\begin{figure}[htbp]
\begin{center}

\includegraphics[scale=0.4]{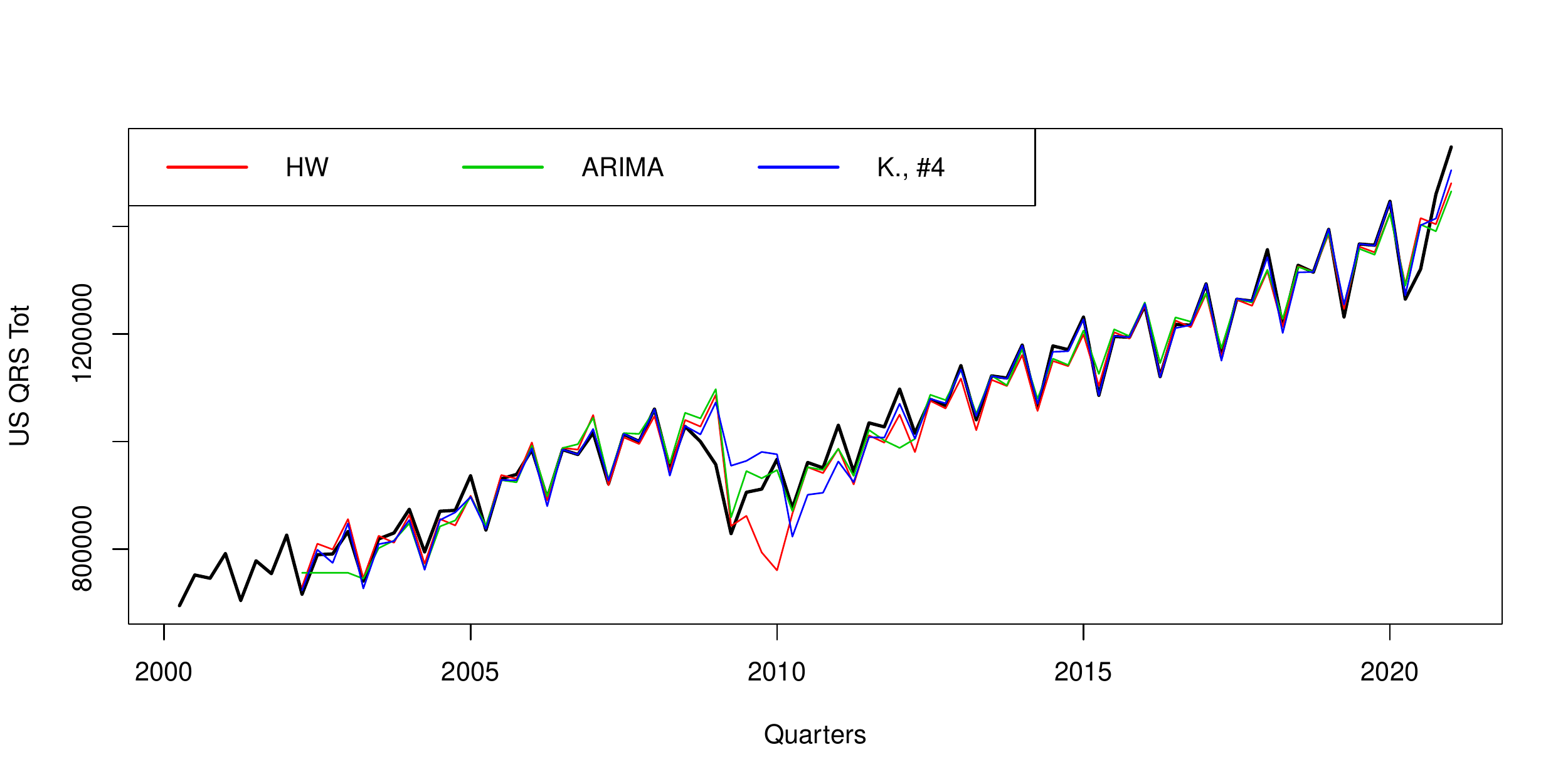}  \includegraphics[scale=0.4]{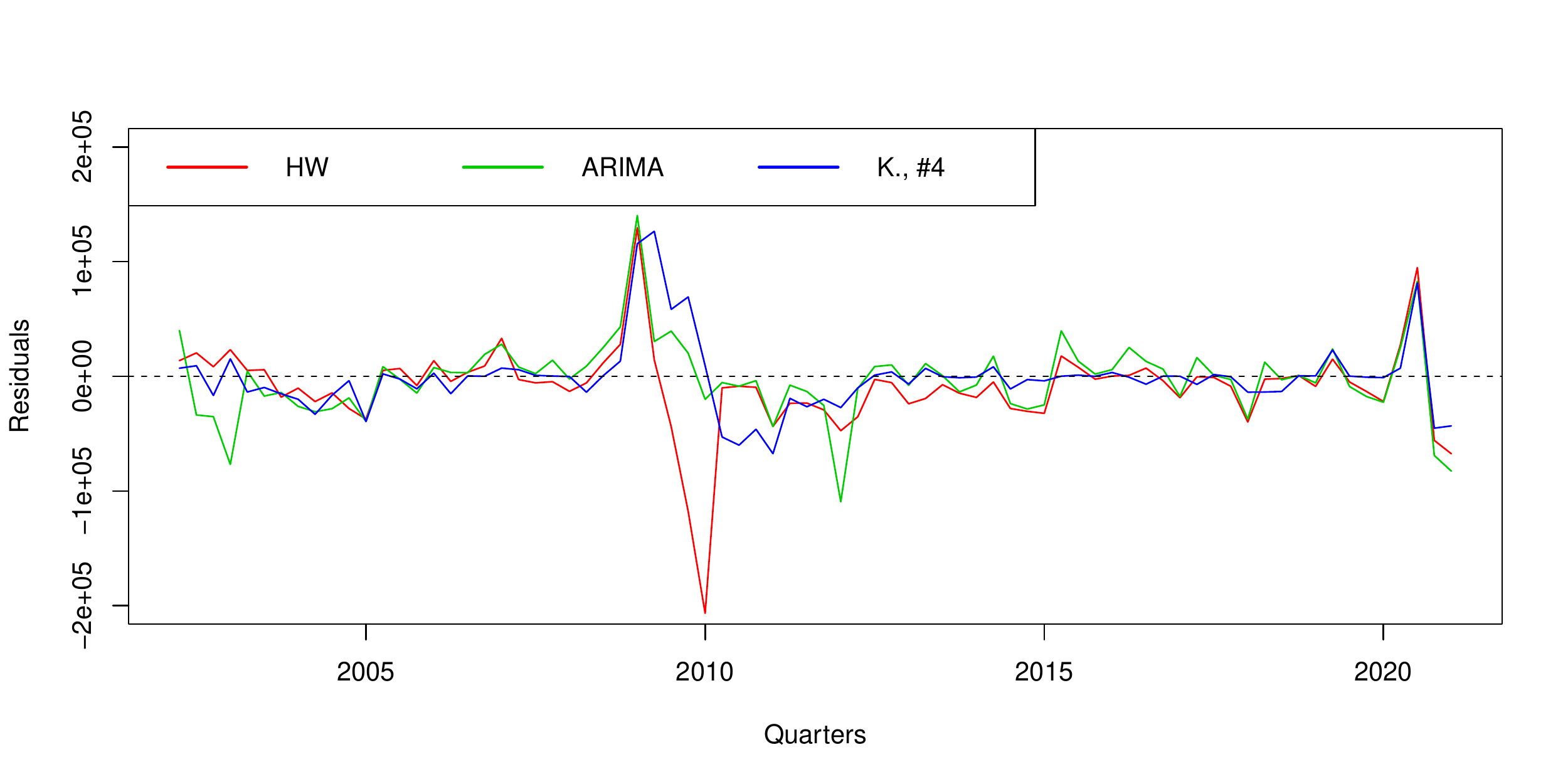}

\includegraphics[scale=0.4]{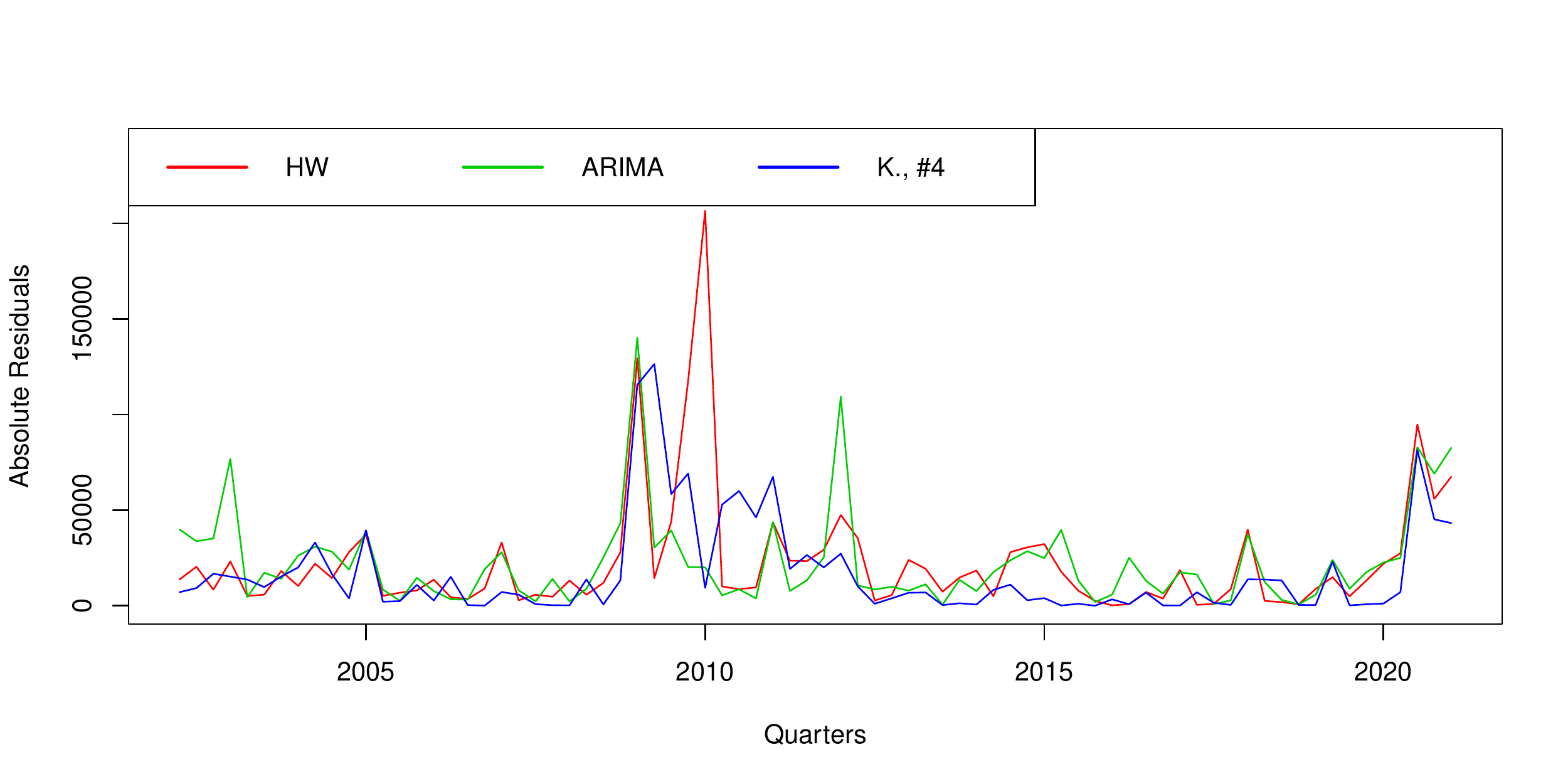} \includegraphics[scale=0.4]{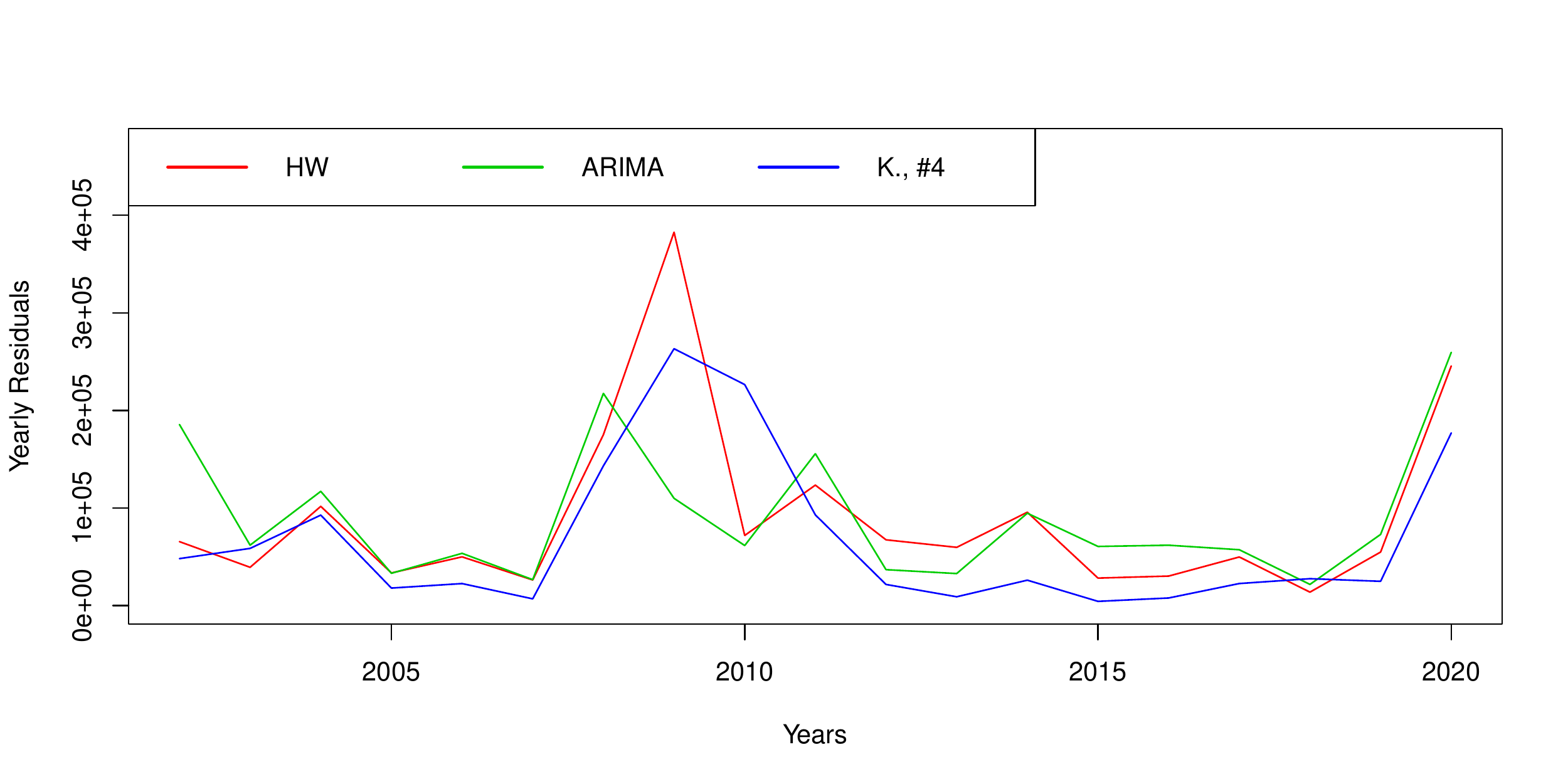}
\caption{US Quarterly Retail Total Sales: Forecasts, residuals, absolute residuals and yearly sum of absolute errors}
\label{}
\end{center}
\end{figure}

\end{Ex}


\section{Application to Stock Market Data}

In this section, we apply the non-seasonal balanced $\delta$-forecasting models to stock market data. Let's consider the stock market data available in Kaggle (\url{https://www.kaggle.com/camnugent/sandp500}) consisting of all S\&P 500 companies historical data from Feb 8, 2008 to Feb 7, 2013. For each stock, at a given day,  five numerical values are given: the opening price, the highest price reached, the lowest price reached, the closing price and the number of shares traded. The data has 505 companies. Out  of these, 467 companies have a complete data, that is, no missing entry for the 1259 daily prices for each of the four variables open, high, low and close.  These 467 companies are used in the following analysis. For each company, let's denote by: $O$ the opening price time series, $H$ the high price reached time series, $L$ the lowest price reached time series, and $C$ the closing price time series. For a given company, we thus have four time series $O = (o_i)$, $H = (h_i)$, $L=(l_i)$ and $C = (c_i)$, where $i=1, \dots, 1259$ (daily prices in dollars for a five year period). Let's define the time series $M$ to be the daily midpoint  price time series, that is, $M = (L+H)/2$, consisting of the of the arithmetic average between the low and high daily prices. We have then $M=(m_i) = ((l_i+h_i)/2)$ where $i=1, \dots, 1259$.  We are interested in the following problems. First, given an S\&P 500 company, forecast the next daily price of a stock and check whether it lies between the lowest and hight prices of that next day. Second, find a method/model that is optimal across all S\&P 500 companies. To answer these two problems, we propose the following: For a given company, let's forecast the last 100 daily prices, and determine the number of forecasts that lie between the lowest and highest prices of that day. We  perform these forecasts using the $\sharp$-, $\flat$- and $\natural$-balanced $\delta$-forecasting functions with $\delta = 92$, that is, each forecast is determined by the previous 92 daily prices (largest number of days for three consecutive months). We have then 
\begin{eqnarray}
\Big({^{\,\,\,\sharp}_{92}}\widehat{x}^{\, \kappa}_{i+1} \Big)_{i=1159, \dots, 1258}&:=& \Big({^\sharp}\Psi^\kappa_{46,  45}(X_{[i-91:i]}, R_{Y_{[i-91:i]}}) \Big)_{i=1159, \dots, 1258}\\
\Big({^{\,\,\,\flat}_{92}}\widehat{x}^{\, \kappa}_{i+1} \Big)_{i=1159, \dots, 1258}&:=& \Big({^\flat}\Psi^\kappa_{46,  45}(X_{[i-91:i]}, D_{Y_{[i-91:i]}}) \Big)_{i=1159, \dots, 1258}\\
\Big({^{\,\,\,\natural}_{92}}\widehat{x}^{\, \kappa}_{i+1} \Big)_{i=1159, \dots, 1258}&:=& \Big({^\natural}\Psi^\kappa_{46,  45}(X_{[i-91:i]}, R_{Y_{[i-91:i]}}, D_{Y_{[i-91:i]}}) \Big)_{i=1159, \dots, 1258}
\end{eqnarray}
For each $i = 1160, \dots, 1259$ and a given optimization criteria $\kappa = 1, \dots, 8$, and for a given company,  let's define  $\sharp$- (resp., $\flat$-, $\natural$) Bernoulli random variables as follows:
\begin{eqnarray}
^\sharp s_i^\kappa:= \left\{\begin{array}{cc}1 & \text{ if } \quad l_i \leq {^{\,\,\,\sharp}_{92}}\widehat{x}^{\, \kappa}_{i}  \leq h_i \\0 & \text{ otherwise}\end{array}\right.\\ {^\flat}s_i^\kappa:= \left\{\begin{array}{cc}1 & \text{ if } \quad l_i \leq {^{\,\,\,\flat}_{92}}\widehat{x}^{\, \kappa}_{i}  \leq h_i \\0 & \text{ otherwise}\end{array}\right.\\
{^\natural}s_i^\kappa:= \left\{\begin{array}{cc}1 & \text{ if } \quad l_i \leq {^{\,\,\,\natural}_{92}}\widehat{x}^{\, \kappa}_{i}  \leq h_i \\0 & \text{ otherwise}\end{array}\right.
\end{eqnarray}
We define also the $\sharp$- (resp., $\flat$-, $\natural$-) score random variable consisting of the number of forecasted values (among 100 forecasts) falling between the lowest and highest prices of that day, that is
\begin{equation}
{^\sharp}S^\kappa:= \sum_{i=1160}^{1259} {^\sharp}s_i^\kappa; \quad  {^\flat}S^\kappa:= \sum_{i=1160}^{1259} {^\flat}s_i^\kappa; \quad  {^\natural}S^\kappa:= \sum_{i=1160}^{1259} {^\natural}s_i^\kappa
\end{equation}

\begin{figure}[htbp]
\begin{center}
\includegraphics[scale=0.55]{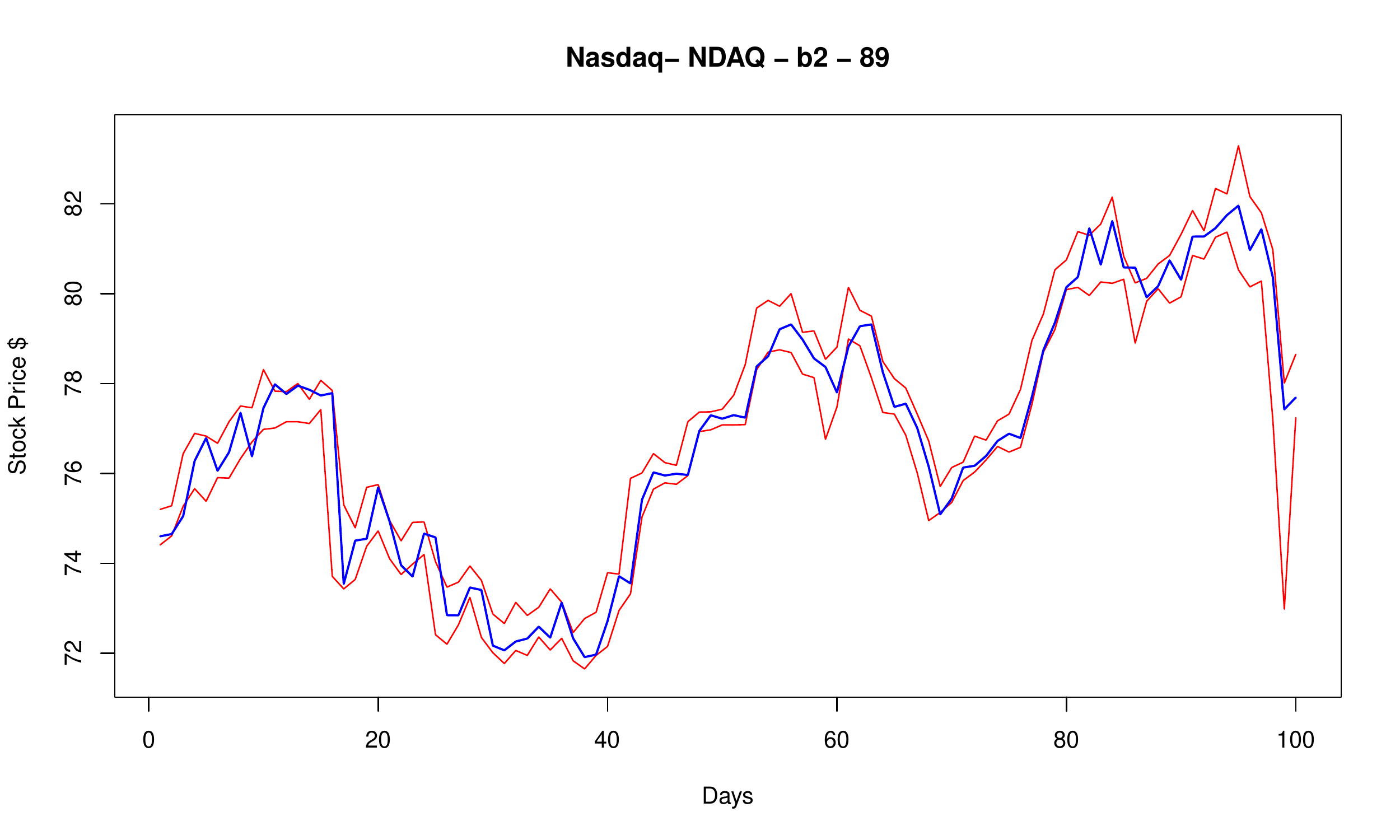}\caption{Forecasting NDAQ 100 daily stock prices using $\flat$-balanced $92$-days-forecasting function with $X = Y = C$ and $\kappa = 2$ leading to a score of ${^\flat}S^2_\text{CC} = 89$ }
\label{s1b5boxplots}
\end{center}
\end{figure}

\begin{figure}[htbp]
\begin{center}
\includegraphics[scale=0.42]{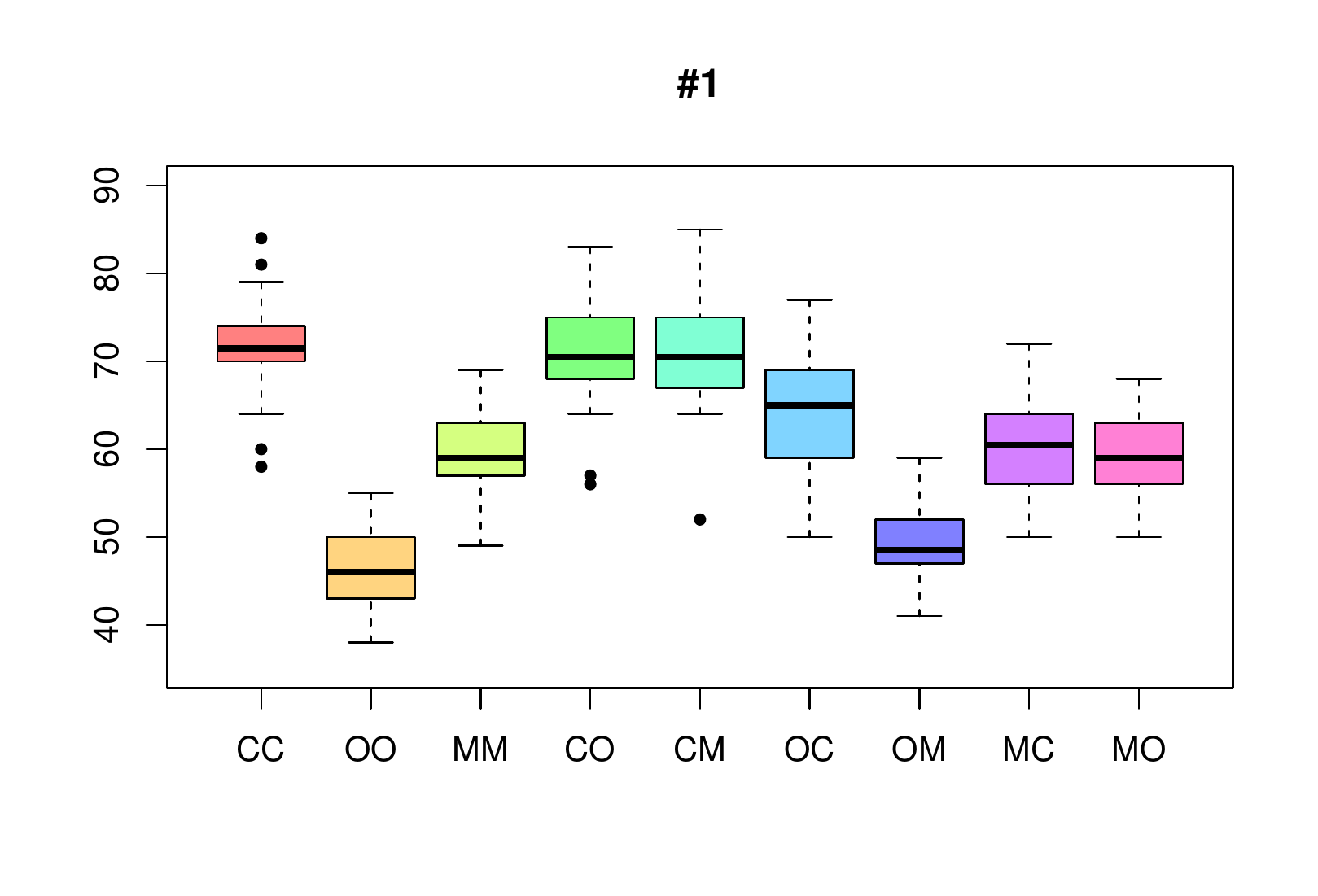} \includegraphics[scale=0.42]{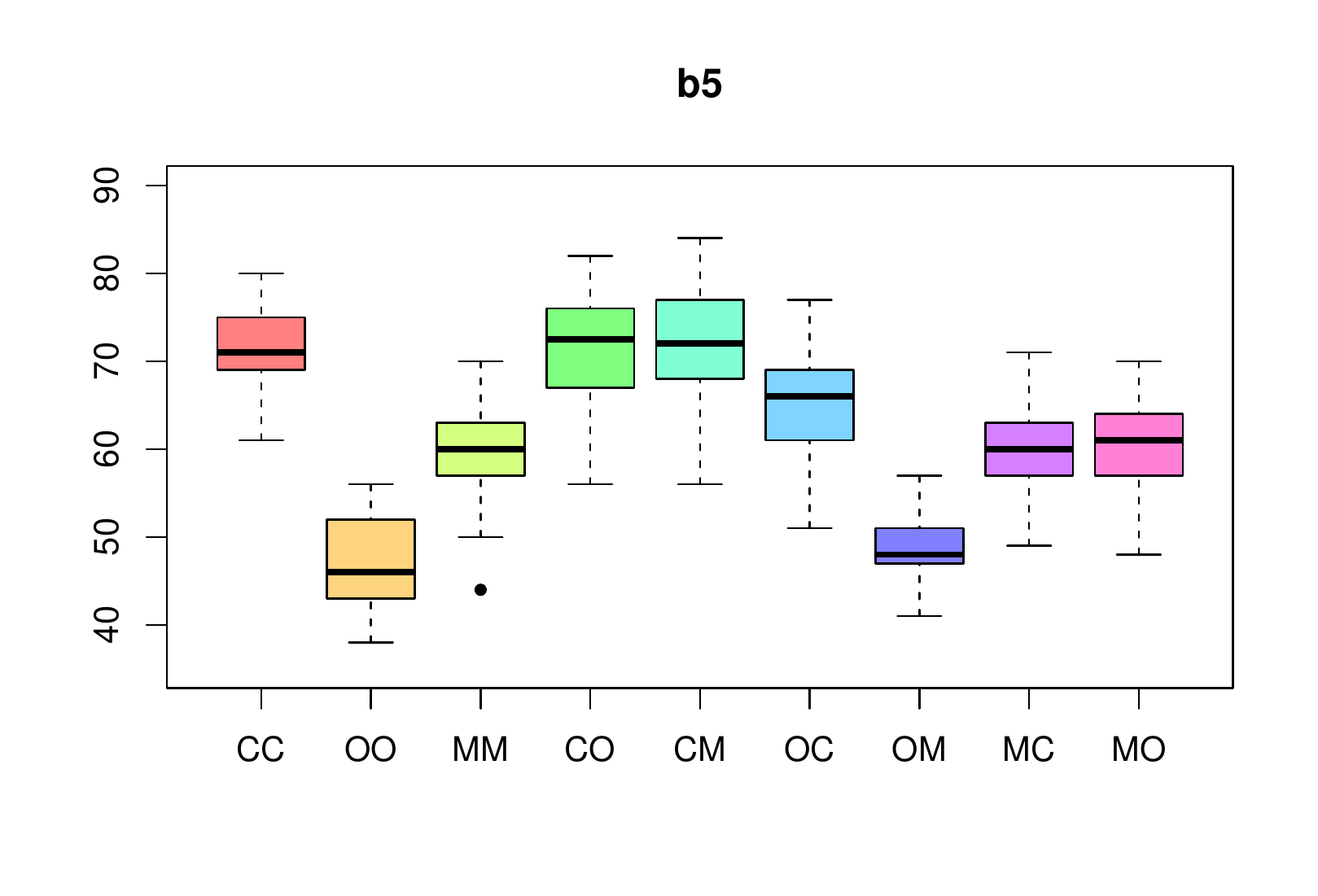}
\caption{Boxplots of the scores  ${^\sharp}S^1$ and ${^\flat}S^5$  of 30 random stocks for various choices of $X$ and $Y$ among $C, O$ and $M$}
\label{s1b5boxplots2}
\end{center}
\end{figure}

Let's select 30 companies at random from the list of 467 companies. For each company, let's determine the $\sharp$- and $\flat$ scores for two optimization criteria $\kappa=1$ and $5$, i.e., ${^\sharp}S^1$ and ${^\flat}S^5$,  of the latest 100 forecasted prices when $X$ and $Y$ are chosen among the closing, opening, and midpoint  time series $C, O$ and $M$. Thus, for each stock, the last 100 forecasted prices ${^{\,\,\,\sharp}_{92}}\widehat{x}^{\, 1}_{i+1}$  and the last 100 forecasted prices ${^{\,\,\,\flat}_{92}}\widehat{x}^{\, 5}_{i+1}$, where $i=1159,\dots, 1258$,  are computed with $(X, Y)$ being one of these nine possible alternatives: $(C, C)$, $(O,O)$, $(M,M)$, $(C,O)$, $(C,M)$, $(O,C)$, $(O,M)$, $(M,C)$ and $(M,O)$. The results are listed in Tables 37 and 38.  The boxplots of these scores are shown in Figure \ref{s1b5boxplots2}. 

One can notice that the highest scores are achieved for $X = C$ and $Y = C, O$ or $M$. In other words, the highest scores are achieved when the time series $X$ used for the forecasts is the closing time series $C$, and the rate of interest and/or discount time series is derived from either the closing, opening, or midpoint time series. For each of the 467 companies, and for each optimization criteria $\kappa = 1, \dots, 8$, we compute 100 $\sharp$-balanced $92$-forecasts with $(X,Y)= (C,C), (CO)$ and $(C,M)$, 100 $\flat$-balanced $92$-forecasts with $(X,Y)= (C,C), (CO)$ and $(C,M)$, and 100 $\natural$-balanced $92$-forecasts with $(X,Y)= (C,C), (CO)$ and $(C,M)$. Therefore, the above analysis is based on 3,362,400 forecasts obtained using 467 companies, 8 optimization criteria $\kappa=1, \dots, 8$, and 9 models $\sharp$-CC, $\flat$-CC, $\natural$-CC , $\sharp$-CO, $\flat$-CO, $\natural$-CO, $\sharp$-CM, $\flat$-CM and $\natural$-CM. We thus obtain nine multivariate data with 467 observations (companies) and eight variables

\begin{center}

\begin{tabular}{ccc}
${^\sharp}S_{\text{CC}} = ({^\sharp}S^1_{\text{CC}}, \dots, {^\sharp}S^8_{\text{CC}})$, & ${^\sharp}S_{\text{CO}} = ({^\sharp}S^1_{\text{CO}}, \dots, {^\sharp}S^8_{\text{CO}})$, & ${^\sharp}S_{\text{CM}} = ({^\sharp}S^1_{\text{CM}}, \dots, {^\sharp}S^8_{\text{CM}})$,\\
${^\flat}S_{\text{CC}} = ({^\flat}S^1_{\text{CC}}, \dots, {^\flat}S^8_{\text{CC}})$, & ${^\flat}S_{\text{CO}} = ({^\flat}S^1_{\text{CO}}, \dots, {^\flat}S^8_{\text{CO}})$, & ${^\flat}S_{\text{CM}} = ({^\flat}S^1_{\text{CM}}, \dots, {^\flat}S^8_{\text{CM}})$,\\
${^\natural}S_{\text{CC}} = ({^\natural}S^1_{\text{CC}}, \dots, {^\natural}S^8_{\text{CC}})$, & ${^\natural}S_{\text{CO}} = ({^\natural}S^1_{\text{CO}}, \dots, {^\natural}S^8_{\text{CO}})$, & ${^\natural}S_{\text{CM}} = ({^\natural}S^1_{\text{CM}}, \dots, {^\natural}S^8_{\text{CM}})$,\\
\end{tabular}
\end{center}

representing the scores of the three times eight  $\sharp$- (resp., $\flat$-, $\natural$) balanced $92$-forecasting functions across 467 companies.  The summary of the $\sharp$-CC multivariate score data is
\begin{center}
\begin{tabular}{r|c|c|c|c|c|c|c|c}
${^\sharp}S_{\text{CC}}^\kappa$  Summary &  $\kappa =1 $ & $\kappa =2 $ & $\kappa =3 $ & $\kappa = 4$ & $\kappa =5 $ & $\kappa =6 $ & $\kappa = 7$ & $\kappa = 8$ \\ \hline
Minimum & 46 &48& 46& 46& 50& 46& 46& 49\\
1$^\text{st}$ Quartile& 67& 68&67& 67& 68& 67& 67& 68\\
Median &72& 72& 71& 71& 72&71& 71& 72\\
Mean & 71.13& 71.50 &71.08 &71.11 &71.55 &71.10& 71.07 &71.53\\
3$^\text{rd}$ Quartile &76& 75& 75& 75&76& 75& 75& 75\\
Maximum &86& 88& 86& 86& 88& 86& 86& 88\\ \hline
Std. Deviation &  6.39 & 5.91  &6.25&  6.29&  5.86&  6.28&  6.28 & 5.92
\end{tabular}
\end{center}

The summary of the $\flat$-CC multivariate score data is
\begin{center}
\begin{tabular}{r|c|c|c|c|c|c|c|c}
${^\flat}S_{\text{CC}}^\kappa$  Summary&  $\kappa =1 $ & $\kappa =2 $ & $\kappa =3 $ & $\kappa = 4$ & $\kappa =5 $ & $\kappa =6 $ & $\kappa = 7$ & $\kappa = 8$ \\ \hline
Minimum & 46& 48& 46& 46& 48& 46& 46& 48 \\
1$^\text{st}$ Quartile & 67& 68& 67& 67& 69& 67& 67& 68  \\
Median&  71&72& 71& 71& 72& 71& 71& 72 \\
Mean &  71.12& 71.56& 71.13& 71.09& 71.58& 71.05& 71.15 &71.58\\ 
3$^\text{rd}$ Quartile & 75& 76& 75& 75& 75.5& 75& 75& 75.5  \\
Maximum &  87& 89& 87& 87& 88& 87& 87& 88 \\ \hline
Std. Deviation & 6.37 &5.87& 6.24& 6.30& 5.89& 6.37 &6.23& 5.88
\end{tabular}
\end{center}
The summary of the $\natural$-CC multivariate score data is
\begin{center}
\begin{tabular}{r|c|c|c|c|c|c|c|c}
${^\natural}S_{\text{CC}}^\kappa$  Summary&  $\kappa =1 $ & $\kappa =2 $ & $\kappa =3 $ & $\kappa = 4$ & $\kappa =5 $ & $\kappa =6 $ & $\kappa = 7$ & $\kappa = 8$ \\ \hline
Minimum & 46 &  48 &  49  & 46 &  50  & 46 &  46 &  48 \\
1$^\text{st}$ Quartile&  67 &     68   &   66    &  67  &    69   &   67   &   67   &   68 \\
Median&71  & 72   &70 &  71   &72 &  71   &71&   72 \\
Mean &  71.09 &71.51& 69.91& 71.06 &71.55 &71.04& 71.01& 71.58 \\
3$^\text{rd}$ Quartile  & 75&   75&   74&    75&    75&    75&    75&    75.5 \\
Maximum &86 &  88 &  87  & 86 &  88 &  86 &  86 &  88 \\ \hline
Std. Deviation & 6.37 &5.88& 6.57& 6.26& 5.88& 6.27& 6.20 &5.89 
\end{tabular}
\end{center}

\begin{figure}[htbp]
\begin{center}
\includegraphics[scale=0.45]{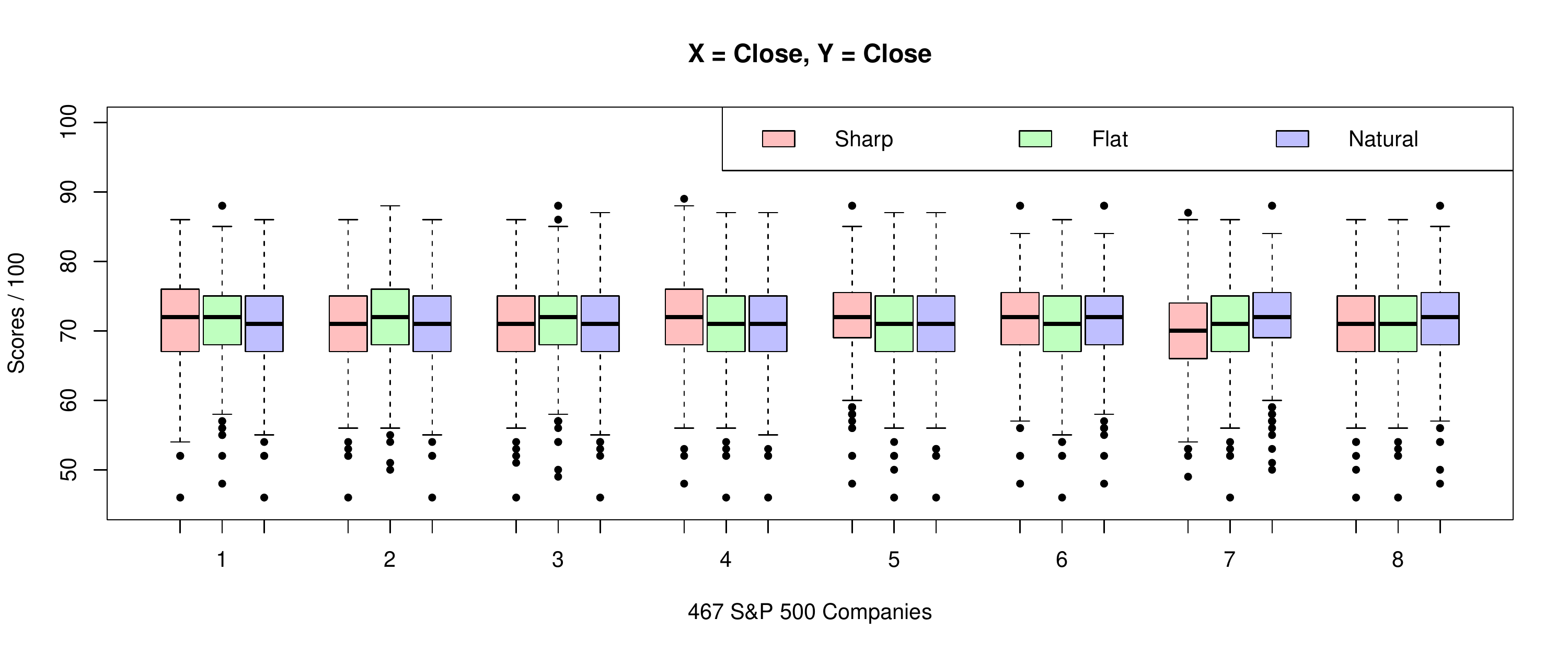}
 \caption{Boxplots of the scores  ${^\sharp}S^\kappa_{CC}$,  ${^\flat}S^\kappa_{CC}$, and ${^\natural}S^\kappa_{CC}$ }
\label{s1b5boxplots}
\end{center}
\end{figure}

The summary of the $\sharp$-CO multivariate score data is
\begin{center}
\begin{tabular}{r|c|c|c|c|c|c|c|c}
${^\sharp}S_{\text{CO}}^\kappa$  Summary&  $\kappa =1 $ & $\kappa =2 $ & $\kappa =3 $ & $\kappa = 4$ & $\kappa =5 $ & $\kappa =6 $ & $\kappa = 7$ & $\kappa = 8$ \\ \hline
Minimum &   50& 48& 51& 51& 47& 49& 49& 47\\
1$^\text{st}$ Quartile & 67& 68& 67& 67& 68& 67& 67& 68   \\
Median& 71&72& 71& 71& 72& 71& 71& 72  \\
Mean &   70.83 &71.41 &70.81 &70.93 &71.38 &70.82& 70.80& 71.38  \\
3$^\text{rd}$ Quartile & 75&75& 75& 75& 75& 75& 75& 75\\
Maximum &   86& 88& 86& 86& 88& 86& 86& 88  \\\hline
Std. Deviation &5.98 &5.94 &6.05 &5.95 &5.89& 6.03 &6.01& 5.94
\end{tabular}
\end{center}

The summary of the $\flat$-CO multivariate score data is
\begin{center}
\begin{tabular}{r|c|c|c|c|c|c|c|c}
${^\flat}S_{\text{CO}}^\kappa$  Summary&  $\kappa =1 $ & $\kappa =2 $ & $\kappa =3 $ & $\kappa = 4$ & $\kappa =5 $ & $\kappa =6 $ & $\kappa = 7$ & $\kappa = 8$ \\ \hline
Minimum &   50& 47& 51& 51& 47& 48& 50& 48  \\
1$^\text{st}$ Quartile & 67& 68& 67& 67& 68& 67& 67& 68   \\
Median&  71& 72& 71& 71& 72& 71& 71& 72  \\
Mean &  70.86 &71.43& 70.92& 70.95 &71.42 &70.86& 70.84 &71.38\\
3$^\text{rd}$ Quartile &  75&76& 75& 75& 75& 75& 75& 75 \\
Maximum & 86& 87& 86& 86& 88& 86& 86& 88  \\\hline
Std. Deviation & 6.09 &5.97 &6.09 &6.01 &5.95& 6.10& 6.04 &5.98
 \end{tabular}
\end{center}

The summary of the $\natural$-CO multivariate score data is
\begin{center}
\begin{tabular}{r|c|c|c|c|c|c|c|c}
${^\natural}S_{\text{CO}}^\kappa$  Summary&  $\kappa =1 $ & $\kappa =2 $ & $\kappa =3 $ & $\kappa = 4$ & $\kappa =5 $ & $\kappa =6 $ & $\kappa = 7$ & $\kappa = 8$ \\ \hline
Minimum &  50  & 48 &  50   &50  & 48  & 49  & 49 &  48   \\
1$^\text{st}$ Quartile &  67 &68 &66& 67 & 68&  67 & 67 &68   \\
Median&71 &  72 &  70   &71  & 72 &  71   &71 &  72   \\
Mean & 70.73& 71.40& 69.74 &70.82 &71.39 &70.76 &70.69& 71.34   \\
3$^\text{rd}$ Quartile &   75 &  75 &  74  & 75   &75  & 75 &  75 &  75     \\
Maximum &  86&   88  & 86  & 86  & 88  & 86 &  86  & 88  \\\hline
Std. Deviation &5.98 &5.90& 6.47& 5.99 &5.87 &6.05& 6.02 &5.93  \end{tabular}
\end{center}

\begin{figure}[htbp]
\begin{center}
\includegraphics[scale=0.45]{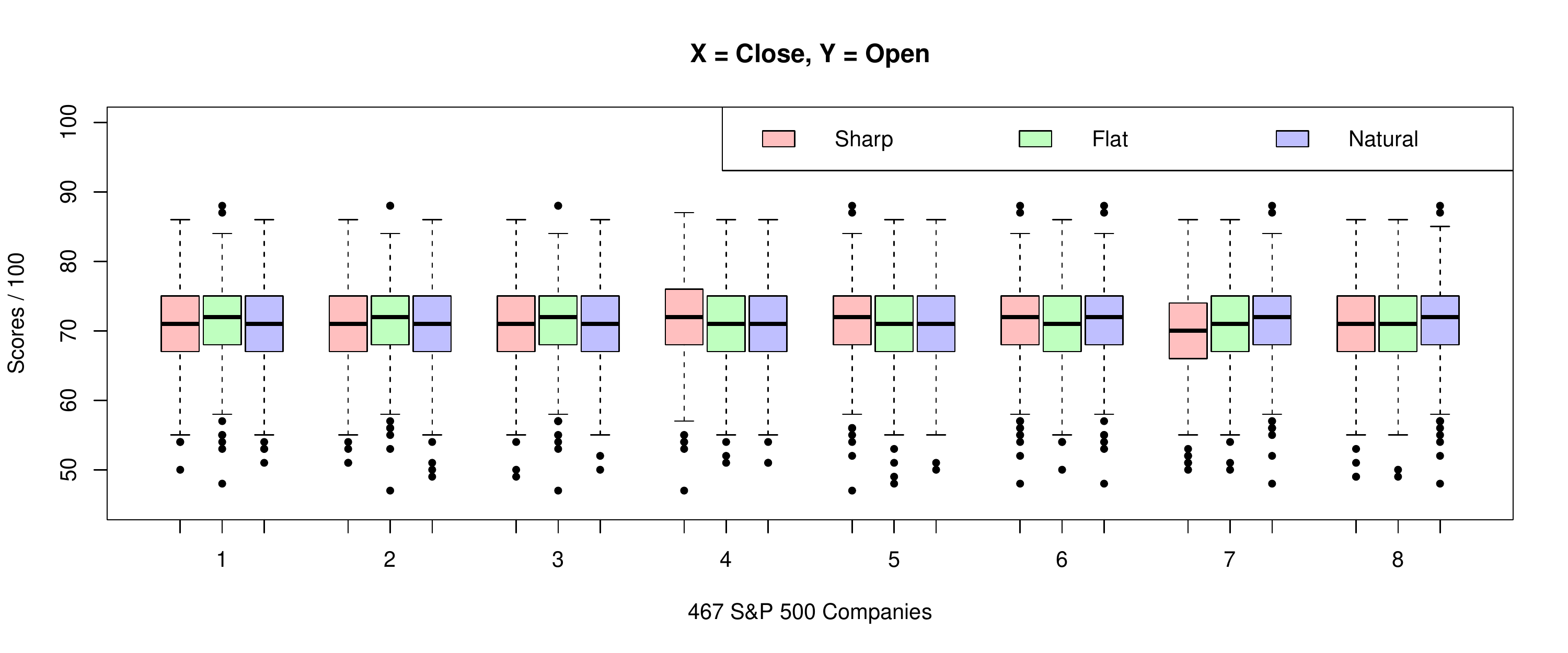}
 \caption{Boxplots of the scores  ${^\sharp}S^\kappa_{CO}$,  ${^\flat}S^\kappa_{CO}$, and ${^\natural}S^\kappa_{CO}$ }
\label{s1b5boxplots}
\end{center}
\end{figure}

The summary of the $\sharp$-CM multivariate score data is
\begin{center}
\begin{tabular}{r|c|c|c|c|c|c|c|c}
${^\sharp}S_{\text{CM}}^\kappa$  Summary&  $\kappa =1 $ & $\kappa =2 $ & $\kappa =3 $ & $\kappa = 4$ & $\kappa =5 $ & $\kappa =6 $ & $\kappa = 7$ & $\kappa = 8$ \\ \hline
Minimum &  49  & 48  & 50  & 50  & 48&   49 &  49 &  48 \\
1$^\text{st}$ Quartile & 67   &   68    &  67   &   67     & 68   &   67  &    67  &    68    \\
Median&  71&   72  & 71 &  71 &  72  & 71 &  71   &72   \\
Mean &  71.18& 71.60 &71.18& 71.18 &71.64 &71.13 &71.20& 71.66  \\
3$^\text{rd}$ Quartile & 76&    76&   76&    75&   75.5  &  75.5 &   76&    76    \\
Maximum & 86 &  86  & 86 &  86 &  86 &  86  & 86 &  86   \\\hline
Std. Deviation &6.32 &6.15 &6.33 &6.22 &6.09 &6.26& 6.29 &6.08 
\end{tabular}
\end{center}

The summary of the $\flat$-CM multivariate score data is
\begin{center}
\begin{tabular}{r|c|c|c|c|c|c|c|c}
${^\flat}S_{\text{CM}}^\kappa$  Summary&  $\kappa =1 $ & $\kappa =2 $ & $\kappa =3 $ & $\kappa = 4$ & $\kappa =5 $ & $\kappa =6 $ & $\kappa = 7$ & $\kappa = 8$ \\ \hline
Minimum &     50  & 48&   51   &50 &  48 &  50  & 49   &48 \\
1$^\text{st}$ Quartile &  67   &   68 &     67     & 67   &   68  &    67    &  67  &    68  \\
Median&  71 &  72 &  71 &  71 &  72  & 71  & 71   &72    \\
Mean &   71.16& 71.63 &71.20 &71.15& 71.65 &71.07& 71.23& 71.68  \\
3$^\text{rd}$ Quartile & 75& 76 & 75 &75 &76 &75 &75& 76      \\
Maximum &  86  & 86 &  86   &86 &  86  & 86&   86&   85 \\\hline
Std. Deviation &6.31& 6.11 &6.30& 6.24& 6.08& 6.28 &6.30 &6.07
\end{tabular}
\end{center}

The summary of the $\natural$-CM multivariate score data is
\begin{center}
\begin{tabular}{r|c|c|c|c|c|c|c|c}
${^\natural}S_{\text{CM}}^\kappa$  Summary&  $\kappa =1 $ & $\kappa =2 $ & $\kappa =3 $ & $\kappa = 4$ & $\kappa =5 $ & $\kappa =6 $ & $\kappa = 7$ & $\kappa = 8$ \\ \hline
Minimum &   50  & 48   &50  & 50 &  48  & 49  & 49   &48  \\
1$^\text{st}$ Quartile &67   &   68  &    66 &     67   &   68  &    67  &    67  &    68  \\
Median&  71 &  72 &  70   &71 &  72  & 71  & 71  & 72   \\
Mean &   71.10 &71.59 &69.80 &71.09 &71.63& 70.99 &71.14 &71.60  \\
3$^\text{rd}$ Quartile & 75  &    76     & 74    &  75 &     76   &   75      &75   &   76      \\
Maximum & 86  & 86  & 86 &  86  & 86  & 86  & 86  & 86   \\\hline
Std. Deviation &6.25 &6.10& 6.49 &6.21& 6.08& 6.25& 6.27& 6.10 
\end{tabular}
\end{center}

\begin{figure}[htbp]
\begin{center}
\includegraphics[scale=0.45]{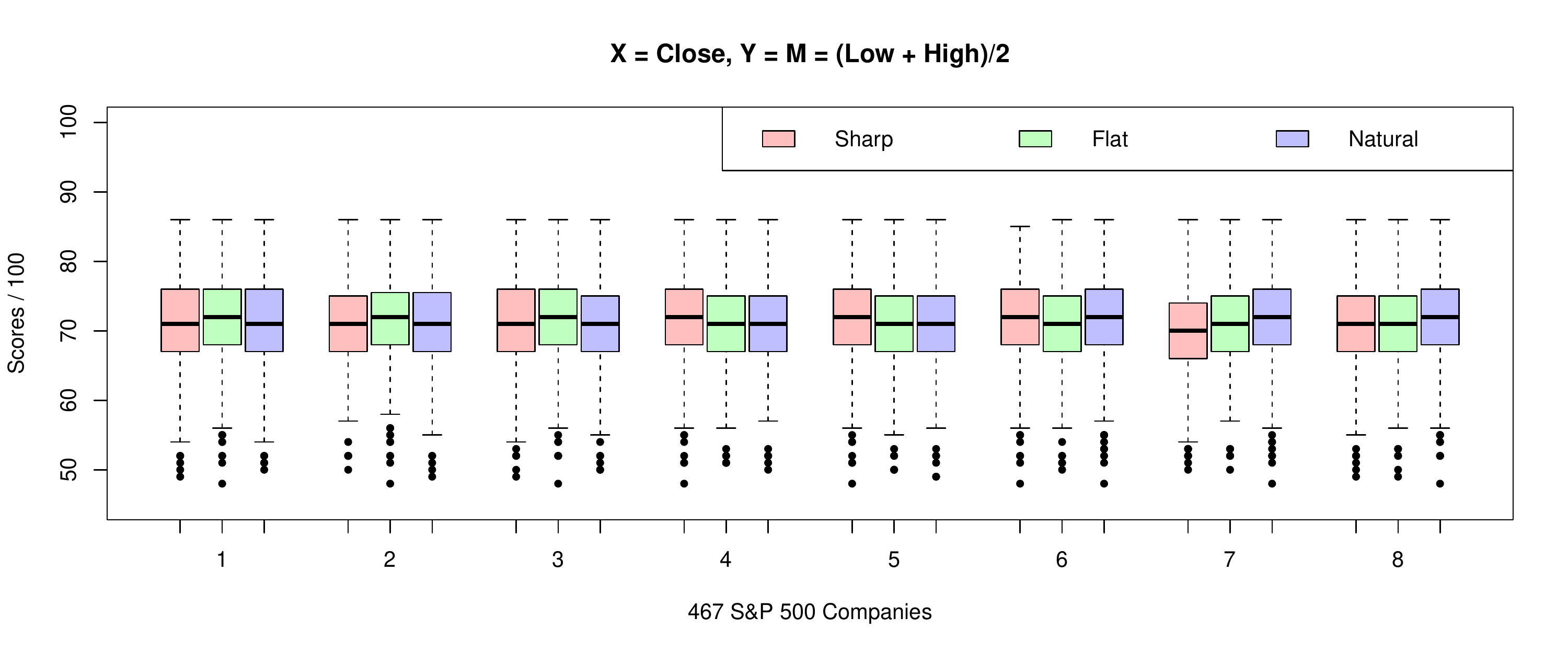}
 \caption{Boxplots of the scores  ${^\sharp}S^\kappa_{CM}$,  ${^\flat}S^\kappa_{CM}$, and ${^\natural}S^\kappa_{CM}$ }
\label{s1b5boxplots}
\end{center}
\end{figure}

From the statistic summary of these nine multivariate score data ${^\sharp}S_{\text{CC}}, {^\flat}S_{\text{CC}}, {^\sharp}S_{\text{CO}}, {^\flat}S_{\text{CO}}, {^\sharp}S_{\text{CM}}$ and ${^\flat}S_{\text{CM}}$, one can see that: the minimum scores varies from 46 to 51, the first quartile scores are between 66 and 69, the median scores are either 71 or 72, the average scores are between 69.74 and 71.68, with optimization  criteria $\kappa=2, 5$ and $8$ being consistently the highest on average,  the third quartile scores are between 74 and 76, and the maximum scores varies from 85 to 89.\\

 For a given  company and optimization criteria $\kappa$, let's denote by $S^\kappa_{\text{max}}$ the maximum score among the nine scores, i.e., we choose the best of the nine methods for each company per optimization criteria. The maximum score $S^\kappa_{\text{max}}$ is then defined as follows: 
\begin{equation}
S^\kappa_{\text{max}}= \max \Big({^\sharp}S^\kappa_{\text{CC}}, {^\flat}S^\kappa_{\text{CC}}, {^\natural}S^\kappa_{\text{CC}}, {^\sharp}S^\kappa_{\text{CO}}, {^\flat}S^\kappa_{\text{CO}}, {^\natural}S^\kappa_{\text{CO}} {^\sharp}S^\kappa_{\text{CM}}, {^\flat}S^\kappa_{\text{CM}}, {^\natural}S^\kappa_{\text{CM}}\Big)
\end{equation}
and hence
\begin{equation}
S_{\text{max}}= (S^1_{\text{max}}, S^2_{\text{max}}, S^3_{\text{max}}, S^4_{\text{max}}, S^5_{\text{max}}, S^6_{\text{max}}, S^7_{\text{max}}, S^8_{\text{max}})
\end{equation}
We have then a tenth multivariate score data $S_{\text{max}}$ with $467$ observations (companies) and 8 variables (optimization criteria $\kappa$). The summary of the $S_{\text{max}}$ multivariate score data is
\begin{center}
\begin{tabular}{r|c|c|c|c|c|c|c|c}
$S_{\text{max}}^\kappa$  Summary&  $\kappa =1 $ & $\kappa =2 $ & $\kappa =3 $ & $\kappa = 4$ & $\kappa =5 $ & $\kappa =6 $ & $\kappa = 7$ & $\kappa = 8$ \\ \hline
Minimum &    52 &  53 &  53 &  52 &  52 &  52  & 52  & 52   \\
1$^\text{st}$ Quartile & 70 &   71  & 70   & 70  &    70   &   70   & 70    &70     \\
Median&   73   &  74  &   74     &74    & 74&     73   &  74  &   74 \\
Mean & 73.37& 73.82 &73.87 &73.38& 73.83 &73.33 &73.44& 73.84    \\
3$^\text{rd}$ Quartile & 77&    77&    78&    77&    77.5  &  77&    77&  78   \\
Maximum &  87  & 89 &  87  & 87  & 88  & 87 &  87  & 88  \\
Std. Deviation &5.77& 5.56 &5.72 &5.73 &5.54 &5.80 &5.54& 5.59
\end{tabular}
\end{center}
Finally, for each company, one can choose not only the best of the nine methods, but also the best optimization criteria, i.e., the best across 72 possible models. The optimal score for a given company is then defined as follows:
\begin{equation}
S_{\text{max}}^\ast = \max_{\kappa=1, \dots, 8} (S_{\text{max}}^\kappa)
\end{equation}
The summary of the $S_{\text{max}}^\ast$ univariate score data is
\begin{equation}
(\text{Min}, Q_1, \text{Median}, \text{Mean}, Q_3, \text{Max}, \text{Std. Deviation}) = (53,    72.50,    76  , 75.34,    79  ,89)
\end{equation}
Note that all $\sharp$-, $\flat$- and $\natural$-scores for $(X, Y) = (C, C), (C, O)$ and $(C,M)$ are normally distributed.

\begin{figure}[htbp]
\begin{center}
\includegraphics[scale=0.45]{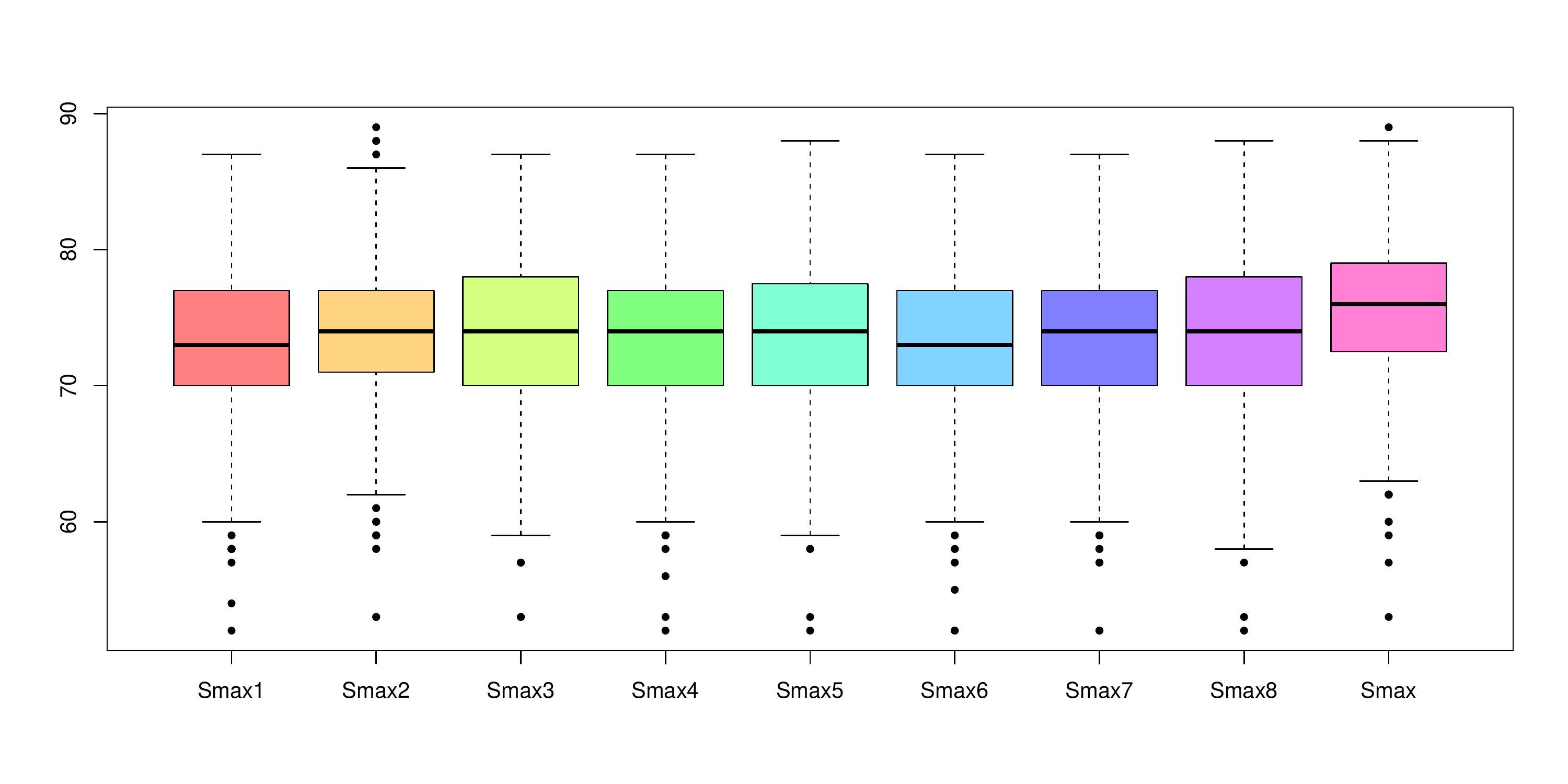}
 \caption{Boxplots of the scores  $S_{\text{max}}^\kappa$ and $S^\ast_{\text{max}}$ }
\label{s1b5boxplots}
\end{center}
\end{figure}

In order to compare the optimization criteria, we define the following random variables: For any given a company,
\begin{equation}
^\sharp g^\kappa =  \left\{\begin{array}{cl}1 & \text{ if } \,  {^\sharp}S^\kappa = {^\sharp S^\ast} \ \\0 & \text{otherwise}\end{array}\right. \hspace{-4mm}; \quad     {^\flat}g^\kappa =  \left\{\begin{array}{cl}1 & \text{ if } \, {^\flat}S^\kappa = {^\sharp S^\ast} \ \\0 & \text{otherwise}\end{array}\right.\hspace{-4mm}; \quad {^\natural}g^\kappa =  \left\{\begin{array}{cl}1 & \text{ if } \,  {^\natural}S^\kappa = {^\sharp S^\ast} \ \\0 & \text{otherwise}\end{array}\right.
\end{equation}
The sum of $^\sharp g^\kappa$ (resp., $^\flat g^\kappa$, $^\natural g^\kappa$) across the 467 companies lead to a number of times the criteria $\kappa$ led to the optimal forecasts, and is defined as
\begin{equation}
^\sharp G^\kappa =  \sum_{\text{all companies}} {^\sharp g^\kappa};  \quad  ^\flat G^\kappa =  \sum_{\text{all companies}} {^\flat g^\kappa};  \quad  ^\natural G^\kappa =  \sum_{\text{all companies}} {^\natural g^\kappa}
\end{equation}
Therefore, 
\begin{center}
\begin{tabular}{c|c|c|c|c|c|c|c|c}
&  $\kappa =1 $ & $\kappa =2 $ & $\kappa =3 $ & $\kappa = 4$ & $\kappa =5 $ & $\kappa =6 $ & $\kappa = 7$ & $\kappa = 8$ \\ \hline
$^\sharp G^\kappa_\text{CC}$ & 160 &175 &132& 141 &193& 153 &130& 194\\
$^\flat G^\kappa_\text{CC}$ & 148 & 170 & 125 & 135 &173 & 146 & 128 & 179 \\
$^\natural G^\kappa_\text{CC}$ & 115 & 151 & 139 & 98 & 154 & 108 & 94 & 156 \\  \hline

$^\sharp G^\kappa_\text{CO}$ & 128 & 196 & 122 & 140 & 199 & 138 & 119 & 200\\
$^\flat G^\kappa_\text{CO}$ & 123 & 193 & 118 & 141 & 183 & 133 & 110 & 183  \\
$^\natural G^\kappa_\text{CO}$ &  110 & 169 & 136 & 108 & 167 & 106 & 90 & 159 \\  \hline

$^\sharp G^\kappa_\text{CM}$ & 144 & 182 & 120 & 127 & 190 & 125 & 134 & 178\\
$^\flat G^\kappa_\text{CM}$ & 134 & 184 & 123 & 124 & 190 & 131 & 133 & 190 \\
$^\natural G^\kappa_\text{CM}$ & 107 & 157 & 118 & 94 & 166 & 90 & 107 & 152 \\  
\end{tabular}
\end{center}
One can notice that the optimization criteria $\kappa = 2, 5$ and $8$ lead consistently to the highest number of optimal forecasts. \\

Let's now perform a principal component analysis on the nine  $\sharp$-, $\flat$- and $\natural$-multivariate score data with $X = C$ and $Y = C, O$ and $M$.  Table \ref{PC1} contains the loadings of the first principal component on each of the nine principal component analyses.  The first principal component loadings are, roughly speaking, just the sum of all eight scores. A company with a positive first principal component score has consistent scores above average across all eight optimization criteria. A company with a negative first principal component scores means that the stock has consistent scores below average across all eight optimization criteria. 

\begin{table}[htp]
\begin{center}
\begin{tabular}{c|r|r|r|r|r|r|r|r|r}
PC.1 & $\sharp$-CC & $\flat$-CC & $\natural$-CC & $\sharp$-CO & $\flat$-CO & $\natural$-CO &$\sharp$-CM & $\flat$-CM & $\natural$-CM \\ \hline
$S^1$ &0.371 &0.371 &0.373 &0.358 &0.363 &0.360& 0.363& 0.363& 0.366\\
$S^2$ &0.330 &0.327 &0.334 &0.342 &0.341 &0.344 &0.343 &0.340 &0.347\\
$S^3$ &0.365 &0.366 &0.340 &0.365 &0.364 &0.342 &0.366 &0.365 &0.327\\
$S^4$ &0.367 &0.368 &0.371 &0.357 &0.356 &0.362 &0.358 &0.360 &0.364\\
$S^5$&0.328 &0.329 &0.334 &0.339 &0.340 &0.342 &0.338 &0.338 &0.345\\
$S^6$&0.366 &0.370 &0.371 &0.361 &0.359 &0.365 &0.359 &0.360 &0.365\\
$S^7$ &0.366 &0.365 &0.366 &0.362 &0.361 &0.366 &0.362 &0.363 &0.368\\
$S^8$&0.331 &0.329 &0.336 &0.343 &0.342 &0.346 &0.338 &0.338 &0.345\\
\end{tabular}
\end{center}
\caption{First principal component loadings of the nine PCAs}
\label{PC1}
\end{table}%

Table \ref{PC2} contains the loadings of the second principal component on each of the nine principal component analyses.  The second principal component loadings are positive for the optimization criteria $\kappa = 1, 3, 4, 6$ and $7$, and are negative for the optimization criteria $\kappa = 2, 5, 8$. Therefore, a company with a second principal component scores near zero is a company that has a consistent scores across all optimization criteria (equally above average, equally average, or equally  below average). A company with a positive second principal component score is a company that has higher scores with optimization criteria $\kappa = 1, 3, 4, 6, 7$ then for $\kappa = 2, 5, 8$.    A company with a negative second principal component score is a company that has lower scores with optimization criteria $\kappa = 1, 3, 4, 6, 7$ then for $\kappa = 2, 5, 8$.

\begin{table}[htp]
\begin{center}
\begin{tabular}{c|r|r|r|r|r|r|r|r|r}
PC.2 & $\sharp$-CC & $\flat$-CC & $\natural$-CC & $\sharp$-CO & $\flat$-CO & $\natural$-CO &$\sharp$-CM & $\flat$-CM & $\natural$-CM \\ \hline
$S^1$ & 0.263  &0.247  &0.164  &0.241 & 0.217 & 0.160 & 0.250 & 0.219  &0.156\\
$S^2$ &-0.472 &-0.477 &-0.446 &-0.466 &-0.463 &-0.447 &-0.458& -0.465 &-0.432\\
$S^3$ &0.263  &0.239  &0.540  &0.260  &0.235  &0.480&  0.249 & 0.230 & 0.508\\
$S^4$  &0.258  &0.270  &0.184  &0.280  &0.302  &0.230&  0.278 & 0.288  &0.212\\
$S^5$ &-0.476 &-0.478 &-0.446 &-0.470 &-0.469 &-0.449 &-0.475 &-0.467 &-0.444\\
$S^6$  &0.234  &0.261  &0.178  &0.267  &0.295  &0.223&  0.250&  0.286  &0.213\\
$S^7$  &0.259  &0.255  &0.176  &0.275  &0.270  &0.205&  0.287 & 0.286 & 0.213\\
$S^8$ &-0.474 &-0.470 &-0.433 &-0.459 &-0.461 &-0.445 &-0.467 &-0.467 &-0.445
\end{tabular}
\end{center}
\caption{Second principal component loadings of the nine PCAs}
\label{PC2}
\end{table}%

\begin{figure}[htbp]
\begin{center}
\begin{tabular}{ccc}
\includegraphics[scale=.3]{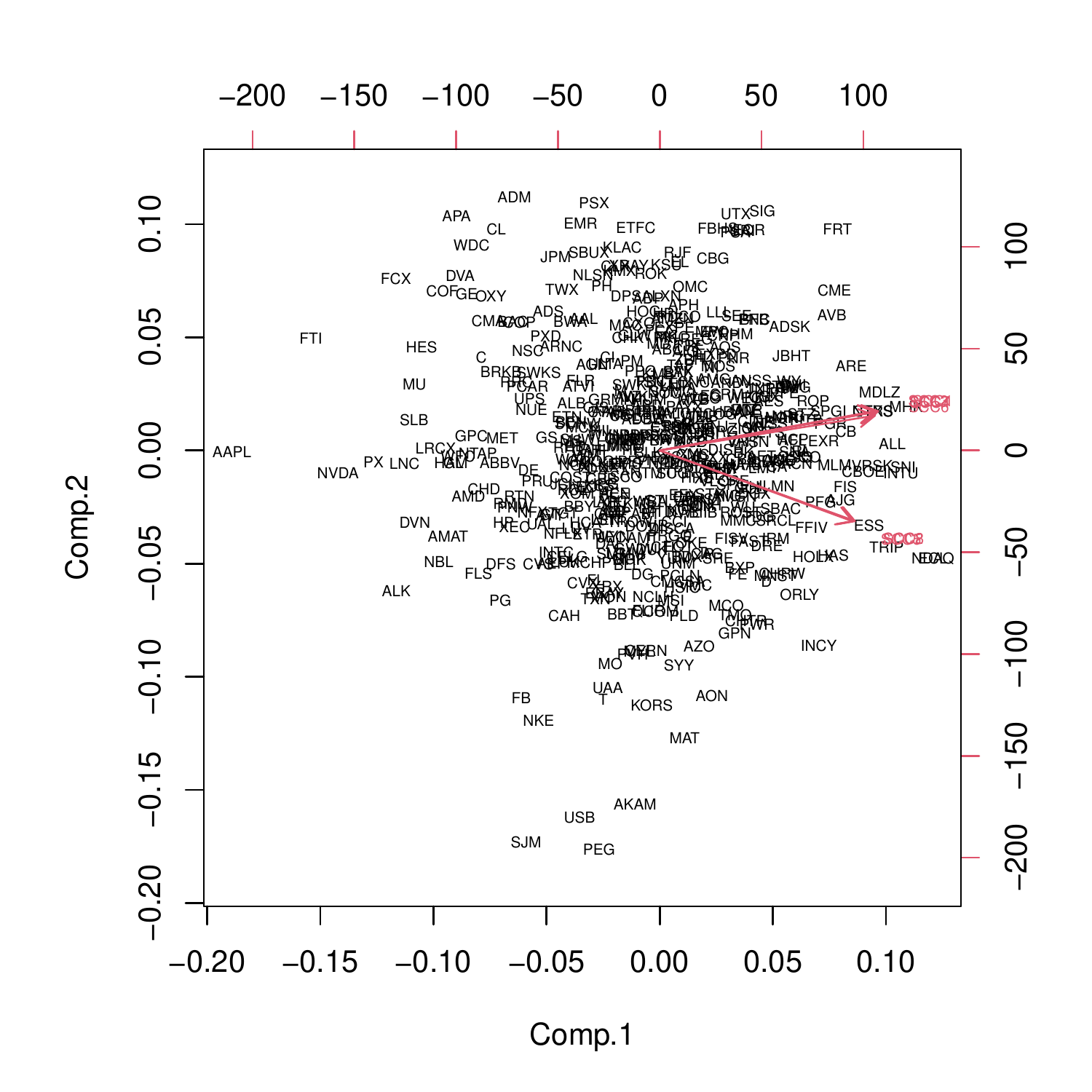}  & \includegraphics[scale=.3]{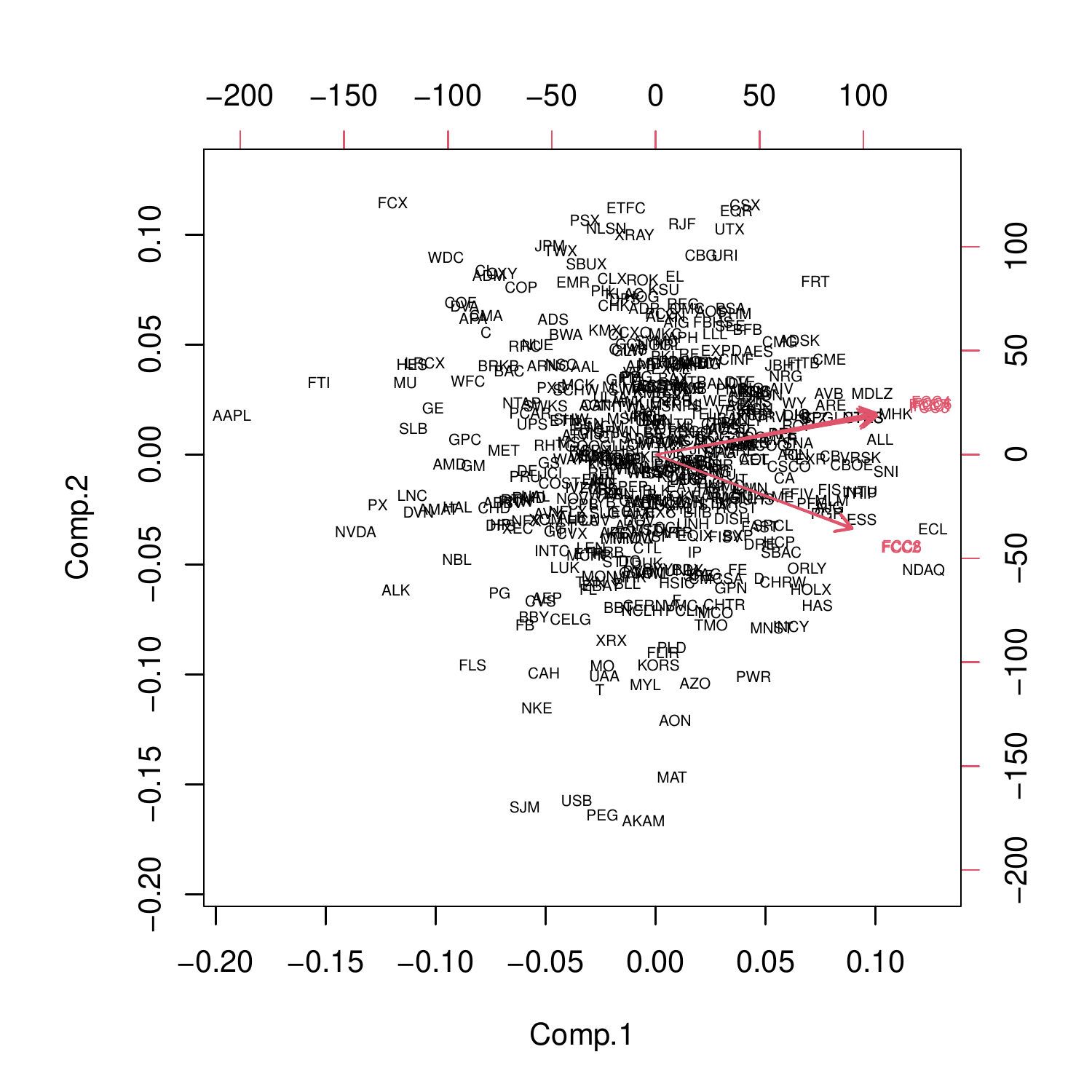} &\includegraphics[scale=.3]{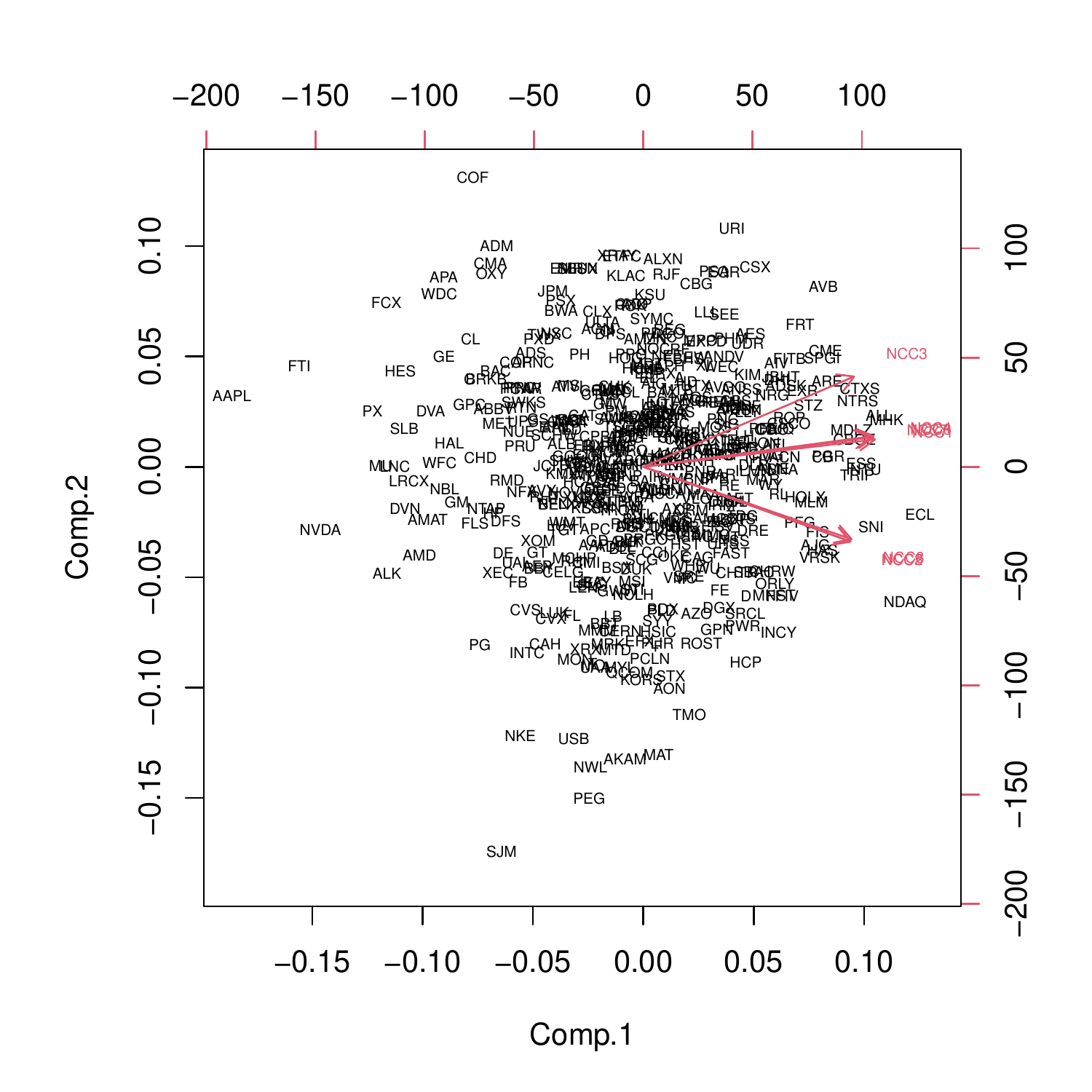}\\
$\sharp$-CC & $\flat$-CC & $\natural$-CC\\
\includegraphics[scale=.3]{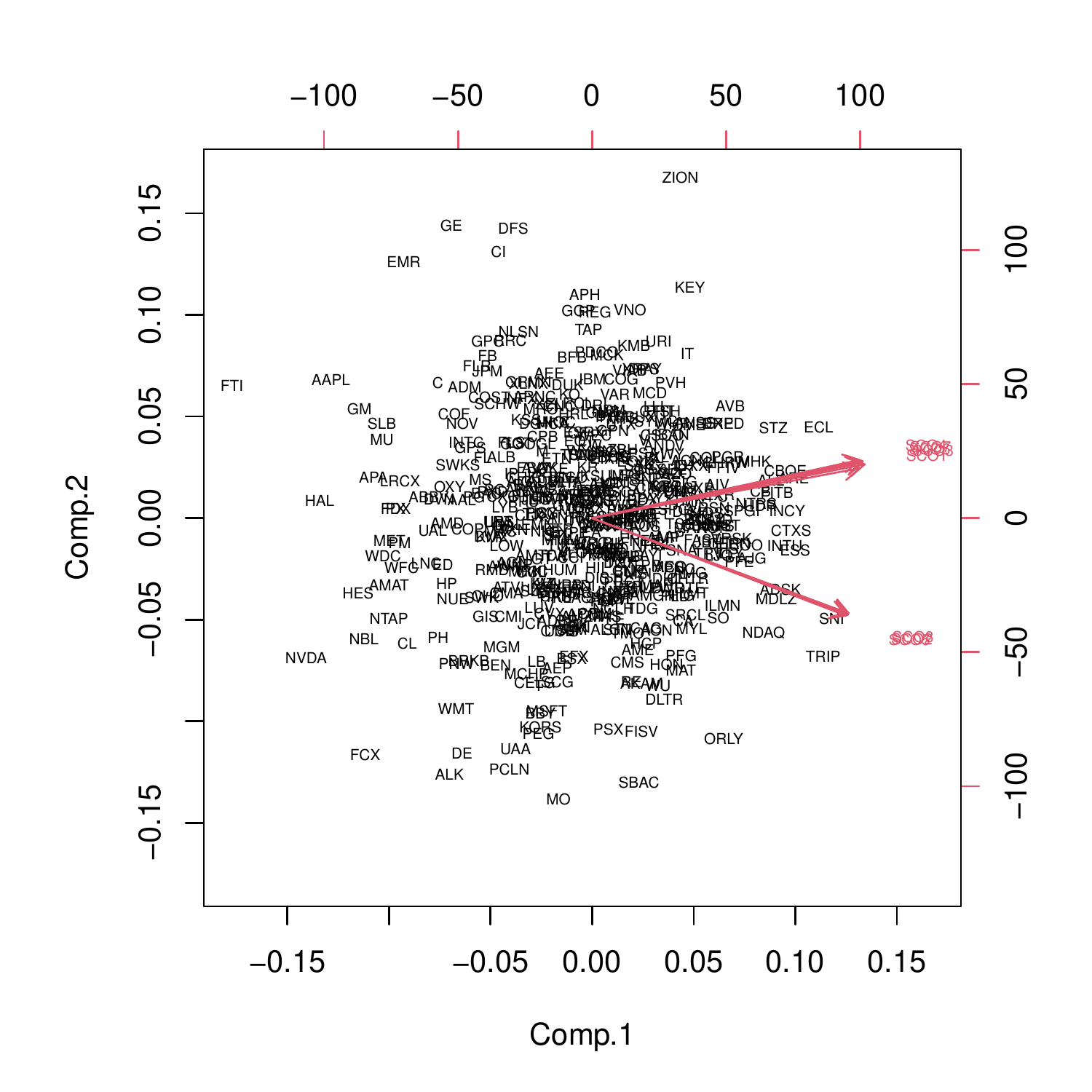}  & \includegraphics[scale=.3]{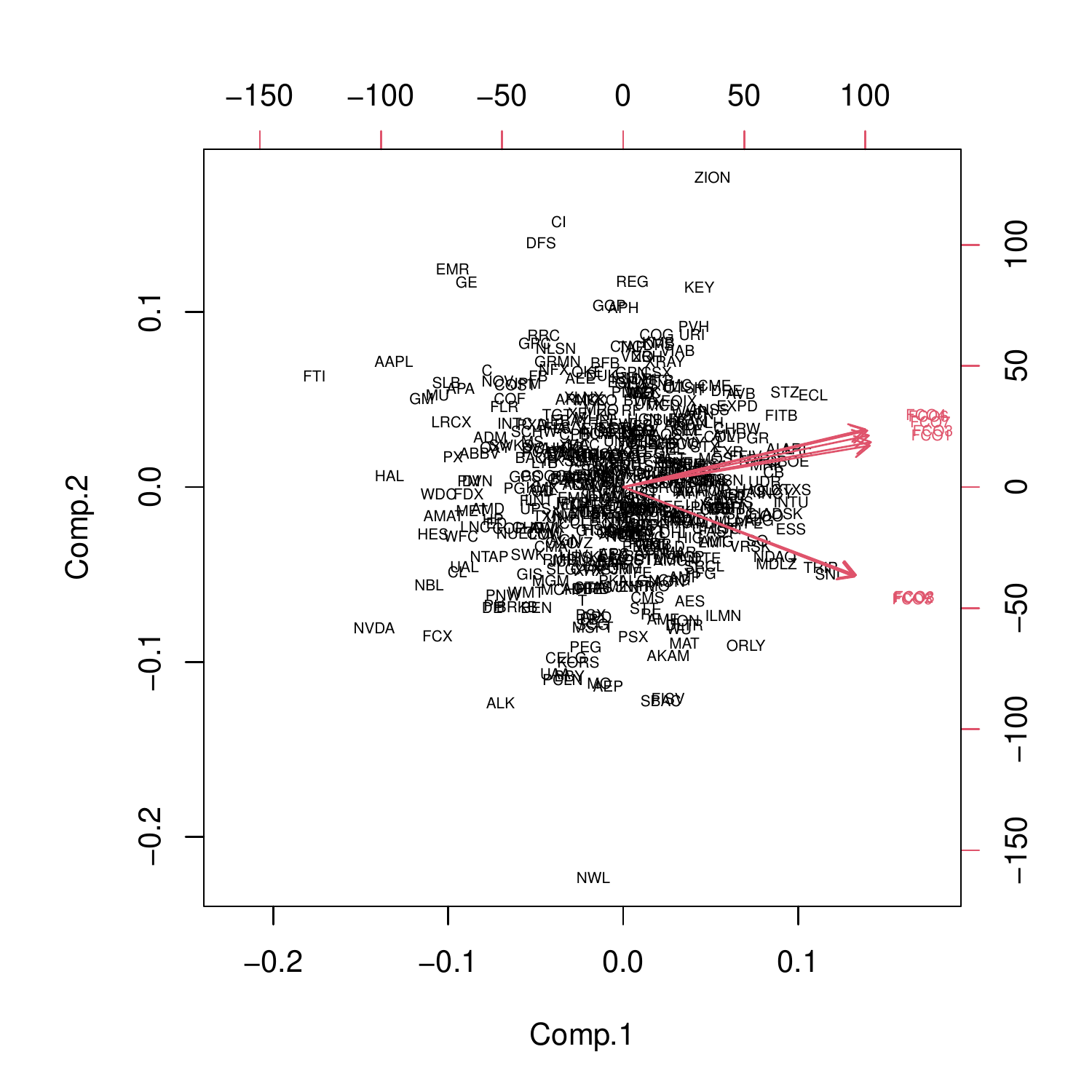} &\includegraphics[scale=.3]{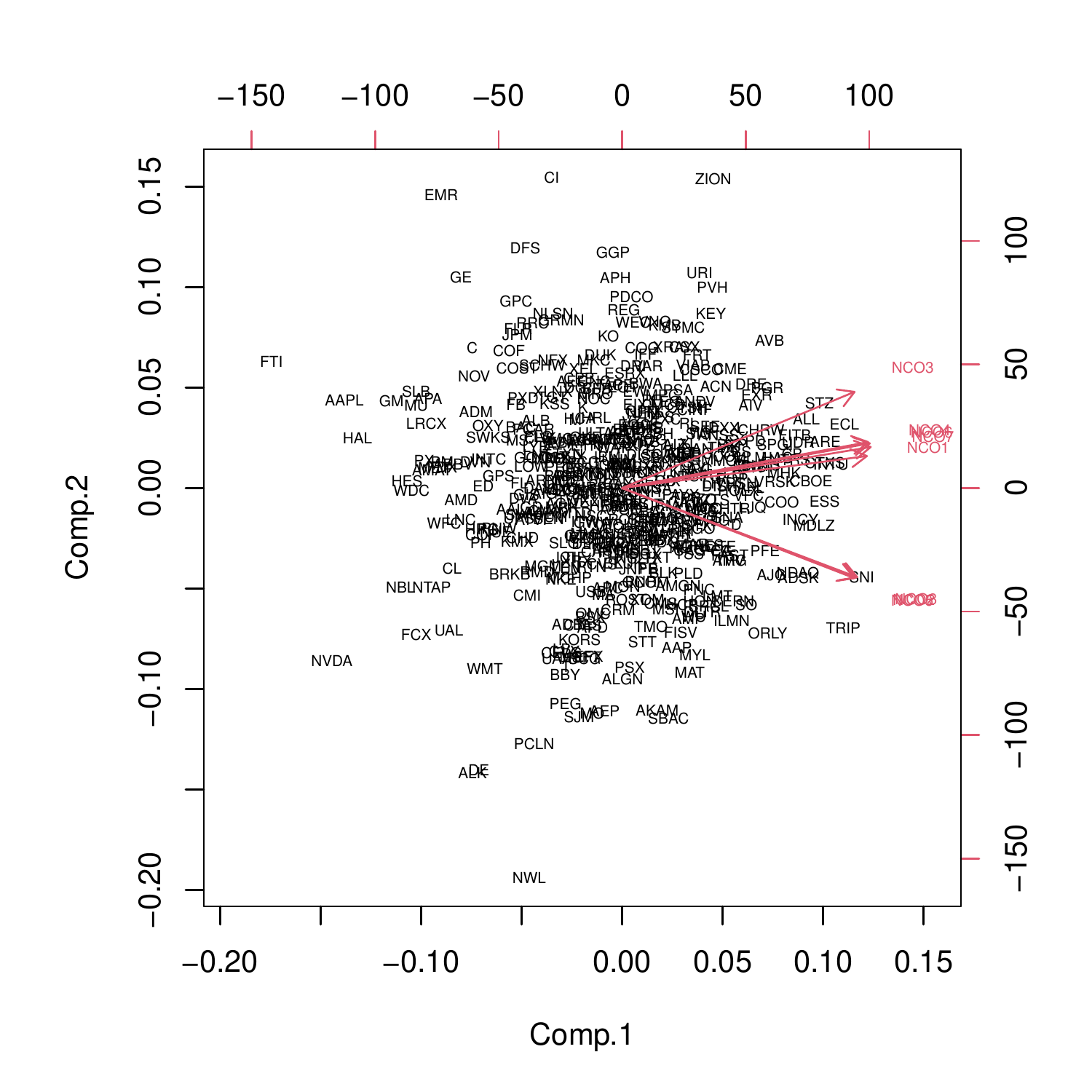}\\
$\sharp$-CO & $\flat$-CO & $\natural$-CO\\
\includegraphics[scale=.3]{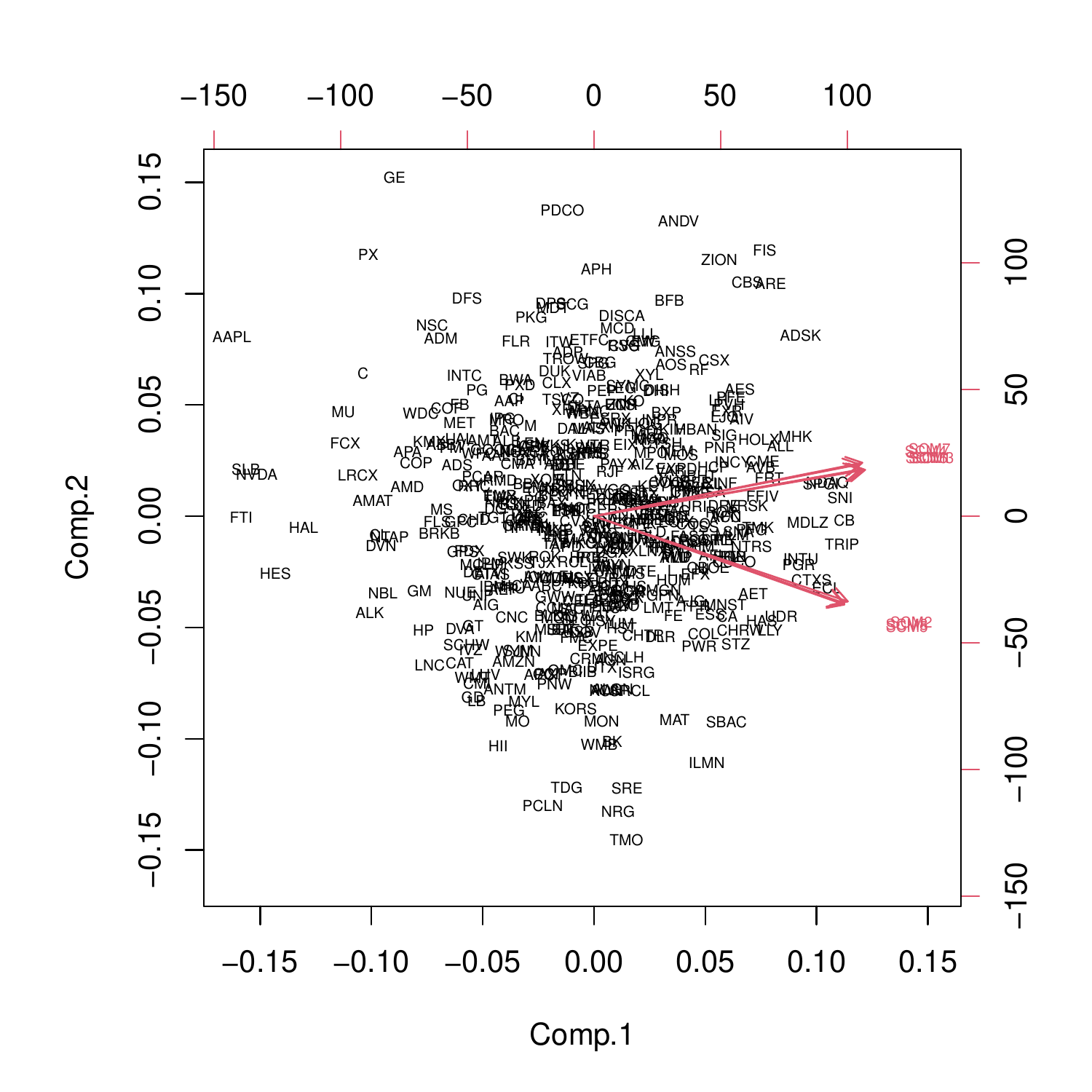}  & \includegraphics[scale=.3]{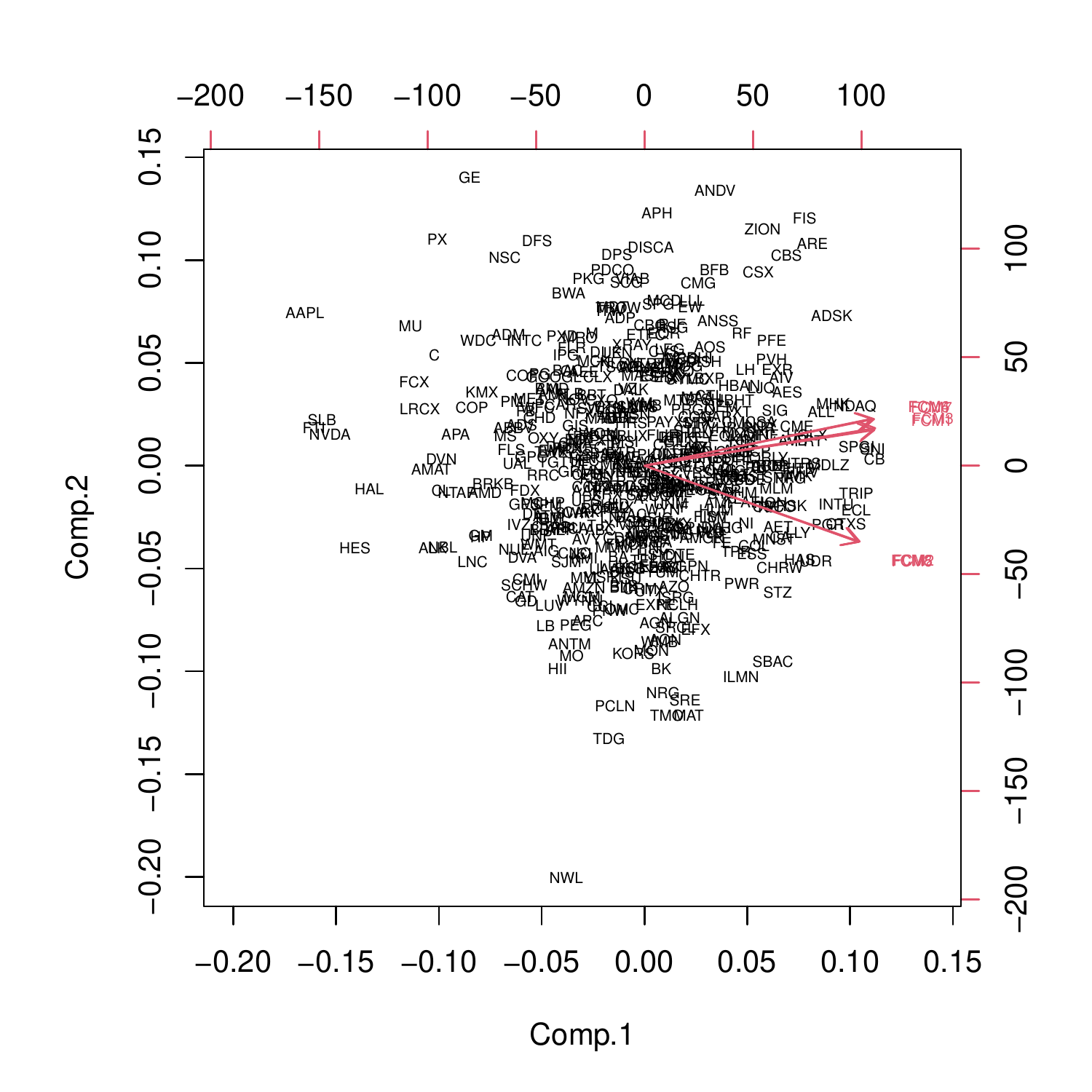} &\includegraphics[scale=.3]{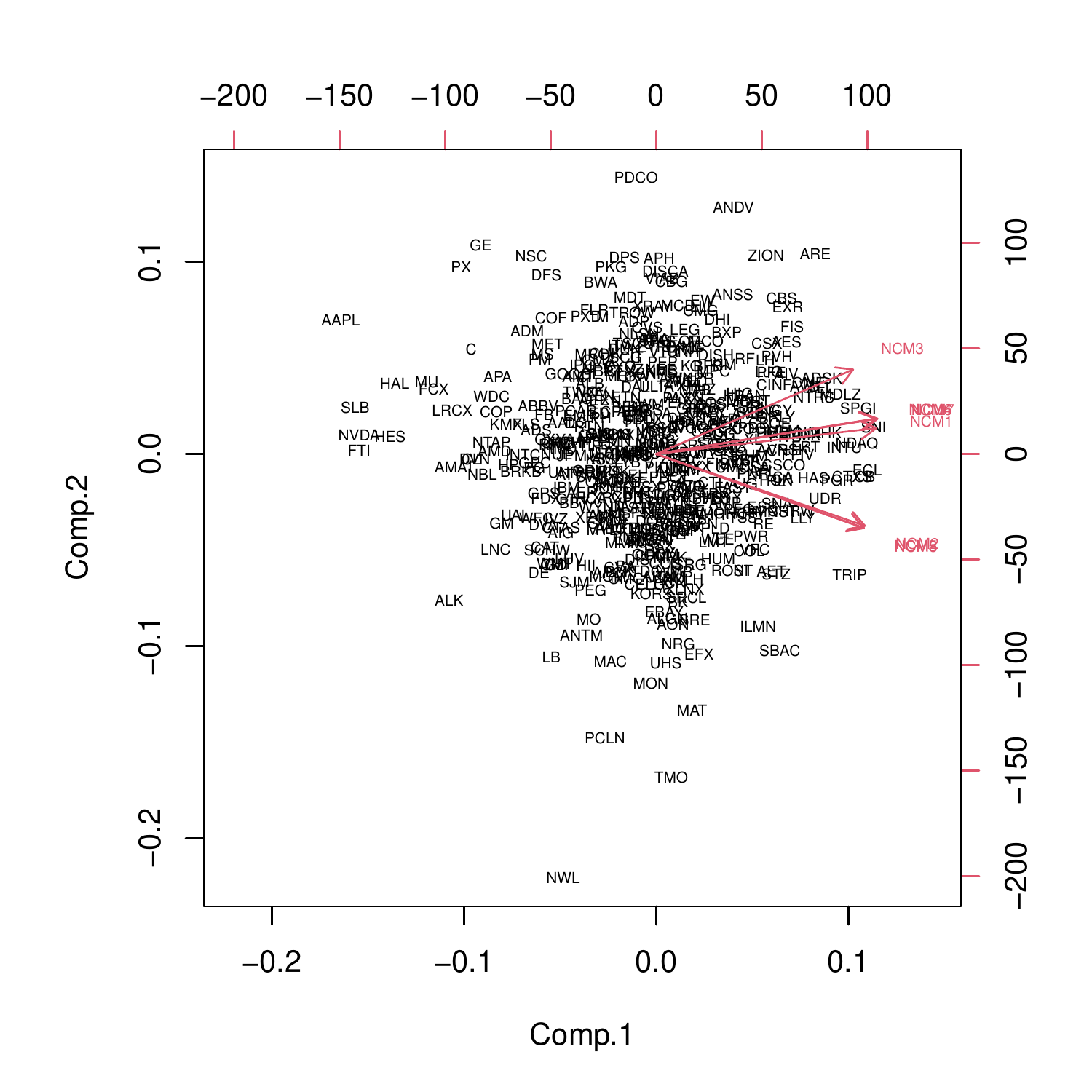}\\
$\sharp$-CM & $\flat$-CM & $\natural$-CM
\end{tabular}
 \caption{PCA Biplots}
\label{biplots}
\end{center}
\end{figure}

Finally, Figure \ref{biplots} contains the biplots of the nine principal component analyses. 

\begin{table}[htp]
\begin{center}
\begin{tabular}{r|c|c|c|c|c|c|c|c|c}
$(X, Y)$& $(C, C)$ & $(O, O)$& $(M, M)$ & $(C, O)$ & $(C, M)$&$(O, C)$&$(O, M)$ &$(M, C)$& $(M, O)$ \\ \hline
ADI&	75	&46	&68	&70	&75	&68	&55	&67	&63\\
ADP	&72	&53	&62	&71	&72	&65	&47	&62	&59\\
AES	&78	&55	&59	&75	&81	&69	&51	&54	&59\\
AMZN&73	&38	&53	&69	&66	&50	&41	&56	&55\\
ARE	&84	&49	&64	&83	&84	&71	&50	&61	&62\\
BAX	&73	&46	&58	&70	&69	&63	&59	&59	&59\\
BSX	&73	&48	&60	&67	&74	&68	&56	&67	&57\\
CI	&69	&44	&62	&69	&68	&66	&52	&53	&60\\
DIS&	73	&46	&57	&71	&68	&57	&45	&62	&56\\
EBAY&65	&50	&64	&68	&71	&56	&45	&64	&63\\
GD	&70	&43	&57	&65	&65	&53	&57	&64	&60\\
HAL	&58	&43	&49	&56	&52	&59	&50	&50	&56\\
HUM	&71	&50	&66	&69	&74	&67	&56	&72	&66\\
KSS	&64	&44	&61	&69	&66	&60	&49	&66	&64\\
MAC&70	&50	&58	&64	&64	&69	&48	&60	&60\\
MO	&65	&43	&56	&64	&64	&69	&47	&54	&51\\
MSI	&71	&42	&57	&73	&70	&60	&47	&53	&55\\
PFE	&79	&47	&57	&79	&80	&71	&52	&65	&59\\
PKG&73	&47	&53	&70	&69	&69	&47	&56	&59\\
PLD	&72	&46	&63	&75	&76	&69	&46	&54	&61\\
PSA	&77	&39	&58	&73	&76	&59	&45	&58	&56\\
PSX	&71	&48	&60	&71	&70	&69	&52	&63	&63\\
SPGI&81	&47	&63	&81	&85	&77	&47	&64	&62\\
TEL&74	&53	&64	&73	&70	&57	&52	&62	&68\\
TRV&79	&53	&54	&78	&75	&71	&48	&58	&59\\
TXN&66	&44	&69	&66	&67	&59	&43	&70	&64\\
VZ	&70	&42	&53	&73	&71	&65	&46	&57	&54\\
WFC	&60	&46	&59	&57	&65	&64	&49	&60	&59\\
XL&	71	&52	&62	&76	&79	&63	&53	&61	&64\\
XRAY&70	&39	&55	&76	&71	&64	&47	&52	&50
\end{tabular}
\end{center}
\caption{$\sharp$-1 multivariate scores of 30 random stocks}
\label{Scores30rs1}
\end{table}%

\begin{table}[htp]
\begin{center}
\begin{tabular}{r|c|c|c|c|c|c|c|c|c}
$(X, Y)$& $(C, C)$ & $(O, O)$& $(M, M)$ & $(C, O)$ & $(C, M)$&$(O, C)$&$(O, M)$ &$(M, C)$& $(M, O)$ \\ \hline
ADI&	75&	46	&66	&72	&75&	68&	49	&67&	66\\
ADP&	68	&55&	63&	72	&67	&63	&50&	63&	59\\
AES	&76	&56&	59	&77	&77	&72	&55&	53	&57\\
AMZN	&69&	38&	50	&73	&70	&54&	45&	55	&49\\
ARE	&80	&44	&61&	82	&77&	71	&48	&60	&63\\
BAX	&71	&43&	59	&71&	69&	63	&55	&53&	57\\
BSX	&72	&56	&64	&73	&73	&69&	55&	69	&65\\
CI&	67&	44	&57&	60&	65&	67	&53	&60&	60\\
DIS	&75&	43&	57	&73	&72	&57&	48&	61&	61\\
EBAY&	71	&53	&64	&67&	75&	63&	44	&63&	65\\
GD	&70	&43	&53	&66&	68&	61	&57&	61	&60\\
HAL	&61&	49&	44&	56	&56	&51	&50&	49&	48\\
HUM	&70	&50&	61&	71&	77	&70	&52	&65&	66\\
KSS	&70&	45&	64&	66&	69	&61	&46&	60&	64\\
MAC	&67	&55	&60	&67&	73&	69	&48&	62&	62\\
MO	&73	&43&	54	&76	&71	&69	&47&	57&	57\\
MSI	&75	&39	&52	&76	&70&	60&	47&	57	&52\\
PFE	&77	&48	&60	&81&	77	&73&	50&	65	&62\\
PKG	&73&	50&	59	&70&	65&	69	&47	&59	&59\\
PLD	&76	&44&	63&	78&	77	&69	&46&	57&	61\\
PSA	&71	&41&	58	&73	&78	&69&	46&	60&	58\\
PSX	&63	&51&	62&	77	&72	&69	&50	&60	&65\\
SPGI&	80&	46&	64	&80&	84&	77	&47&	58	&64\\
TEL	&73	&52	&65	&76	&74&	56&	51	&63	&66\\
TRV	&77	&54	&58	&79	&76&	72&	50&	61	&63\\
TXN	&71	&44	&70	&67	&68	&58&	48&	71	&70\\
VZ	&69	&41&	53&	70&	67	&65&	41&	58&	48\\
WFC	&61	&47&	61&	62	&64	&64&	48&	62&	61\\
XL	&74	&55&	59	&74&	78&	63	&51	&51	&61\\
XRAY&	66	&41	&61&	71	&67	&64	&47	&62	&51 
\end{tabular}
\end{center}
\label{Scores30rb5}
\caption{$\flat$-5 multivariate scores of 30 random stocks}
\end{table}%




\end{document}